NASA Innovative Advanced Concepts (NIAC) Phase II
80HQTR18NOA01 – 18NIAC-A2

# DIRECT MULTIPIXEL IMAGING AND SPECTROSCOPY OF AN EXOPLANET WITH A SOLAR GRAVITY LENS MISSION

FINAL REPORT

Slava G. Turyshev[1], Michael Shao[1], Viktor T. Toth[2], Louis D. Friedman[3],
Leon Alkalai[1], Dmitri Mawet[4], Janice Shen[1], Mark R. Swain[1], Hanying Zhou[1],
Henry Helvajian[5], Tom Heinsheimer[5], Siegfried Janson[5], Zigmond Leszczynski[5], John McVey[5],
Darren Garber[6], Artur Davoyan[7], Seth Redfield[8], and Jared R. Males[9]

[1]*Jet Propulsion Laboratory, California Institute of Technology,
4800 Oak Grove Drive, Pasadena, CA 91109-0899, USA*
[2]*Ottawa, Ontario K1N 9H5, Canada*
[3]*The Planetary Society, Pasadena, CA 91101 USA*
[4]*California Institute of Technology, Pasadena, CA 91125, USA*
[5]*The Aerospace Corporation, El Segundo, CA*
[6]*NXTRAC Inc., Redondo Beach, CA 90277*
[7]*Department of Mechanical and Aerospace Engineering, University of California, Los Angeles*
[8]*Wesleyan University, 45 Wyllys Ave, Middletown, CT 06459, USA*
[9]*Department of Astronomy & Stewart Observatory, University of Arizona, Tucson AZ 8572, USA*

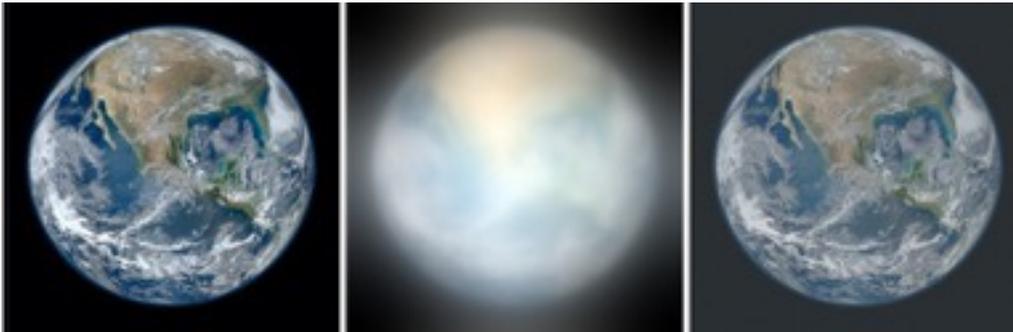

A meter-class telescope with a coronagraph to block solar light, placed in the strong interference region of the solar gravitational lens (SGL), is capable of imaging an exoplanet at a distance of up to 30 parsecs with a few 10 km-scale resolution on its surface. The picture shows results of a simulation of the effects of the SGL on an Earth-like exoplanet image. Left: original RGB color image with a (1024×1024) pixels; center: image blurred by the SGL, sampled at an SNR of ~$10^3$ per color channel, or overall SNR of $3\times10^3$; right: the result of image deconvolution.

March 18, 2020

**Executive Summary: Innovations and Advanced Concepts Enabled**

Direct multipixel imaging of exoplanets requires significant light amplification and very high angular resolution. With optical telescopes and interferometers, we face the sobering reality: i) to capture even a single-pixel image of an "Earth 2.0" at 30 parsec (pc), a ~90 kilometer (km) telescope aperture is needed (for the wavelength of $\lambda = 1$ μm); ii) interferometers with telescopes (~30 m) and baselines (~1 km) will require integration times of ~$10^5$ years to achieve a signal-to-noise ratio, SNR=7 against the exozodiacal background. These scenarios involving the classical optical instruments are impractical, giving us no hope to spatially resolve and characterize exolife features.

To overcome these challenges, in our NIAC Phase II study we examined the solar gravitational lens (SGL) as the means to produce *direct* high-resolution, multipixel images of exoplanets. The SGL results from the diffraction of light by the solar gravitational field, which acts as a lens by focusing incident light at distances >548 AU behind the sun (Figure 1). The properties of the SGL are remarkable: it offers maximum light amplification of ~$10^{11}$ and angular resolution of ~$10^{-10}$ arcsec, for $\lambda = 1$ μm. A probe with a 1-m telescope in the SGL focal region (SGLF), namely, in its strong interference region, can build an image of an exoplanet at 30 pc with 10-km scale resolution of its surface, which is not possible with any known classical optical instruments. This resolution is sufficient to observe seasonal changes, oceans, continents and surface topography.

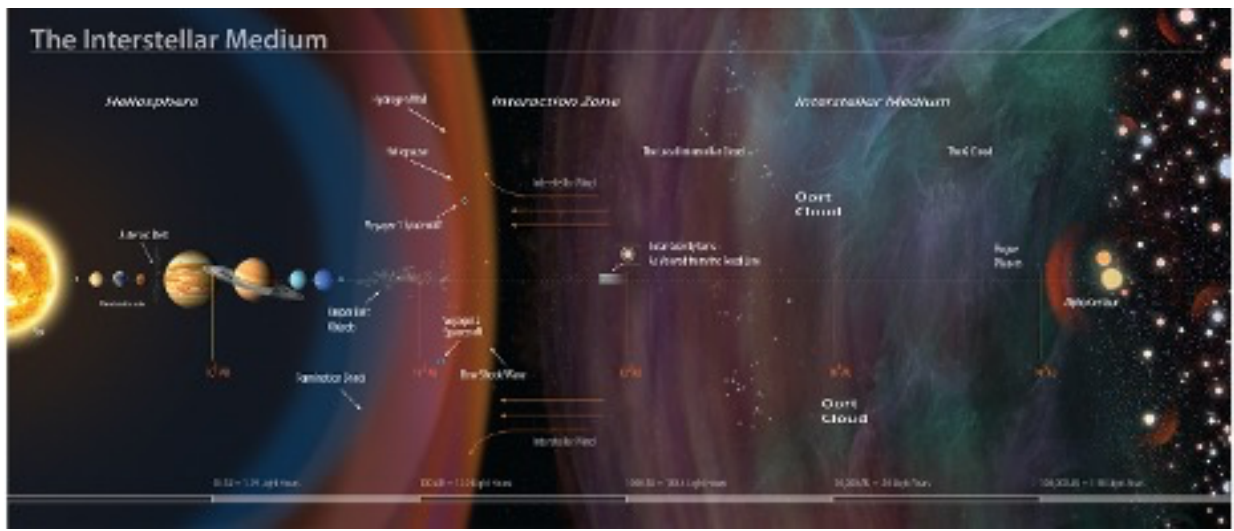

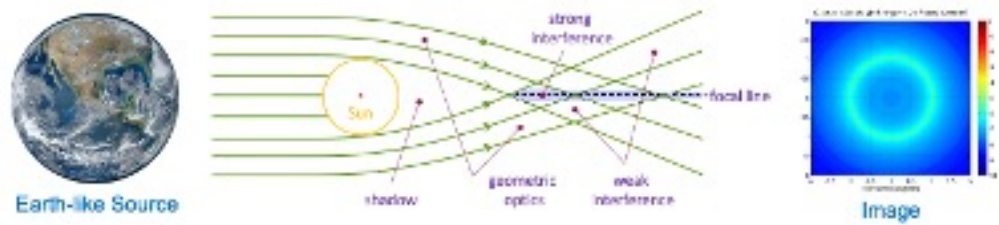

Figure 1. Our stellar neighborhood and the optical regions of the solar gravitational lens
(Turyshev 2017, Turyshev & Toth, 2017, Turyshev et al., 2018, KISS 2015, Turyshev & Toth, 2020c).

We reached and exceeded all objectives set for our Phase II study: We developed a new wave-optical approach to study the imaging of exoplanets while treating them as extended, resolved, faint sources at large but finite distances. We designed coronagraph and spectrograph instruments needed to work with the SGL. We properly accounted for the solar corona brightness. We developed deconvolution algorithms and demonstrated the feasibility of high-quality image reconstruction. We identified the most effective observing scenarios and integration times.

As a result, we are now able to estimate the SNR for light from realistic sources in the presence of the solar corona. We have proven that multipixel imaging and spectroscopy of exoplanets up to 30 pc are feasible. By doing so, we were able to move the idea of applications of the SGL from a domain of theoretical physics to the practical mainstream of astronomy and astrophysics. Under a Phase II NIAC program, we confirmed the feasibility of the SGL-based technique for direct imaging and spectroscopy of an exoplanet, yielding technology readiness level (TRL) of TRL 3.

We have developed a new mission concept that delivers an array of optical telescopes to the SGL focal region and then flies along the focal line to produce high resolution, multispectral images of a potentially habitable exoplanet. Our multisatellite architecture is designed to perform concurrent observations of multiple planets and moons in a target exoplanetary system. It allows for a reduction in integration time, to account for target's temporal variability, to "remove the cloud cover".

In this Report we describe the mission architecture and the relevant technology steps, which we can begin today, that would allow the launch of a Solar Gravity Lens Focus mission by 2028-2030. The new architecture developed in this study uses smallsats (<100 kg) with solar sails to fly a trajectory spiraling inward toward a solar perihelion of 0.1-0.25 AU and then out of the solar system on a nearly radial-out trajectory at 15-25 AU/year. Our design goal is 25 AU/year, to permit reaching the SGLF region in <25 years. A long time, but less than it took Voyager to reach the heliopause at less than a fifth of the distance of our goal in the interstellar medium (ISM). We would reach the heliopause and enter the ISM in ~7 years, compared to the ~40 years of Voyager. Today we are technologically ready to seize the unprecedented opportunity of using the SGL with a mission transit time only ~2.5× longer than the transit time of New Horizons to Pluto.

The SGLF CONOPS uses multiple small satellites in an innovative "string-of-pearls" (SoP) architecture where a pearl consisting of an ensemble of smallsats is periodically launched. As a series of such pearls are launched (to form the "string") they provide the needed comm relays, observational redundancy and data management needed to perform the mission. For example, if pearls are launched annually, then they will fly outward towards and then along the SGLF at 20 AU intervals. By employing smallsats using AI technologies to operate interdependently, we build in mission flexibility, reduce risk, and drive down mission cost. This makes possible concurrent investigations of multiple exosolar systems by launching strings towards multiple exoplanet candidates.

We used physics-based analytical tools to define the trajectories towards, about, and outward from the sun towards the SGLF. Solar sail propulsion brings each smallsat to perihelion via an inspiral trajectory from Earth, accelerates the spacecraft towards the SGLF target, and is used to remove residual injection errors (done via the NASA's DSN) during exit from the inner solar system.

Each pearl is targeted to and aligned with the SGL that will exist when the pearl arrives at the focal line (>548 AU). Once the solar sails are no longer useful, they can be ejected or repurposed. Subsequently, the $\Delta V$ requirements (~315 m/s including barycentric motion) are provided by onboard propulsion that must be highly efficient and long-lived. We identified commercial entities that have applicable technology (some flying, TRL9) that could be adapted for the SGL mission.

We addressed the position, navigation and timing (PNT) requirements by extending the capabilities of the NASA's Deep Space Network (DSN) to include distributed onboard star-trackers and X-ray pulsar sensors. The design of the spacecraft (s/c) utilized a concurrent engineering methodology tool. Four different constructs were analyzed, resulting in s/c units of ~30 kg in CBE mass (+15% contingency), a solar sail of 400 m$^2$/kg ratio and distributed satellite functionality where the downlink, science and PNT functions are distributed among the spacecraft within the pearl.

Power for the smallsats far from the Sun is an issue. Two forms of power system designs have been evaluated (RTG only and RTG+ radiation hardened battery). The tradeoffs show that the RTG + battery offer the best option, given the continued development of radiation hardened batteries. We also analyzed self-induced contamination and the effects on critical sensors given the ~50-year mission. We have established a technology roadmap for the evolution of the SW/HW needed for onboard computation. The analysis includes the artificial intelligence (AI) and machine learning capabilities needed to accomplish the SGL mission. Finally, we have identified a list of the technology areas where improvement would further reduce the mission risk.

We concluded that most of the technologies for SGLF mission either already exist (rideshare/cluster launch, sailcraft, RF/optical comm, all at TRL9), or are at intermediate levels of readiness: Sail materials (TRL 2-3), thermal management in solar proximity (TRL7), swarm operations (TRL5), terabit onboard processing (either FPGA or GPU, TRL 9/7), CONOPS (TRL7). What is missing is the system approach to assemble all these technologies for autonomous operations in deep space (TRL3). There is a clear path on how to close this gap, maturing the SGLF concept to TRL 4-5.

This affordable architecture design reduces cost in many ways: 1) It cuts the cost of each participant by enabling multiple participants (space agencies, commercial firms, universities, etc.) broad choices of funding, building, deploying, operating, and analyzing system elements. 2) It delivers economies of scale in an open architecture designed for mass production to minimize recurring costs. 3) It drives down the total mass (and thereby both NRE and recurring costs) by using smallsats. 4) It uses solar arrays of realistic size (~16 vanes of $10^3$ m$^2$) to achieve high velocity at perihelion (~150 km/s). 5) It applies maturing AI technologies to allow virtually autonomous mission execution, eliminating the need for operator-intensive mission management, (6) It reduces launch costs by relying on "rideshare" opportunities to launch the smallsats, avoiding the costs of large dedicated launchers, and 7) the SoP approach makes possible concurrent and affordable investigations of multiple exosolar systems by launching strings towards multiple exoplanet candidates.

Our SGLF mission concept proposes three innovations: i) a new way to enable exoplanet imaging, ii) use of smallsat solar sails to go further and faster at lower cost into the interstellar medium, and iii) an open architecture to take advantage of swarm technology in the future. It enables entirely new missions, providing a great leap in capabilities for NASA and the greater aerospace community. It lays the foundation for fast transit (>20 AU/yr) and exploration of our solar system and beyond (outer planets, moons, Keiper Belt Objects (KBOs), and interstellar objects/comets).

The results of our NIAC Phase II study are encouraging as they lead to a realistic design for a mission that will be able to image exoplanets in our stellar neighborhood. It could allow exploration of exoplanets relying on the SGL capabilities decades, if not centuries, earlier than possible with other extant technologies. The architecture and mission concepts for a mission to the SGL, at this early stage, are promising and should be explored further.

TABLE OF CONTENTS:

# 1 INTRODUCTION

## 1.1 Objectives and Expected Significance

We are standing at the threshold of a major discovery: The age-old question, "are we alone in the Universe?", may be answered within our lifetime. Extensive evidence indicates that planets capable of harboring life are ubiquitous in our galaxy and are a standard phenomenon of a typical stellar evolution. Hot topics, such as discoveries of many exoplanets orbiting nearby stars, efforts to understand the conditions that trigger, stimulate, and guide planetary formation processes, ignite their atmospheres and life-promoting conditions, as well as the development of techniques needed to find and study the new planets are at the focus of multiple ongoing science efforts. As a result, we are rapidly approaching the day when a major newspaper will open with a headline: "*The first habitable Earth-like exoplanet is discovered!*" What do we do the next day? How are we going to explore this alien world? Can we do anything today to prepare for this extraordinary event?

Direct detection of light reflected by a small, distant object moving in close proximity to its parent star is a formidable undertaking (Traub & Oppenheimer, 2010). The angular size of a planet is very small, requiring very large apertures or interferometric baselines. The light received from the exoplanet is also extremely faint. Advanced coronagraphic techniques are required to block the starlight (Cash, 2011; Crill & Siegler; 2017). The planet rides on a noisy background, thus detecting it requires excessively long integration times together with exquisite pointing stability. These challenges make direct high-resolution imaging of an exoplanet with a conventional telescope or interferometer a very difficult, if not impossible task.

Fortunately, Nature has presented us with a powerful "instrument" that we have yet to explore and learn to use for imaging purposes. This instrument is the solar gravitational lens (SGL), which exists as a consequence of the solar gravitational field focusing and greatly amplifying light from faint, distant sources of significant scientific interest, such as a habitable exoplanet.

According to Einstein's general theory of relativity (Einstein, 1915; Einstein, 1916), the gravitational field induces refractive properties on spacetime (Fock, 1959), with a massive object acting as a lens that is capable of bending the trajectories of incident photons (Turyshev, 2008, Turyshev & Toth, 2017, 2019b). As a result, gravitationally deflected rays of light passing from two sides of the lensing mass converge at a focus (Figure 1). Like an imperfect optical lens, a gravitational lens suffers from spherical aberration, with the bending angle inversely proportional to the impact parameter of the light ray with respect to the lensing mass. Therefore, such a lens has no single focal point but a semi-infinite focal line (FL).

Although all bodies in the solar system may act as lenses, only the Sun is massive and compact enough for the focus of its gravitational deflection to be within the range of a realistic mission. Its focal region is broadly defined as the area beyond ~547.8 AU. The line connecting the center of a distant object and that of the Sun begins to form the FL at this distance on the opposite side of the Sun. A spacecraft positioned beyond this heliocentric distance could use the SGL to magnify light from an exoplanet on the opposite side of the Sun (Eshleman, 1979, Turyshev & Toth, 2017).

By focusing light from a distant source, the SGL provides major brightness amplification (~$10^{11}$ at λ=1 μm) and extreme angular resolution (~$10^{-9}$ arcsec) in a narrow field of view. A modest telescope at the SGL could be used for direct imaging of an exoplanet (Turyshev & Toth, 2017). While all currently envisioned NASA exoplanetary concepts aim at getting a single pixel of an exoplanet, a mission to the SGL opens up the possibility for *direct* imaging with $10^3 \times 10^3$ pixels and high-resolution spectroscopy of an Earth-like planet up to a distance of 30 parsecs (pc) with resolution of ~10 km on its surface, enough to see its surface features and signs of habitability. Such a capability is truly unique and was never studied before in the context of a realistic mission.

In our NIAC[1] Phase I study (Turyshev et al., 2018a), we considered the architecture and design for an outer solar system mission that will be able to exploit the SGL's remarkable optical properties and to provide an astronomical facility that is capable of *direct* high-resolution imaging and spectroscopy of a potentially habitable exoplanet. Although theoretically feasible, the engineering aspects of building such a facility involving spacecraft at the large heliocentric distances involved have not been addressed before. Our NIAC Phase I effort addressed this question, establishing path towards a mission to the SGL.

## 1.2 Phase I Study – Key Points

The main objective of our Phase I effort was the development of the instrument/mission requirements, and to study a set of mission architectures to detect the signs of habitability of an exoplanet via remote imaging and spectroscopy. The following two major tasks have been accomplished:

<u>Task I: Development of the system and mission requirements</u> to guide the preliminary design concepts and formulate key mission, system, and technology operation requirements:

- The SGL's optical properties (Turyshev & Toth 2017) led to a solar <u>coronagraph</u> design capable of blocking sunlight to the level of the solar corona at a given position of the Einstein ring. The design resulted in a coronagraph with $2 \times 10^{-7}$ suppression, meeting the requirements (Shao, Zhou & Turyshev, 2017; Zhou, 2018). The 10% coronagraphic throughput initially called for a 2-m telescope. To reduce the size, weight and cost of the telescope we identified the driving instrument/mission design parameters: i) a more advanced occulter mask, ii) an external starshade solar coronagraph, iii) operating at larger heliocentric distances.
- To demonstrate imaging with the SGL, we investigated the application of <u>deconvolution</u> and have shown that the SGL allows for a megapixel-class image of an exoplanet. We also estimated the effectiveness and integration time required for direct deconvolution. Our estimates show that a high-resolution image of an exoplanet is possible with ~2 years of integration time by a direct deconvolution approach, suggesting exciting solutions to the imaging problem.
- We addressed the question of finding and studying life indicators based on spectroscopic and imaging data. With a respectable <u>spectroscopic</u> SNR of $10^3$ in 1 sec, we conclude that the signal will be sensitive to disturbances in the atmosphere of an exoplanet. It will be able to detect methane, oxygen and likely other molecules.
- We addressed the fact that with minimal impact to the mission design and architecture the same mission may also be able to image all exoplanets orbiting the same star.
- We determined the required delta-V for the spacecraft transiting the SGLF. This included the motion of the barycenter, that would cause comparable lateral motion of the SGLF and the delta-V needed to stay on the SGLF of the target planets as they orbit the parent star. The resulting delta-V is a very manageable ~315 m/sec for a ten-year data collection period.
- We developed an end game navigation protocol so that the navigational spacecraft in the pearl will use their on board instruments to home in on the SGLF of the parent star (that is much brighter and wider than those of the planets) using it as a beacon guiding them to the SGLF of the first target planet.

<u>Task II: Identification and study of alternative mission architectures:</u> Initially keeping the design envelope broad to allow assessment of key mission, system, and technology drivers:

---

[1] NASA Innovative Advanced Concepts (NIAC): https://www.nasa.gov/directorates/spacetech/niac/index.html

- We investigated CONOPS of spacecraft flying towards and then along the SGLF for detecting, tracking, and studying the brightness of the Einstein ring around the Sun. We formulated the requirements for such a mission to deliver healthy and capable spacecraft to heliocentric distances beyond 650 AU, place them on a controlled trajectory, and aim and operate the telescope that exploits the optical properties of the SGL. We conceptually designed a set of instruments and onboard data collection, compression and transmission capabilities needed for the mission.
- We developed the navigation and guidance concepts for all phases of operations. Our baseline approach relies on optical comm/nav, utilizing lasers and precision optical astrometry. We considered a set of instruments and onboard capabilities needed for unambiguous detection/study of life on another planet.
- We considered several mission concepts involving various numbers and sizes of spacecraft. We identified the design trade parameters that could lead to a robust mission and improve its performance. By making the spacecraft small, we enabled the use of a solar sail to act as the propulsion system to inject each spacecraft towards the SGLF, relying on high $I_{sp}$ long-duration thrusters for the remaining needed delta-V.
- We explored an architecture that relies on a pair of spacecraft (s/c) connected by a boom (or tether) of variable length (DeLuca, 2017). We also explored the role of s/c swarms or clusters to reduce the navigational/maneuver requirements to capture images at higher speeds. We considered the effect of the proper motion of the exoplanet, its orbital motion, and rotation on the imaging spacecraft requirements from the perspective of the necessary delta-V and data collection constraints.

Details of the results can be found in the Phase I final report (Turyshev et al., 2018a). The general conclusion is that it is feasible to conduct a mission to the SGL location not only for exoplanet imaging but also for the purpose of finding the elementary constituents of life on the exoplanet.

Our Phase I study results demonstrated the feasibility of the SGL imaging mission, providing us with a solid foundation for this effort. We have identified the next steps to improve our understating of the entire mission design envelope. We addressed these objectives in Phase II.

## 1.3 Approach for Phase II

During Phase II, we continued to explore the topics for a robust SGL mission, including refinement of the mission architectures by taking them through simulations and design trades. Specifically, we considered the following eight major topics:

1. We investigated the science operations in support of the primary objectives of high-resolution imaging and spectroscopy. We explored the ways of detecting photons from the Einstein ring, collecting them in a 4-dimensional data cube, processing and deconvolution. Insights on the image formation improved the mission concepts, yielding realistic mission requirements.
2. We studied direct and rotational image deconvolution approaches in a realistic setting, including effects of planetary rotation, varying planetary features (i.e., time-variable clouds, seasons), telescope pointing errors, etc. We combined these approaches in a more generalized deconvolution simulation. We examined the nonuniformity of light distribution within the Einstein ring, which may yield additional information for the deconvolution process.
3. We studied an instrument comprising a swarm of small spacecraft, each moving at a slightly different trajectory but parallel to the instantaneous SGLF optical axis. Such an instrument would rely on the light collection capabilities enabled by a formation flying architecture, taking full advantage of the SGL amplification and differential motions.

4. Given the enormous amplification of the SGL, we studied the possibility of spectroscopy of the exoplanet, even spectropolarimetry. We considered the possibility of producing a spectrally resolved image over a broad range of wavelengths, providing a powerful diagnostic for the atmosphere, surface material characterization, and biological processes on an exoplanet.
5. We addressed the issue of imaging when accounting for the time variability of the SGL system, which results from the solar motion with respect to the barycentric coordinate reference system (BCRS) (Turyshev & Toth, 2015). We investigated how the spacecraft could raster an exoplanet image as its optical axis moves on a $10^4$ km orbit. We also studied the relevant station keeping aspects needed to ensure pointing stability.
6. We studied mission design: i) flight system and science requirements; ii) key mission, system, operations concepts, and technology drivers; iii) description of mission and small craft concepts to reach and operate at the SGL; and iv) study instruments and systems for the SGL spacecraft, including power, comm, navigation, propulsion, pointing, and coronagraph. To conduct mission architecture trade studies aiming at PNT requirements for the SGL mission.
7. We conducted trade studies with a set of key driving parameters: a) heliocentric distance along the SGLF, b) telescope aperture, c) integration time, d) detector type and sensitivity, e) coronagraph/starshade performance, etc. We performed trades between a single telescope vs. a distributed smallsat system. A small telescope has limited capabilities but opens up the possibility of sending multiple spacecraft.
8. We identified the cost drivers needed to make the concept affordable – specifically to employ many low cost spacecraft, achieve large dedicated launch vehicles by adopting "rideshare" principles, eliminate costly ground-based personnel by artificial intelligence (AI)-driven autonomy, and distribute the cost among many stakeholders over the program life cycle.
9. We studied the use of amplified light from the parent star to navigate the spacecraft.

To date, all results look promising, both for getting there and for capturing high-resolution images with spectral content. Technological considerations with regards to mission architecture, instruments, comm, etc. also look feasible. The mission has the potential of being the most (and perhaps only) practical and cost-effective way of obtaining kilometer scale resolution of an exoplanet.

As a result, we now understand the complexities of the capture/creation of direct images of an exoplanet with the SGL. The cost of such a mission is yet to be studied, as a post-Phase II effort.

**1.4   Phase II Study Highlights**

We now have a good understanding of the SGL's optical properties and image formation process with the SGL (Turyshev & Toth 2019ab, 2020abc). We identified i) key mission requirements; ii) novel and unique means to obtain a high-resolution, multispectral images of identified likely habitable exoplanets; iii) new robust distributed architecture of interplanetary smallsats. The Phase II work extended the results of the Phase I effort with the necessary refinements.

The scientific merit and technical feasibility of utilizing the SGL for imaging exoplanets is documented in this report. A unique attribute of the SGL is that the s/c designs and concept of operations (CONOPS) are identical for the planetary systems of any candidate parent star. This contrasts favorably with classical NASA missions in which the target is one-of-kind (e.g. Saturn, Europa, Enceladus, etc.) and therefore the needed flight system and associated ground software and personnel are custom built to accomplish that science objective. In the case of the SGL, we design a set of spacecraft and CONOPS that are "agnostic" to the target star system.

Because of the impressive progress in the search for exoplanets, there will soon be a list of attractive exoplanet targets, all of which could be imaged by flying sets of identical spacecraft to the relevant SGLF. This is because, unlike missions to solar system destinations, all the SGLF missions are sent to the solar gravity line of the target – all of which are identically distant from Earth (>550 AU) and differ only by the celestial direction of the target. The locus of targets is a sphere that defines the starting point of all SGL missions, independent of the actual distance to the star. Figure 2 describes such a capability. From an aim point on the SGL virtual sphere and because of the $10^{11}$ magnification of the SGL, we can capture images of exosolar systems up to 100 ly from the Sun and beyond. No matter how distant the parent star, the image is collected by flying to the SGL, that starts outward from the sun at ~550 AU.

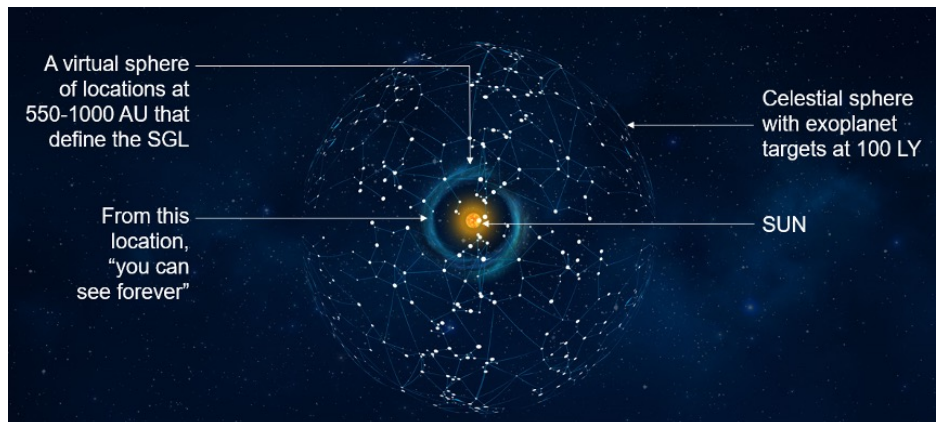

Figure 2. Our stellar neighborhood with notional targets.

The technology documented in these pages describes an enterprise architecture that exploits this unique SGL targeting feature, to conduct SGL missions to exosolar system targets chosen by scientific merit. Moreover, the architecture is cost-effective, permits multiple stakeholders to be involved including international participation and is designed around the concept that utilizes "self-learning" for improvement and upgradeability of hardware and software for more efficient operations. The key concepts of this architecture are:

1. A string-of-pearls architecture in which small s/c are grouped into "pearls" of 10-20 s/c sent to fly together towards, and then along, the Solar Gravity Lens Focal Line (SGLF) of a parent star and its planets. A pearl arrives at the focal point (the start of the SGLF) after some 20+ years of flight, and upon arrival, collects observational data for some 20 years.
2. Within each pearl, functionality is shared (and reallocated in flight) among the spacecraft to increase reliability, drive down costs and minimize the mass of each spacecraft.
3. By launching these pearls on an approximately annual basis, we create the "string", with pearls spaced along the string some 20-25 AU apart throughout the timeline of the mission. So that later pearls have the opportunity to incorporate the latest advancements in technology for improved capability, reliability, and/or reductions in size/weigh/power which could translate to further cost savings.
4. The first pearl arriving at the SGL conducts the pre-planned guidance-navigation-collection-data management aspects of the mission. It transmits its data and operational experience to Earth, where the CONOPS is optimized based upon lessons learned in the initial phase. These lessons are then used to enhance the science return of the first pearl, and to optimize the mission protocol of each pearl that follows (approximately one year behind). This optimization process continues as each pearl moves along the SGL – thereby making best use of total capability of the instruments, the board resources, and the data compression and transmission.

5. The string also serves as a pearl-to-pearl communications relay in which selected s/c in each pearl have responsibility for intra-pearl data management and inter-pearl data transfer.
6. By using an ensemble of small spacecraft, the architecture does not depend upon expensive, dedicated launch vehicles – it employs the "rideshare" principle that is now being increasingly used for smallsat flights to LEO and GEO and will become routine for flights to cislunar space as the Moon-Mars and commercial traffic to cislunar space expands.
7. The spacecraft in a pearl are flown into cislunar space on various rideshare opportunities and then aggregate themselves using the solar sails into pearls – which are then deployed towards the perihelion point and from there to the SGL of a candidate parent star. This approach allows the launching, deployment and flight of pearls on concurrent missions to several stars of interest. The only difference between flights to one star and another will be the selected aim point at the perihelion (controlled by the inclination of the orbit down to the perihelion, and the time of perihelion passage.
8. The SGL mission architecture eliminates the need for a billion dollar highly specialized s/c design with costly dedicated launchers – a basic small sat mission design with rideshare access to space will be able to go to any and all stars of interest. The nonrecurring costs are amortized over many concurrent and/or sequential parent star imagery missions. The costs can be shared among many interested parties that wish to participate in "new worlds" discovery.

This is our Final Report for our NIAC 2018 Phase II investigation, which is structured as follows: In Section 2 we discuss the mission and instrument requirements. Our imaging approach is based on the optical properties of the SGL. We describe the concept of operations (CONOPS), instrumental design, as well as the direct deconvolution. We also present considerations for target selection and anticipated properties.

Section 3 highlights our mission design studies capable of reaching the focal region of the SGL and operating with the image volume. We discuss the study approach and relevant mission tradeoffs. We present and discuss various mission concepts that were considered during the study, including a single spacecraft, a "string-of-pearls" (SoP) approach based on solar sail technology.

In Section 4 we present the technology drivers and approach to technology maturation. We discuss current status in various technology areas relevant to missions to the SGL. We present technology gaps, identify risk and describe mitigation strategies.

In Section 5 we summarize results we obtained during Phase II. We also present recommendations and approach for transition strategy to realize missions to SGL.

## 2  IMAGING CONTINENTS & SIGNS OF LIFE ON AN EXOPLANET

### 2.1  Why do we need the SGL?

The challenges of direct detection of exoplanets are well known and are related to the fact that the planets are not self-luminous. They are small, very distant and are moving in a very highly light-contaminated environment (Traub & Oppenheimer, 2010; Wright & Gaudi, 2013). The thought of resolved images of exoplanets elevates this problem to the next level by requiring prohibitively large telescopes or interferometric baselines. For instance, to image our Earth from the distance of 30 pc with a modern diffraction-limited telescope, we would need a telescope aperture of ~ 90 km (in combination with an aggressive coronagraph (Angel, 2003), Figure **3**), which is not practical.

Using optical interferometers for this purpose would not only involve many km-scale variable interferometric baselines with telescope apertures of several tens of meters but will also require integration times of several hundred million years to reach a reasonable signal-to-noise ratio (SNR) of ~7. Clearly, an imaging approach relying on conventional astronomical techniques is not feasible. However, the age old human desire to see alien worlds that may exist in the form of terrestrial exoplanets in our stellar neighborhood, especially those whose imaging and spectroscopy could show the presence of current life (Seager, 2010), motivates us look for alternative approaches and consider the SGL as the only realistic means to overcome these challenges.

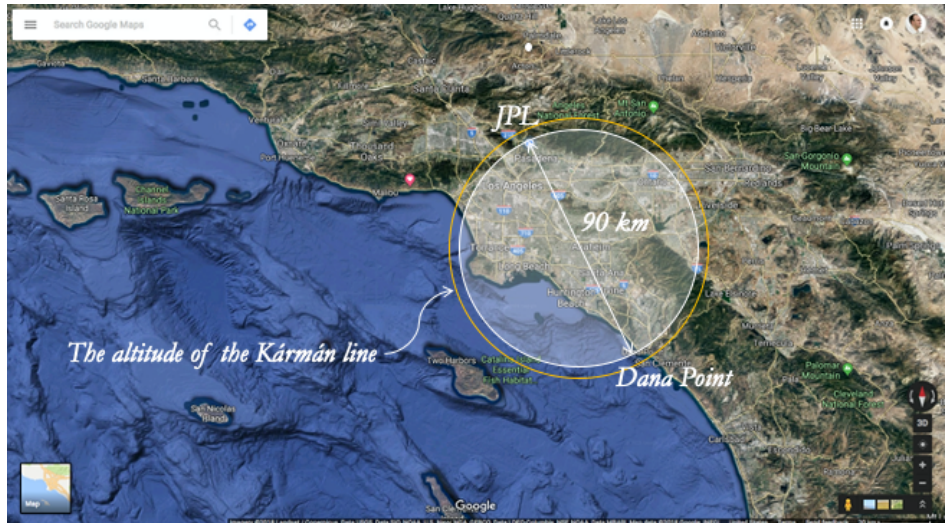

Figure 3. The tyranny of the diffraction limit: To make a 1-pixel image of an exo-Earth at 100 light years, one needs a telescope with a diameter of ~90 km. The relevant scales in Los Angeles area are shown.

In pursuit of this objective, we examined the SGL as the means to produce *direct* high-resolution, multipixel images of exoplanets (Turyshev, 2017; Turyshev & Toth, 2017, 2018, 2019ab, 2020abc). The SGL results from the diffraction of light by the solar gravity field, which acts as a lens by focusing incident light at distances >548 AU behind the sun. The properties of the SGL are remarkable: it offers light amplification of ~$10^{11}$ and angular resolution of ~$10^{-10}$ arcsec (Turyshev, 2017; Turyshev & Toth, 2017, 2018). It allows for direct imaging of an exoplanet at 30 pc using a single 1-m telescope, achieving 100×100-pixel resolution in 12 months, which is not possible otherwise. This is sufficient to observe seasonal changes, oceans, continents, and surface topography (Turyshev & Toth, 2020c).

## 2.2 Why now?

Our analysis suggests that in addition to the fact that the SGL provides a set of unique capabilities for exoplanet investigations, the mission to the SGL is an effort that humanity may be able to develop and implement in the very near future. In fact, the current technology status indicates that we now are at the sweet spot for the mission to SGL, which is based on:

- We understand the optical properties of the SGL as well as the observational physics, the phenomenology and the data collection process. No flight testing needed for this; a laboratory technology demonstration may be developed to demonstrate the basic principles.
- Candidate exoplanets are being discovered in large numbers; in a decade there will be a significant number of a exoplanets that could have life-supporting conditions.
- Deep space missions like Voyager 1[2] and 2, Pioneer 10[3] and 11, and New Horizons[4] shows that deep space is "friendly" to s/c, demonstrating that we can fly and operate robotic spacecraft to larger heliocentric distances. Note that New Horizons proved the effectiveness of hibernation to bridge the gap between Earth and the target that is directly applicable to the SGL.
- Needed technology is mature or rapidly maturing, including i) spacecraft miniaturization; ii) solar sailing, iii) low cost access to space via ride sharing (Isakowitz & Schingler, 2020), iv) AI and machine learning, v) autonomy and repurposing, vi) high temperature operations near the sun (Parker Solar Probe[5] and ESA Solar Orbiter[6]).
- Most of the development and operational costs can be shared including flight prototypes.

---

[2] https://en.wikipedia.org/wiki/Voyager_1
[3] https://en.wikipedia.org/wiki/Pioneer_10
[4] https://en.wikipedia.org/wiki/New_Horizons

In summary, the science and technology readiness analyses show that a mission to the SGL may be implemented in the next decade. Given the fact that a set of realistic targets will be ready in the same timeframe, our mission concept enables an exciting opportunity to see and study life on exoplanets within our lifetimes. This makes our effort of studying the SGL and the mission very relevant and timely.

### 2.3 Optical properties of the SGL

#### 2.3.1 Diffraction of light in the gravitational field of the Sun

A wave-theoretical description of the SGL (Turyshev, 2017; Turyshev & Toth, 2017) demonstrates that it possesses a set of rather remarkable optical properties. Specifically, by naturally focusing light from distant, faint sources, the SGL amplifies their brightness by the enormous factor of $\sim 2GM/(c^2\lambda) \sim 10^{11}$ (for $\lambda = 1$ μm). Moreover, the SGL has extreme angular resolution of $\lambda/D_0 \sim 10^{-10}$ arcseconds (with $D_0$ being the diameter of the Sun,) making it exceptionally well-suited for imaging distant objects.

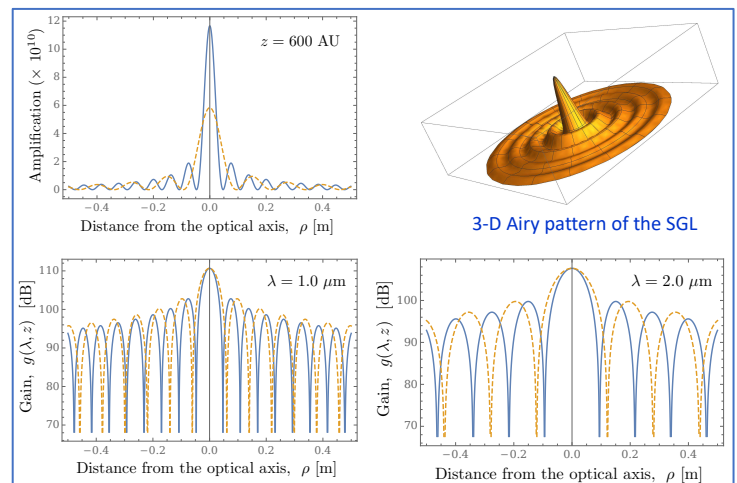

Figure 4. Optical properties of the SGL (Turyshev & Toth, 2017). Up-Left: Amplification of the SGL. Up-Right: Point spread function. Bottom: Gain of the SGL as seen in the image plane as a function of observational wavelength.

The optical properties of the SGL are now better understood, which enables the design of an astronomical telescope, i.e., to describe the point-spread function (PSF), resolution, magnification, plate scale, etc., with some of them given in Figure 4. The Phase I effort allowed us to develop a comprehensive understanding of the image formation by the SGL and the technology needs to conduct a realistic mission, including data collection and image deconvolution.

---

[5] https://en.wikipedia.org/wiki/Parker_Solar_Probe
[6] https://en.wikipedia.org/wiki/Solar_Orbiter

During Phase II, we studied the optical properties of the SGL which exists because of the natural ability of the solar gravitational field to cause the diffraction of the electromagnetic (EM) waves that travel in the close proximity of the Sun (Einstein, 1936; Eshleman, 1979). After passing by the Sun the wavefront that envelops the Sun develops a concave form with the closest parts of this wavefront beginning to move inward and towards the optical axis: an imaginary line connecting the center of the Sun and the point source (see Figure 5). The parts of the wavefronts that just graze the Sun meet each other while intersecting the optical axis at heliocentric distance beyond $R^2_{sun}/2r_g$ ~ 547.8 astronomical units (AU); the parts of that same front that travel farther from the Sun will meet at larger distances. The two opposing wavefronts interfere with each near the optical axis of the SGL creating the strong interferometric region – the region of our primary interest where the images of distant sources are created. A spacecraft with a modest telescope and a coronagraph to block the solar light will be able to observe the Einstein ring formed around the Sun.

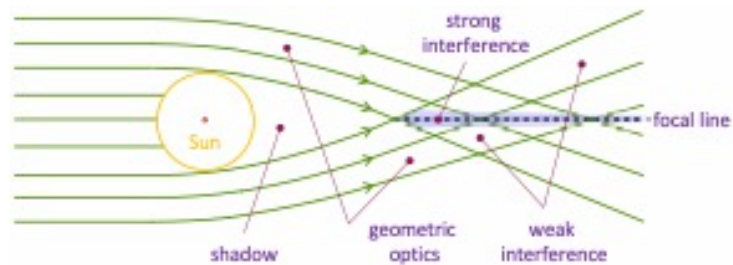

Figure 5. The different optical regions of the SGL.

In Refs. (Turyshev 2017; Turyshev & Toth, 2017; Turyshev & Toth, 2018) we developed a wave-optical treatment of the SGL by considering diffraction of the EM waves in the monopole gravitational field produced by the Sun. Thus, in (Turyshev, 2017; Turyshev & Toth, 2017) we have demonstrated that, once an EM wave passes by the Sun, the diffracted light forms four regions with various optical properties (see Figure 5): i) the shadow region, where no incident light exist, ii) the geometric optics region, where only the incident light is present, iii) the weak interference region, where, in addition to the incident wave, one also finds a scattered wave; and finally iv) the strong interference region, where two waves with nearly equal optical paths are present, resulting in the strong light amplification of $4\pi\, r_g/\lambda \sim 10^{11}$ and angular resolution of $\lambda/2R_{sun}$ ~0.5 nano-arcseconds (nas), both for $\lambda = 1$ µm. In (Turyshev & Toth, 2018) we studied the diffraction of light propagating in vacuum in the presence of a spherical obscuration produced by a compact body both in the absence and in the presence of its gravitational field.

In (Turyshev & Toth, 2019a), we considered the properties of the SGL in the presence of the solar corona. For that, we studied the diffraction of the EM wave that are passing through the solar corona. We have found that the impressive optical amplification and angular resolution of the lens are severely affected for wavelengths longer than ~1 mm, to the point that these SGL advantages almost vanish, thus confirming results of (Turyshev & Andersson, 2002).

On the other hand, we have demonstrated that the propagation of light at optical wavelengths, $\lambda \approx$ 1 µm and shorter, is practically unaffected by plasma in the solar corona. These results showed that the available wave-optical description of the SGLs optical properties may be used to describe image formation process for faint sources (Turyshev et al., 2018ab; 2019a). This work had opened the way to consider the SGL for imaging unresolved point sources, which was the first step towards imaging of more complex objects.

### 2.3.2 Treatment of extended sources

The entire image of an Earth-like planet at 30 pc is compressed by the SGL into a cylinder with a diameter of ~1.3 km in the vicinity of the focal line. The telescope, acting as a single-pixel detector while traversing this region, can build an image of the exoplanet with kilometer-scale resolution of its surface.

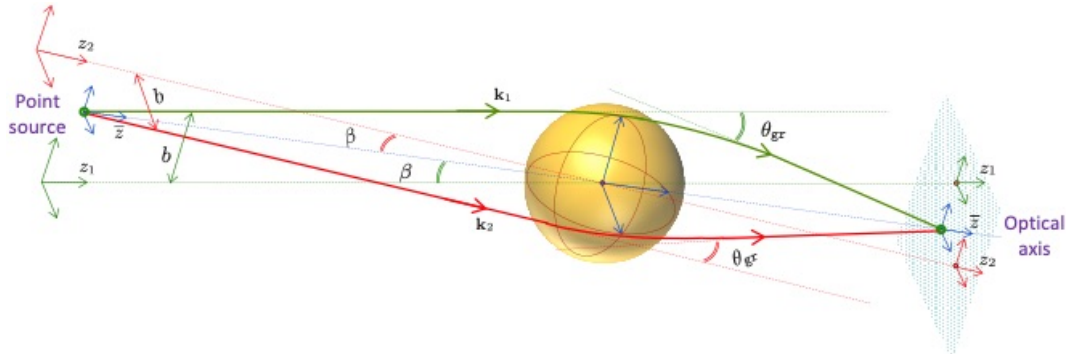

Figure 6. The three-dimensional geometry of the SGL, focusing light from a point source located at a finite distance. Two rays of light with wavevectors $k_1$ and $k_2$ are shown. The rays move in different planes, which intersect along the optical axis. Note that the z-axis is no longer uniquely defined. However, the optical axis z-bar is unique and preserves the axial symmetry.

The next step was to consider properties of SGL for extended sources (Turyshev & Toth, 2019b, 2020a). We modeled an exoplanet as an extended object that is at a large, but finite distance from the Sun (see Figure 6, Figure 7). In (Turyshev & Toth, 2019b) we developed a wave-optical theory of the SGL for such sources (previously unavailable). While considering the optical properties of the SGL in this case, we realized that extended sources present an interesting challenge that relates to blurring of the images. Because of the properties of the SGL PSF (Figure 4), this blur results in mixing light received from many widely separated areas of the surface of the source and distributing it across the entire image. In (Turyshev & Toth, 2020a) we addressed that issue by discussing separately the directly imaged region and that due to the rest of the source (see Figure 8). Using this approach, we studied photometric imaging with the SGL for both point and extended sources and developed analytical expressions needed to treat them.

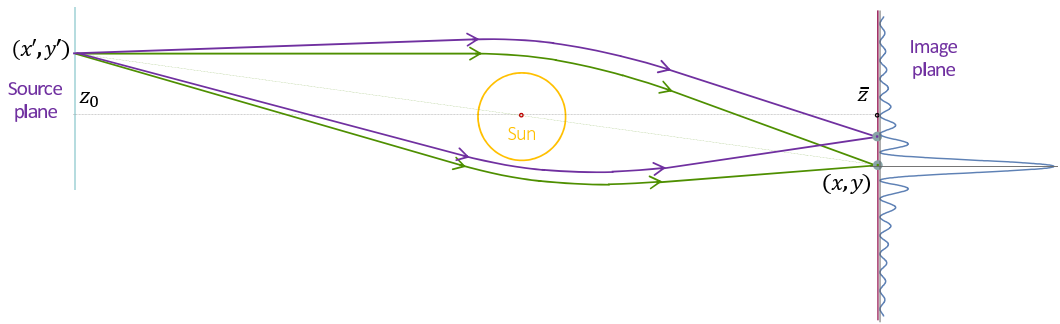

Figure 7. The geometry of imaging a point source with the SGL. A point source with coordinates $(x',y')$ is positioned on the source plane, at the distance $z_0$ from the Sun. The SGL image plane is at the heliocentric distance $z$. Rays with different optical paths produce a diffraction pattern in the SGL image plane that is observed by an imaging telescope.

We considered an exoplanet as an extended source of radius, $R_\oplus$, that is located at a large, but finite distance $z_0$ from the Sun (Figure 7). The image of this object is formed in the strong interference region of the SGL at the heliocentric distance of $z > R^2_{sun}/2r_g$. There is no single focal point of the SGL, but a semi-infinite focal line. The SGL, being a convex lens, compresses the source forming the image of the exoplanet within the volume occupied by a cylinder with a diameter of $r_\oplus=(z/z_0) R_\oplus \sim 1.34\times10^3$ $(z/650$ AU$)(30$ pc$/z_0)$ m. Placing a spacecraft in any of the image planes within the cylinder allows to take the data that may be used to form an image of a distant faint object.

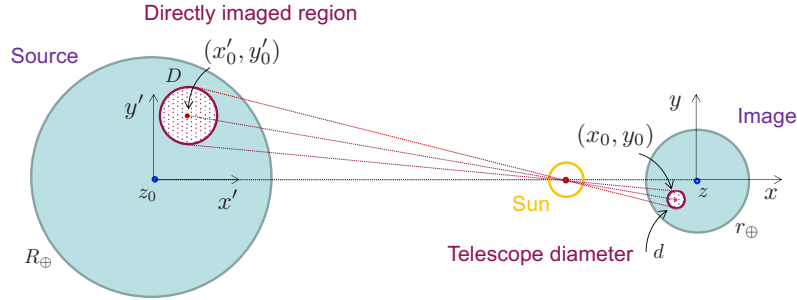

Figure 8. Imaging of extended resolved sources with the SGL. The SGL is a convex lens, producing inverted images of a source.

As was shown (in Turyshev, 2017; Turyshev & Toth, 2017), the point-spread function (PSF) of the SGL has the form $\propto J_0^2(kr\sqrt{2r_g/z})$, where $\rho$ is the deviation from the optical axis and $J_0$ is the zeroeth Bessel-function of the first kind. For large $\rho$, this PSF behaves as $\propto 1/\rho$, which is different from the typical PSF of a thin lens that is given by $\propto (2J_1(\alpha\rho)/(\alpha\rho))^2$, which, for large $\rho$, behaves as $\propto 1/\rho^3$. The $1/\rho$-diminishing behavior of the SGL's PSF, which characterizes the SGL's spherical aberration, results in the fact that the telescope, although it points toward the directly imaged region, collects light from the areas far from this region. This extra signal from the rest of the exoplanet results in the image blurring, whose intensity is much higher than that received from the directly imaged region.

Reconstructing the original, unblurred image from this light, or deconvolution, entails knowledge of the convolution operator that characterizes the blurring effect, and applying its inverse to the collected signal. This process is significantly affected by noise and decreases the SNR of the recovered image. Ultimately, the image quality depends on the SNR achieved per image pixel for each position of the imaging telescope on the image plane. Evaluating the SNR that may be achieved in this case was one of our primary objectives.

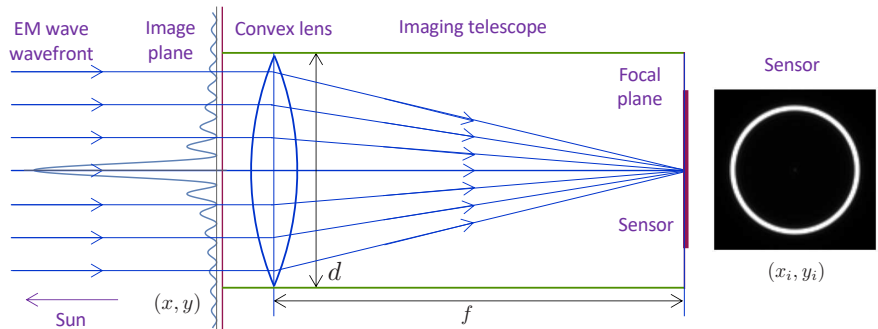

Figure 9. Imaging a point source with the SGL with a telescope. The telescope is positioned on the optical axis that connects the source and the Sun and it "sees" the Einstein ring. The telescope is represented by a convex lens with a diameter $d$ and a focal length $f$. Positions in the SGL image plane, $(x,y)$, and the optical telescope's focal plane, $(x_i,y_i)$, are also shown.

### 2.3.3 Imaging of point and extended sources with an optical telescope

We next addressed the imaging of realistic sources with an optical telescope that is positioned within the image plane in the strong interference region of the SGL (Turyshev & Toth, 2019b, 2020ab). For that, we considered the intensity distribution pattern of this signal that appears in the focal plane of the optical telescope (see Figure 9), as a function of the telescope's displacement form the optical axis of the SGL.

In (Turyshev & Toth, 2020c), we studied the image formation process with the SGL in the case of extended, resolved sources. We developed expressions for the intensity distribution at the focal plane of an imaging telescope. We also derived the power of the signal to be measured by an imaging sensor within the annulus that contains the Einstein ring, as seen by a diffraction-limited optical telescope on its focal plane. We also considered the contribution of the solar corona brightness to the total energy deposited on the focal plane of the optical telescope. This allowed us to estimate the SNR for imaging of various realistic exoplanetary sources.

To describe the image of faint objects with the SGL, we take an imaging telescope and position it in the image plane in the strong interference region of the SGL (see Figure 9). As we have shown (Turyshev & Toth, 2020c), the signal received from the directly imaged region forms the Einstein ring in the focal plane of the optical telescope. The signal from the rest of the planet (i.e., blur) will be also deposited at the Einstein ring and will significantly exceed in brightness the signal from the directly imaged region. Therefore, the total signal collected at each telescope position in the image plane will be dominated by the blur signal received from the entire target. This is the signal that we will use to recover the true image of the exoplanet by deconvolution.

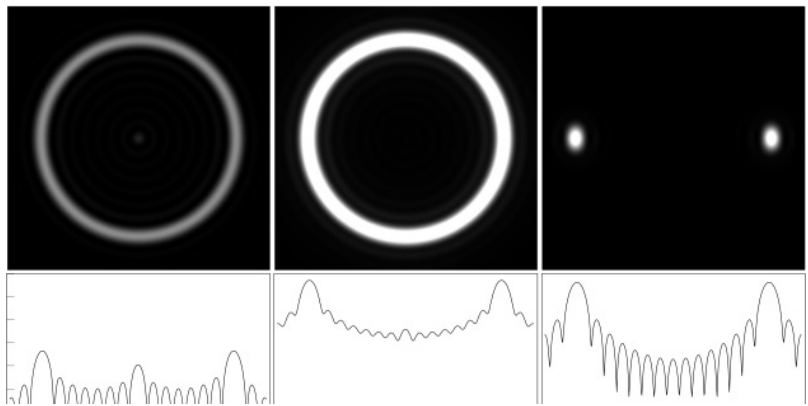

Figure 10. Top row: Density plots simulating images that appear in the focal plane of the optical telescope. Left: the directly imaged region. The brightness of this image is exaggerated to ensure that the Einstein ring and diffraction artifacts remain visible. Center: Light from the rest of the source. This is the dominant light contribution, yielding a much brighter Einstein ring with less prominent diffraction artifacts. Right: image contamination due to a nearby source of light, showing light from another uniformly illuminated disk of the same size, offset horizontally by 10 radii. Bottom row: corresponding dimensionless intensities depicted on a decimal logarithmic scale. The contribution from the directly imaged region is $O(10^3)$ less than the contribution from the rest of the source. Contribution from a nearby object is of similar intensity but confined to narrow sections of the Einstein ring.

As a result of these multiple recent efforts, we now have all the tools necessary to study realistic SNRs that we may be able to reach while imaging exoplanets, while treating them as extended, resolved, faint sources that a located at large but finite distance from us. The main motivation for our work was the need to improve the knowledge of the optical properties of the SGL and to evaluate its advantages for imaging of nonluminous, faint, extended, resolved, and distant sources. This allowed us to describe realistic observing scenarios and appropriate observing conditions.

## 2.4 Sensitivity estimates for imaging with the SGL

### 2.4.1 Photon fluxes from realistic targets

In (Turyshev & Toth, 2020c), we estimated the signals that could be expected from realistic targets when they are imaged with the SGL. We considered a planet identical to our Earth that orbits a star identical to our Sun. The total flux received by such a target is the same as the solar irradiance at the top of Earth's atmosphere, given aa $I_0 = 1{,}366.83$ W/m$^2$. Approximating the planet as a Lambertian sphere illuminated from the viewing direction yields a Bond spherical albedo of $2/3\pi$, and the target's average surface brightness becomes $B_s = (2/3\pi)\, \alpha\, I_0$, where we take Earth's broadband albedo to be $\alpha = 0.3$ and assuming that we see a fully-illuminated planet at 0 phase angle.

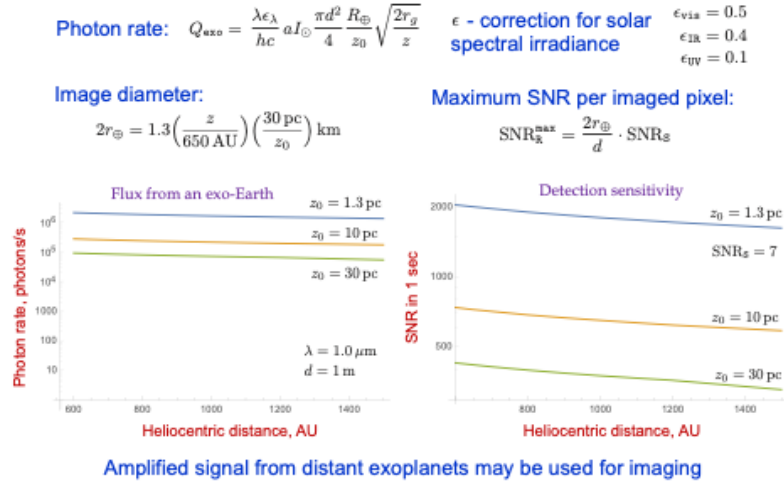

Figure 11. Photon fluxes and signal-limited SNR for an exoplanet located at distances of 1.3pc, 10pc, and 30pc from the Sun.

Assuming that the planet is positioned at $z_0 = 30$ pc away from the Sun, we estimated the signal from the planet as $Q_{\text{planet}} = 8.01\times10^4\, (d/1\text{ m})^2\, (650\text{ AU}/z)^{1/2}\, (30\text{ pc}/z_0)\, (\lambda/1\ \mu\text{m})$ photons/s. This estimate translates to a flux of $2.40\times10^5$ photon/s for an exoplanet at $z_0 = 10$ pc and $1.85\times10^6$ photon/s for at $z_0 = 1.3$ pc. Figure 11 summarizes the photon fluxes and the relevant signal-limited SNR (i.e., no noise) as a function of heliocentric distance.

Using these estimates, we compared the performance of a conventional telescope against one aided by the SGL. The angular resolution needed to resolve features of size D in the source plane requires a telescope with aperture $d_D \sim 1.22\,(\lambda/D)\, z_0 \sim 1.19\times10^5$ km = $18.60\, R_\oplus$, which is not realistic. The photon flux of a d=1m telescope from such a small area on the exoplanet yields the value of $1.97 \times 10^{-8}$ photons/s, which is extremely small. Comparing this flux with $Q_{\text{planet}}$ received with the SGL, we see that the SGL, used in conjunction with a d = 1 m telescope, amplifies the light from the directly imaged region (i.e., an unresolved source) by a factor of $\sim 3.38\times10^9\, (d/1\text{m})(650\text{ AU}/z)^{1.5}(z_0/30\text{ pc})^2$. This estimate justifies using the SGL for imaging of faint sources.

### 2.4.2 Solar corona brightness as a noise source

We investigated the most significant source of noise, the solar corona, and have shown that it is possible to obtain a detailed image of a distant exoplanet with integration times consistent with a realistic mission. For that we have developed semianalytical models of the deconvolution process that was needed to understand the impact of deconvolution on the noise that may be present in the measurements. It was shown that deconvolution amplifies the noise, thus reducing the sensitivity.

The Einstein ring corresponding to a distant target, as observed from a position in the SGL focal region, is seen through the bright solar corona, which represents an important noise contribution that must be considered. Noise from the solar corona can be mitigated by letting as little light from the corona to reach the instrument as possible. This is achieved by employing a suitably designed solar coronagraph, needed in any case to block direct light from the Sun, but which can also be used to reduce the noise from the solar corona.

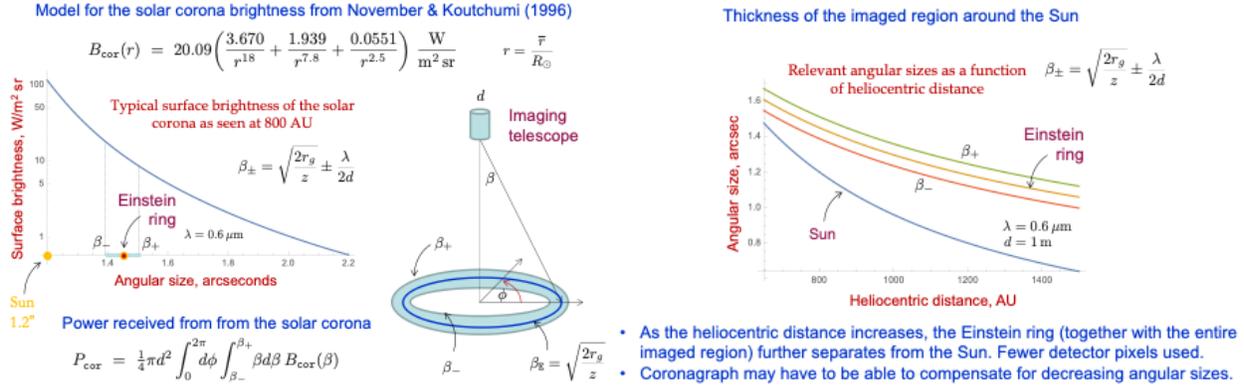

Figure 12. Modeling the contribution of the solar corona brightness on the imaging measurements conducted with the SGL.

Solar coronagraphy was invented by Lyot (Lyot, 1932) to study the solar corona by blocking out the Sun and reproducing solar eclipses artificially. Coronagraphs are also considered to block out light from point sources, such as the host star of an exoplanet imaged with conventional telescope (Traub & Oppenheimer, 2010). The SGL coronagraph is different, as it needs to block light from the Sun and the solar corona, leaving visible only those areas where the Einstein ring appears.

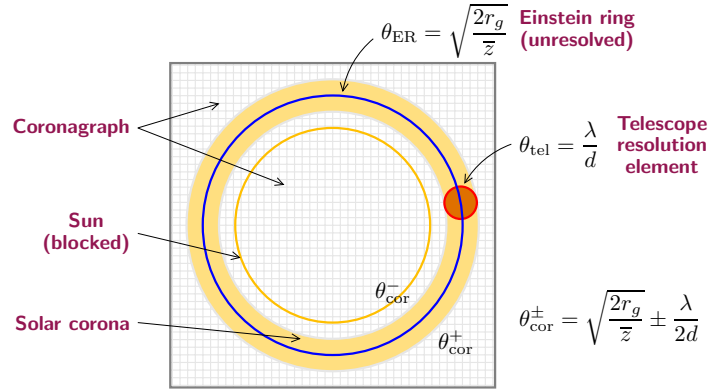

Figure 13. The annular coronagraph concept: The coronagraph blocks light from both within and outside the Einstein ring. The thickness of the exposed area is determined by the diffraction limit of the optical telescope at its typical observational wavelength (Turyshev & Toth, 2020c).

The already available design for the SGL coronagraph (Zhou, 2018) rejects sunlight with a contrast ratio of ~$10^7$. At this level of rejection, light from the solar disk is completely blocked to the level comparable to the brightness of the solar corona. Taking a further step, we consider two possible coronagraph concepts: A conventional coronagraph (which we call a "disk coronagraph") that blocks light only from the solar disk and the solar corona up to the inner boundary, $\beta_-$, of the $\lambda/d$ annulus centered on the Einstein ring, and a coronagraph that also blocks light outside the outer boundary, $\beta_+$, of the $\lambda/d$-annulus centered at the Einstein ring (the "annular coronagraph"). Figure 12 describes the relevant sizes and observing configuration.

Compared to the disk coronagraph, the annular coronagraph (see Figure 13) reduces the noise contribution from the solar corona by an additional ~10%. As the solar corona is quite bright compared to the Einstein ring, the use of an annular coronagraph is preferred for an SGL imaging instrument. Thus, in the estimates that we develop for the corona contribution, we assume an annular coronagraph design.

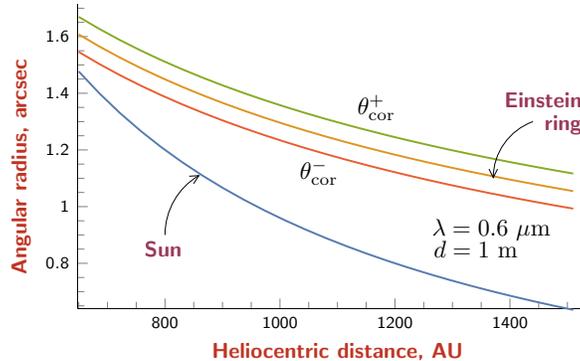

Figure 14. Angular sizes of the Sun and the diffraction-limited view of the Einstein ring as functions of heliocentric distance. As the heliocentric distance increases, the Einstein ring (together with the entire imaged region) further separates from the Sun.

In (Turyshev & Toth, 2020c) we have developed a corona model that may be applied to estimate the corona noise contribution for imaging with the SGL. In particular, we estimated that for the relevant heliocentric ranges the corona photon flux is at the level of

$$Q_{\rm cor} = 2.29 \times 10^9 \left[1 + 0.79 \left(\frac{650\text{ AU}}{\bar z}\right)^{5.1} + 0.05 \left(\frac{\bar z}{650\text{ AU}}\right)^{2.65}\right]\left(\frac{d}{1\text{ m}}\right)\left(\frac{650\text{ AU}}{\bar z}\right)^{4.4}\left(\frac{\lambda}{1\text{ }\mu\text{m}}\right)^2 \text{ photons/s.}$$

Assuming that the contribution of the solar corona is removable (e.g., by observing the corona from a slightly different vantage point) and only stochastic (shot) noise remains, using $Q_{\rm planet}$, we estimate the resulting SNR per second in the solar corona dominated regime as

$$\text{SNR}_c = \frac{Q_{\rm planet}}{\sqrt{Q_{\rm cor}}} = \frac{1.68\epsilon(\rho_0)}{\sqrt{1 + 0.79\left(\frac{650\text{ AU}}{\bar z}\right)^{5.1} + 0.05\left(\frac{\bar z}{650\text{ AU}}\right)^{2.65}}}\left(\frac{d}{1\text{ m}}\right)^{\frac{3}{2}}\left(\frac{30\text{ pc}}{z_0}\right)\left(\frac{\bar z}{650\text{ AU}}\right)^{1.7}\sqrt{\frac{t}{1\text{ s}}}.$$

It is noteworthy to consider the behavior of this SNR with respect to the several parameters involved: (1) It does not depend on the wavelength. This is because for this estimate, we assumed the presence of an annular coronagraph. The width of the annulus of such a coronagraph is $\sim\lambda/d$, canceling out the wavelength dependence. (A disk coronagraph would increase the noise contribution from the corona by ~10%.) (2) Within heliocentric ranges of interest, the SNR improves almost linearly with the heliocentric distance. Although the angular size of the Einstein ring decreases as $\propto 1/z^{0.5}$, the plasma contribution diminishes much faster, as $\propto 1/z^{4.4}$ (see Figure 14). Combination of these two factors results in the overall $\propto z^{1.7}$ behavior of the SNR. (3) The SNR has a rather strong dependence on the telescope aperture, behaving as $\propto d^{1.5}$. Again, this is due to our use of the annular coronagraph in deriving the estimate of the solar corona signal.

### 2.4.3 Signal-to-noise estimates for realistic sources

During Phase II we studied imaging with the SGL in a context of a realistic deep space mission (Turyshev & Toth, 2020c). We considered an exoplanet that is 30 pc away and accounted for the zodiacal background, solar corona brightness, spacecraft jitter, realistic losses, etc. We assumed a coronagraph suppression of $10^{-6}$. With these assumptions, we estimate that a 1-m telescope, operating beyond 650 AU would allow reaching a post-deconvolution SNR of 7 in ~1 year, yielding an image of this target with (100×100)-pixel resolution.

Creating a megapixel image requires ~$10^6$ separate measurements. For a typical photograph, each detector pixel within the camera is performing a separate measurement. This is not the case for the SGL. Only the pixels in the telescope detector that image the Einstein ring measure the exoplanet, and the ring contains information from the entire exoplanet, due to the blur of the SGL and also to the relative distribution of different regions of the exoplanet to different azimuths of the ring.

What is encouraging is that temporal variability in the cloud cover helps the deconvolution. Assuming $N$~50 observations of every pixel, clouds "disappear" after ~10 observations. If spectroscopic data is also used, we can reduce this issue and "see through" the clouds. The deconvolution of the data from several s/c allows to see the surface of the Earth in a few months of data.

By studying direct deconvolution, we have shown that with a 1-m telescope we would need ~1 year to build a (100×100)-pixel image with SNR~7. Two factors that can reduce the integration time by a factor of up to $10^2$ are i) the number of image pixels, $N$, and ii) the telescope diameter. The higher the desirable resolution, the longer the integration time, which scales as $T \sim N D^4$, where $D$ is the distance to the target. Another scaling law is related to the telescope diameter, $d$. A telescope with double the diameter will collect four times as many photons; its diffraction pattern will be twice narrower and, thus, it will collect half as many corona photons. As a result, the integration time scales as $T \sim 1/d^3$. Thus, a larger image of $10^3 \times 10^3$ pixels of an exoplanet at 25 pc may be produced in ~6 years with a 2-m telescope. This time may be reduced if there are time-varying features of predictable periodicity on the planet's surface or in its atmosphere. Also, the integration time is reduced by a factor of ~$1/n$ if we fly $n$ imaging spacecraft. Other factors to consider: i) the rotational motion of a crescent exoplanet could help in rotational deconvolution, ii) the heliocentric distance (i.e., the solar separation of the Einstein ring) affects noise.

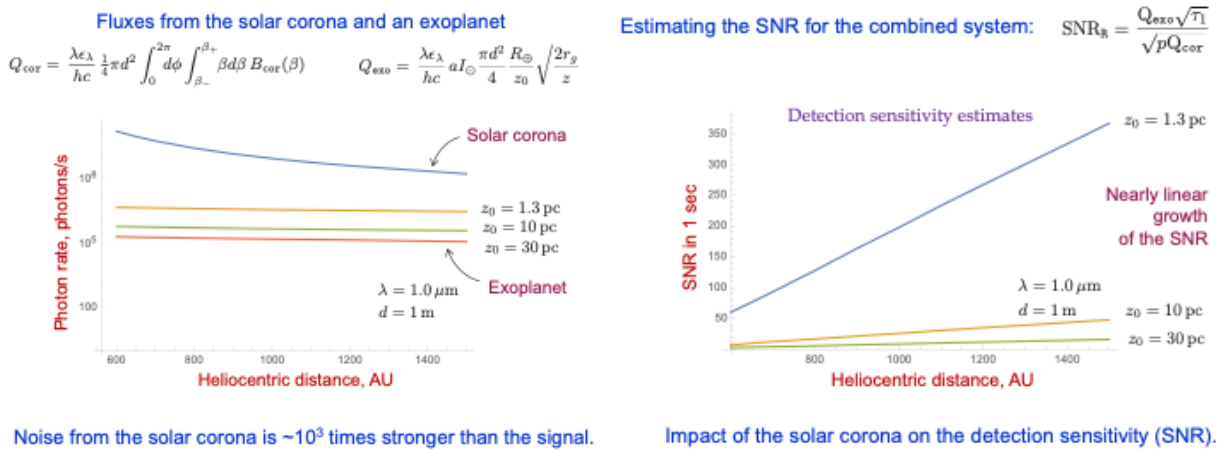

Figure 15. Photon fluxes, SNRs in the presence of the solar corona brightness.

The simulation tools using the rotational and direct deconvolutions developed during Phases I and II (not available any previous studies of the SGL), are the key innovative features of our proposal. Not only do they demonstrate that the SGL enables the high-resolution imaging and spectroscopy, they provide important insight into the SGL's optical properties. As such, these analysis tools will be used to inform our upcoming instrument/mission designs, and relevant trade studies.

The key mission and instrument requirements are driven by the process of image formation. Toward that goal, we identified a number of key driving parameters that would allow us to do a trade study needed to identify the most optimal mission design, namely i) heliocentric distance, ii) aperture of the telescope, iii) integration time, iv) detector sensitivity and its type, v) coronagraph or starshade performance, etc. Such a trade space is another mission-related innovation. To assess the value of the estimates obtained in the processing section, we need to consider them with the realistic imaging scenario. We considered the power received from the directly imaged region together with the power received from the rest of the planet and that from on off image pointing.

We observe that, given the assumed uniform source brightness, $B_s$, the image plane is uniformly illuminated with the signal whose power at each unique telescope position on the image plane (i.e., image pixel), $P_{dir}$, corresponding to the photon flux, $Q_{dir}$. However, in addition to this signal and depending on the telescope's position in the image plane $\rho_0$, the PSF of the SGL contributes a very strong blur signal from the rest of the exoplanet. The power of this blur signal, $P_{blur}$, corresponds to the corresponding photon flux, $Q_{blur}$. Furthermore, if the telescope moves outside the image, it will observe no contribution from $Q_{dir}$, but will continue to detect a signal corresponding to that for off image pointing, $P_{fp.off}$ and flux $Q_{fp.off}$ estimated by the same expressions as $P_{fp.blur}$ and $Q_{fp.blur}$. This signal is recovered during deconvolution process that we briefly address next.

### 2.4.4 Deconvolution and integration time. Lessons learned for a SGLF mission

We investigated the image formation process with the SGL. For that, we analyzed the EM field originating from an extended, resolved source and received in the focal plane of an imaging telescope, represented by a thin convex lens. Our estimate for the SNR of the deconvolved signal can be directly compared against simulated exoplanet image reconstruction at various levels of noise. Since the PSF of the SGL is known, convolution and deconvolution of a simulated image is a relatively straightforward process (Toth & Turyshev, 2020c).

We estimated the convolved signal received from a uniformly illuminated source and measured at a particular location in the image plane. In the presence of the solar corona, we obtain an estimate for the $SNR_R$ of the deconvolved image in the presence of the solar corona which has the from $SNR_R \geq (10/\sqrt{N})Q_{planet}/\sqrt{Q_{cor}}$, where $N$ is the total number of pixels in the image. This expression yields the following per-pixel integration time, $t_{pix}$, in the presence of the solar corona noise:

$$t_{pix} \leq 10^{-2} N \frac{Q_{fp.cor} SNR_R^2}{Q_{fp.blur}^2} =$$

$$= 3.54 \times 10^{-3} N\, SNR_R^2 \left[1 + 0.79 \left(\frac{650\ AU}{\bar{z}}\right)^{5.1} + 0.05 \left(\frac{\bar{z}}{650\ AU}\right)^{2.65}\right] \left(\frac{1\ m}{d}\right)^3 \left(\frac{z_0}{30\ pc}\right)^2 \left(\frac{650\ AU}{\bar{z}}\right)^{3.4}\ s.$$

This result suggests that for $d = 1$ m it could take up to $3 \times 10^3$ sec of integration time per pixel to reach the $SNR_R = 7$ for an image of $N = 100 \times 100 = 10^4$ pixels. For $z = 650$ AU, this translates into $t_{tot} = t_{pix} N \sim 1$ year of total integration time needed to recover the entire $100 \times 100$-pixel image of an exoplanet at 30 pc. Using for this purpose a larger telescope, $d = 2$ m, the per-pixel integration time drops to 390 sec, reducing the integration time required to recover an image with the same number of pixels to < 1.5 months of integration time. Use of a 5 m telescope implies a per-pixel integration time of ~150 s, for a total integration time of ~110 days for a $250 \times 250$-pixel image.

Collecting more data at different temporal periods will allow us to account for the diurnal rotation of the exoplanet and its variable cloud cover. To compensate for the diurnal rotation, we may also benefit from a multitelescope architecture that can reduce the total integration time (Turyshev et al., 2018), while matching the temporal behavior of the target. However, if the direct spectroscopy of an exoplanet atmosphere is the main mission objective, this can be achieved with a single spacecraft. We emphasize that direct imaging and spectroscopy of an exoplanet at such resolutions are impossible using any of the conventional astronomical instruments, either telescopes or interferometers. The SGL is the only means to obtain such results.

In Figure 16, we show the results of a simulated convolution of an Earth-like exoplanet image with the SGL PSF and subsequent deconvolution. The top row depicts the result of deconvolution of a monochrome image of the target, using modest image resolution (128 × 128 image pixels). We estimate that an image of this quality is achievable with less than 1.5 years of cumulative integration time even for a source at a distance of 30 pc, using only a single 1-m telescope, situated at 650 AU from the Sun.

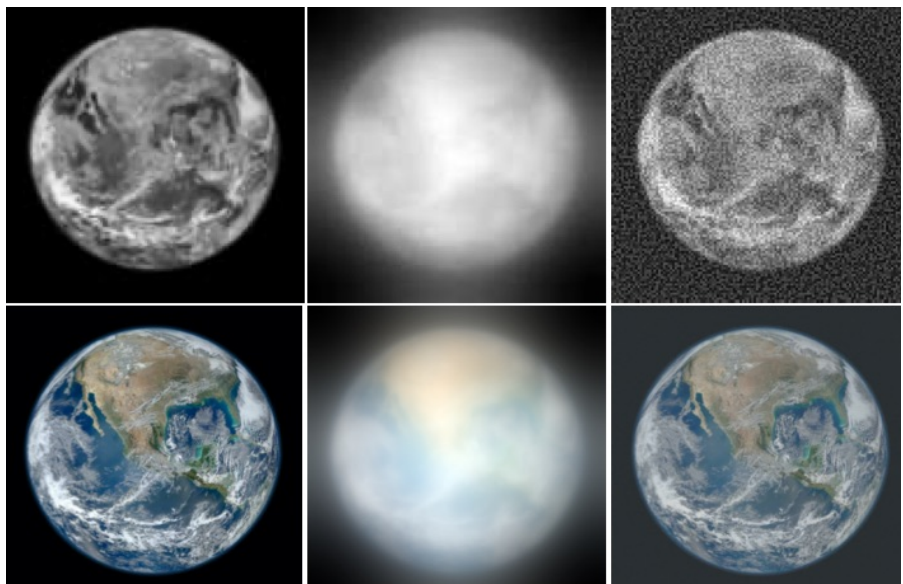

Figure 16. A simulation of the effects of the monopole solar gravitational lens on an Earth-like exoplanet image. Top row left: a monochrome image, sampled at 128x128 pixels; center: blurred image; right: deconvolution at SNR~64, corresponding to an image sampling depth of 6 bits. Bottom row, left: original RGB color image with a 1024x1024 pixel resolution; center: image blurred by the SGL, sampled at an SNR of ~$10^3$ per color channel, or overall SNR of $3\times10^3$; right: the result of image deconvolution.

Clearly, the SNR and the resulting image quality can be much improved by using a larger telescope, conducting an observational campaign at a greater distance from the Sun, and of course, using multiple instruments. A much more ambitious image reconstruction is depicted in the bottom row of in Figure 16: a high-resolution (megapixel) RGB color image of an exoplanet. An image of this quality requires a sampling of (1024 × 1024) image pixels in three color channels. Yet even this is within the realm of the feasible if we consider a target at $z_0 = 3$pc, observed through the SGL using 2.5-m telescopes at $10^3$ AU from the Sun. The cumulative integration time needed to obtain this image is less than 20 years; with two or more instruments, it may be possible to collect the necessary data over the course of less than a decade.

These estimates demonstrate that utilizing the SGL to obtain a good quality resolved image of an exoplanet of interest within 30 pc from the Earth is firmly within the realm of the possible.

Concluding, we note that the properties of the exoplanet (size, distance, albedo, parent star brightness, etc.), telescope parameters (aperture size, optical throughput, etc.), coronagraph parameters (annular vs. disk, contrast ratio, etc.), increasing heliocentric distance (as the spacecraft travels along the optical axis), use of multiple telescopes, spectral filtering and other factors may improve the SNR estimates. However, already at this level, the results are promising, justifying further studies. Such work is ongoing and results, when available, will be reported elsewhere.

## 2.5 Requirements and Design Considerations for a Mission to the SGLF

### 2.5.1 Telescope

Given our work during Phase II (Turyshev & Toth, 2020c), we see a need for a 1-m telescope (or larger) for imaging with the SGL. The larger telescope allows for shorter integration time, thus reducing the total time to collect the data needed for high-resolution imaging and spectroscopy. Given the size of the instrument, a monolithic telescope has the advantages, as compared to a segmented one. Thus, having access to a 1-m monolithic telescope is our baseline choice for the mission to SGL, as was discussed in (Turyshev et al., 2018).

Clearly, a segmented mirror achieved by the in-space telescope assembly would offer more light collecting power, which is a much-preferred option. In space assembly is a technique that currently is rapidly evolving (Stahl et al., 2005; Mukherjee et al., 2019). One can envision assembly of a large optical telescope with various pieces delivered on several smallsats to assembly region, say beyond the orbit of Jupiter, using modular approach demonstrated by NovaWurks.[7] These various pieces, that could be made in a LEGO-style, are then assembled, while performing proximity operations on a fast-moving hyperbolic trajectory exiting from the solar system. This concept is currently investigated for a potential space deployment and results, when available, will be reported.

### 2.5.2 Coronagraph

As the instrument ultimately determines the size of the spacecraft and, thus, its motion in the image plane, we began our efforts with a coronagraph design. We require the coronagraph to block solar light to the level set by the solar corona brightness at a given position of the Einstein ring.

At 1 μm, the peak light amplification of the SGL is ~$2\times10^{11}$ (equivalent to an increase of 27.5 mag), so an exoplanet, initially is seen as an object of 32.4 mag, becomes a ~4.9 mag object. When averaged over a 1-m telescope aperture (reducing light amplification to ~$2\times10^{9}$), it would be 9.2 mag object, still sufficiently bright. However, also present in the image are photons from the solar corona, the residual solar light, and the zodiacal light. Light contribution from the parent star must also be considered, even though its light is focused thousands of kilometers away from the exoplanet optical axis, due to the very slow fall-off of the SGL PSF; however, light from such an off-image source only affects narrow sections of the Einstein ring and can therefore be accounted for during image processing.

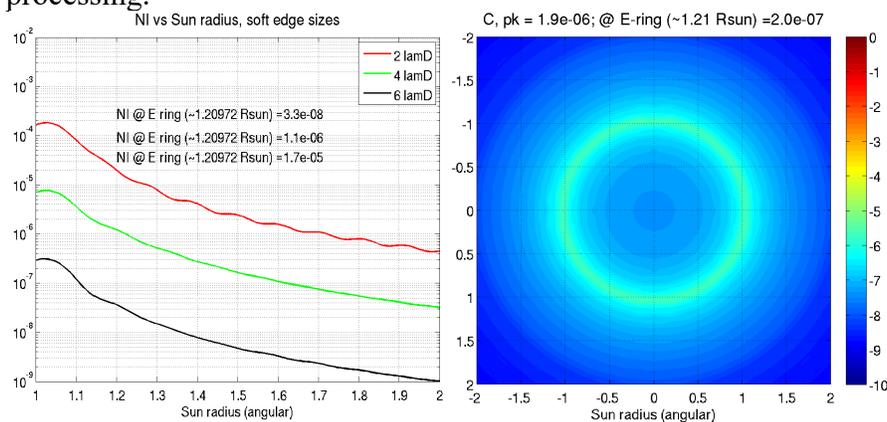

Figure 17. Left: Gaussian soft edge has a great impact on light suppression ability of the coronagraph. Right: Simulated coronagraph performance showing the solar light suppression by $2\times10^{-7}$, sufficient for imaging with the SGL.

---

[7] https://www.novawurks.com/

To validate our design assumptions, during Phases I/II, we performed a preliminary coronagraph design and simulations. Suppressing the Sun's light by a factor of $10^{-6}$ when imaging with the SGL is significantly less demanding than the modern-day exoplanet coronagraphs which aim to suppress the parent star's light by a factor of $10^{-10}$ to detect an exoplanet at least as a single pixel. The coronagraph bears greater similarities to coronagraphs used in solar astronomy rather than starshades used in exoplanet imaging (Cash, 2011). A traditional solar coronagraph uses sharp edge masks in both the focal plane and pupil plane. A more complex mask is needed if a classic Lyot coronagraph is used as in our case. Specifically, a mask flat with radius of the Sun, having a Gaussian soft edge profile achieved better rejection than a sharp-edge mask.

We evaluated the performance of the coronagraph with a Fourier-based diffraction modeling. The Sun is modelled as a collection of incoherent point sources with its corona (~$1/r^3$ power law profile). Design parameters include telescope size, distance to the SGL, occulter mask profile, and Lyot mask size. The FWHM of the Gaussian soft edge, has a significant impact on the coronagraph's performance (Figure 17).

Defining contrast as brightness normalized to peak brightness without coronagraph, we achieved a total planet throughput of ~10% (meaning that only 10% of planetary light reaches the sensor.) Figure 17 shows the contrast at the image plane after the coronagraph. At a contrast of $2\times10^{-7}$, the leaked solar light is ~5 times lower in intensity than the corona suggesting that we could have a realistic instrument design satisfying the stated objectives for imaging with the SGL. Our preliminary results suggest that heliocentric distances of 750–900 AU and a ~2-m telescope are needed to provide an optimal combination to suppress solar light.

- **The instrument:**
  - A diff.-limited high-resolution spectrograph, enabling Doppler imaging techniques;
  - Given the rapid development of coronagraphic capabilities, we can assume that direct imaging will provide spectro-photometric characterization of the exo-Earth.
- **Coronagraph is needed to block the light from our Sun:**
  - A conventional coronagraph would block just the solar light, but we want the coronagraph to transmit light only at the Einstein ring (where the planet's light is).
  - When averaged over a 1m telescope the planet is sufficiently bright;
  - To derive an image with the SGLF, including solar corona brightness (the parent star will be resolved), zodiacal light, instrument, and s/c systematics;
- **Parent start light contamination is not an issue**
  - Parent star is resolved from the planet ~0.01 AU away from the optical axis, making the parent star contamination issue negligible.
- **Perhaps several small spacecraft?**
  - We could rely on a swarm of small spacecraft, lunched together each moving at a slightly different trajectory parallel to the optical axis.

Figure 18. Instrumental concept for imaging with the SGL.

When viewed from greater heliocentric distances, the Einstein ring appears further away from the Sun, resulting in the rapid increase of its brightness relative to the solar corona. This option opens the design trade in which we would offload some of the difficulties in achieving the required optical performance on the mission architecture and designs needed to reach the deeper space regions. Alternatively, we may consider an external coronagraph (e.g., starshade-like) or a hybrid internal/external coronagraph. The use of a starshade could allow for significant reduction in aperture size of the telescope providing an opportunity for a swarm of small s/c to the considered for the task. General coronagraph requirements (applicable to both disk and annular approaches) and concept of their operations are summarized in Figure 18.

### 2.5.3  Choosing the targets and preliminary knowledge of the target

The occurrence rate of Earth-sized terrestrial planets in the habitable zones (HZ) of Sun-like (FGK) stars remains a debated quantity (Torres et al., 2015). Estimates range from 2% (Foreman-Mackey et al., 2014) to 22% (Catanzarite & Shao 2011, Petigura et al., 2013). The Simbad database lists 1688 F stars, 5,309 G stars, and 8,589 K stars within 30 pc. Taking the lowest estimates, we expect ~280 terrestrial planets in the HZ of a star within 30 pc to be detected in the near future. Once such a planet is discovered, significant observational resources will be devoted to study it.

- We want to image Earth 2.0, around a G star, which is not transiting:
  – Once habitability is confirmed ("big TPF" for spectra), the next step is to image it.
- We will rely on astrometry, RV, spectroscopy, and direct imaging to obtain:
  – orbital ephemeris: to ~mas accuracy and precision;
  – rotation: from temporal monitoring of the spectroscopy;
  – atmosphere: temperature, structure, chemical composition, and albedo, from non-spatially-resolved spectroscopy;
  – understanding of cloud & surface properties from Doppler imaging.
- This information will help us to point the s/c:
  – Time to reach 550 AU ~10 years, enough to observe the parent star's location ~100 times with 1 $\mu$as precision, so that its position would be known to 0.1 $\mu$as;
  – The parent star's position would be known to ~45 km at a distance of 30 pc;
  – Orbital period to <1% ⇒ the semi-major axis is known to ~0.7% (~1 million km);
  – If face-on, the radial distance to ~1 million km, with tangential error ~6 larger;
  – Earth's diameter is 13,000 km, so we will search the (80 × 500) grid on the sky;
  – Once SGLFM detects the planet ⇒ scan a smaller area to define the "edges".

Figure 19. The a priori properties of the target.

Most likely, we will want to image Earth 2.0, around a G star, which is not transiting. An SGL mission could follow a "big TPF" (i.e., terrestrial planet finder) that will observe an exoplanet around a G star and measure its spectra. We should be very confident that the target is habitable. A spacecraft at the SGL would be the next major step, possibly the biggest step in the 21st century for exoplanet exploration.

- Imaging is done on a pixel-by-pixel basis:
  – The image of an exo-Earth occupies ~(1.3km×1.3km) area from the optical axis.
  – Each pointing corresponds to a different impact parameter: 1 image ⇔ 1 pixel.
  – Between the adjacent pixels the impact parameter changes, brings light from adjacent surface areas on the planet ⇒ a raster scan moving the spacecraft;
  – To build a ($10^3 \times 10^3$) pixels image, we would need to sample the image pixel-by-pixel, while moving in the image plane with steps of ~1 km/$10^3$ = 1 m:
    • Pointing: Inertial navigation and 3 laser beacon spacecraft in heliocentric orbit in the plane of the Einstein's ring (for precision pointing & comm).
  – Contamination from the parent star is negligible for an SGL scenario.
- Exoplanet imaging requires several key technologies that are challenging:
  – determination of an exoplanet astrometric orbit at ~10 nas,
  – motion & stabilization of the s/c over millions of pointings with limited power.
- Perhaps even spectroscopy or even spectro-polarimetry of the exoplanet?
  – Potentially a spectrally resolved image over a broad range of wavelengths: atmosphere, surface material characterization, biological processes.

Figure 20. Imaging approach with the SGL

Once we know of a terrestrial HZ planet so close to our own, we posit that significant resources will be devoted to characterizing the planet and its system using the conventional techniques above. The knowledge we gain from this will include: i) orbital ephemeris, to ~mas accuracy and precision, ii) detailed knowledge of the atmosphere, including temperature, structure, chemical composition, and albedo, all inferred from spatially unresolved spectroscopy; iii) estimates of rotation rate, gained from temporal monitoring of the photometry, iv) some understanding of cloud and surface properties from Doppler imaging (Crossfield et al., 2014) or rotational deconvolution.

The planetary orbit would have to be measured in 3D, using either astrometry and/or RV (radial velocity) measurements together with direct imaging. It may also be inclined so that it transits, providing a radius.

The SGL mission would begin after the discovery of an Earth-like exoplanet. It could take ~20 years of "cruise" for the spacecraft to reach 548 AU. During those 20 years, the parent star's location would be observed with 1 μas precision at least 100 times, so that its position would be known at 0.1 μas level. The parent star's position would be known to ~45 km at 100 ly. The orbital period of the planet would be known to <1% meaning that the semimajor axis is known to ~0.7% or ~1 million km. If the planet is in a face-on orbit, we will know the radial distance to ~1 million km, but the error in the tangential direction will be ~6 times larger. The diameter of the Earth is ~13,000 km, so that the area on the sky we must search is an (80 × 500) grid. Once the SGL telescope detects the planet, it would scan a much smaller area to define the "edges" of the planet. Astrometry of the star when planet was discovered would have measured its mass, that plus its size give us the density of the planet. With this critical information about the target, we may proceed with a mission design.

### 2.5.4 Imaging with the SGL

The image of an exoplanet at 30 pc is compressed by the SGL to a cylinder with a diameter of ~1.3 km in the immediate vicinity of the focal line. (This diameter corresponds to the Einstein ring around the Sun with the same thickness of 1.3 km.) Imaging the exoplanet with $10^3 \times 10^3$ pixels requires moving s/c in the image plane in steps of 1.3 km/$10^3$ ~1.3 m. So, each ~1 m pixel in the image plane correspond to $13 \times 10^3$ km/$10^3$ ~ 10 km pixel sizes on the surface of the planet.

The challenge comes from recognizing the fact that the PSF of the SGL is quite broad (see Figure 4), falling off much slower than the PSF of a typical lens. For any particular pixel on the image plane, this leads to admixing the light from many surface pixels adjacent to that on the instantaneous FL into that one image pixel. This admixing results in a significant image blurring.

To overcome this challenge, imaging must be done on a pixel-by-pixel basis by measuring the brightness of the Einstein ring at each of the image pixels. With the knowledge of the PSF one can use modern deconvolution tools that allow for efficient reconstruction of the original image. However, for this process to work, a significant signal to noise ratio (SNR) is required. Luckily, the SGL's magnification leads to SNR of over $5 \times 10^2$ in 1 sec, sufficient for a nearly noise-less deconvolution. (We discuss results of our simulations of such an image reconstruction in Sec. 2.4.4.)

Light contamination from the parent star is a major problem for all modern planet-hunting concepts. Despite to the SGL's ultra-high angular resolution (~0.1 nas) and a very narrow FOV, the parent star still contributes light to the Einstein ring, but as the image of the parent star is centered ~$10^4$ km away from the optical axis, this contribution is limited to narrow segments of the Einstein ring, which can be removed during image processing.

### 2.5.5 Trajectory requirements

Accounting for the motion of the Sun in the solar system BCRF (Figure 21), consider the plate scale. The SGL reduces the size of the image of the exoplanet at 30 pc by a factor of ~10,000 at 650 AU. An orbital radius of 1 AU becomes ~$1.5 \times 10^4$ km and an orbital velocity of 30 km/s translates into 3 m/s. Solar gravity accelerates the Earth at 6 mm/s$^2$. Consequently, the imager s/c needs to accelerate at ~1 μm/s$^2$ to move in a curved line mimicking the motion of the exoplanet.

Even if the probe mass is ~1,000 kg, the corresponding force is 1 mN, which may be achieved with electric propulsion. It is even more reasonable for a ~100 kg notional s/c, assumed for this study. Whether it is feasible for interplanetary propulsion, electric propulsion may be sufficient for maneuvering to sample the pixels in the Einstein ring needed for imaging.

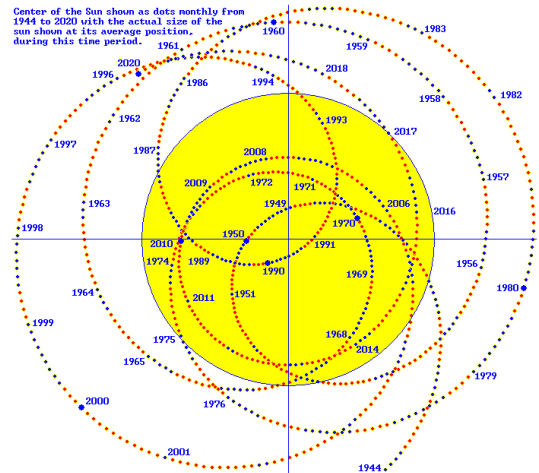

Figure 21. Solar motion in the BRCS as seem from the distance of 10 pc (i.e., solar wobble).

We investigated the attitude control of the SGL s/c. For that we considered the use of 3-axis stabilized spacecraft with a few microarcsecond pointing knowledge and stability in combination with a set of laser beacons in the inner solar system. We studied the motion in the image plane needed to sample ($10^3 \times 10^3$) image pixels. This analysis allowed us to formulate and study the image reconstruction requirements that led us to formulate the key mission and instrument requirements. The also helped us to analyze image formation processes and to derive realistic mission requirements with relevant architecture trades.

### 2.5.6  Towards mission design

Our analysis of the relevant SNR has suggested that the solar corona is the largest source of systematic noise for imaging of faint sources with the SGL. Thus, the mean corona signal must be measured and subtracted photometrically perhaps with a dedicated spacecraft that hosts an identical set of instruments including telescope and coronagraph. As the spacecraft moves outwards from the solar system, its coronagraph may have to be able to compensate for the decreasing angular sizes of all the objects involved – the sun and the Einstein ring, etc.

- Pointing precision (between three objects):
  – Needs to be maintained to ~ few µas for proper operation of the SGL.
  – Knowledge is needed at 1 µas level, control is at the ~100 µas.
  – The motion is unfortunately complex (1-m of motion at 600 AU ~ 1 µas of angle seen from Earth)
- Simple motions (straight lines):
  – Motion of the target star around the galaxy; the Sun around the galaxy
- More complex motions:
  – Motion of the exoplanet around its host star (Keplerian)
  – Motion of our Sun around the solar system barycenter.
    • Dominated by the orbits of Jupiter, Saturn.
    • Jupiter ⇒ 75 million m motion of the Sun (12 yr orbit)
    • Saturn ⇒ 50 million m of motion of the Sun (29 yr)
    • Earth  ⇒ 450,000 m (1 yr)
- Propulsion system must compensate for the reflex motion of the Sun
  – Due to most of the planets in the solar system motion over a short time may be just a straight line.

Figure 22. Trajectories for SGL spacecraft

For a given number of desirable pixels, more distant (fainter) planets will require longer integration times resulting in a longer imaging mission phase. In fact, image quality improves with time allowing for more repeat scanning of the same pixel and also as spacecraft moves to further heliocentric ranges the SNR increases. These factors allow for an improved image reconstruction.

It is desirable for the duration of the imaging mission phase to be on the order of 10 years. This would translate into much increased image quality and temporal resolution of the atmospheric and surface processes occurring on the target exoplanet.

- The mean corona signal must be measured and subtracted photometrically
  - perhaps identical telescope & coronagraph on a separate spacecraft?
  - Coronagraph may have to be able to compensate for decreasing angular sizes
- For a given number of pixels, more distant planets will require longer integration times resulting in a longer imaging mission phase
  - Image quality improves with time: 1) more repeat scanning of the same pixel and also 2) spacecraft moves to further heliocentric ranges with higher SNR
  - Duration of the imaging mission phase: 10 years
- To improve measurement sensitivity we are pushed towards:
  - larger heliocentric ranges: 700-1000 AU
  - larger (effective) aperture: 1-2 m (perhaps formed with smaller telescopes)
  - more imaging spacecraft: 2 is good, more would be better
- Per pixel integration (dwell) time for N = 100 pixels at 800 AU:
  - @ 800 AU:   31 sec (1.3pc), 30 min (10pc), and 4.5 hour (30pc)
  - @ 1000 AU:  13 sec (1.3pc), 13 min (10pc), and 2.0 hour (30pc)
- Observational scenario
  - Closer exoplanets: slow movements of spacecraft in the image plane
  - Farther exoplanets require pointing: point, integrate, slew to the next pixel

Figure 23. Mission implications for SGLF.

To improve measurement sensitivity, we are pushed towards: 1) larger heliocentric ranges of 700-1,000 AU, 2) larger (effective) apertures of 1-2 m (perhaps formed with smaller telescopes), and 3) more imaging spacecraft where $n = 2$ is good, but $n > 2$ would be even better (the integration time will be $1/n$ times shorter.) In fact, the more spacecraft would observe the target, the better.

If we are to make an image with 100 linear pixels, the per pixel integration (dwell) time @ 800 AU would be 31 sec (for a target at $z_0 = 1.3$ pc), 30 min (for $z_0 = 10$ pc), and 4.5 hours ($z_0 = 30$ pc). This time decreases as we move outwards. Thus, at 1,000 AU: we would have to integrate for 13 sec ($z_0 = 1.3$ pc), 13 min ($z_0 = 10$ pc), and 2.0 hours ($z_0 = 30$ pc). This effect should be incorporated in the mission scenario. As far as observational scenario is concerned, for closer exoplanets, the spacecraft may move slowly in the image plane. However, more distant exoplanets require pointing and dwell time. Thus, mission must be able to point, integrate, slew to the next pixel. These considerations will have to be incorporated in the ultimate mission planning.

With our proposed mission concept (Section 3) we can deliver an array of optical telescopes to the focal region of the SGL and then fly along the focal line to produce high resolution, multispectral images of a potentially habitable exoplanet. Our multiple smallsat architecture is designed to perform concurrent observations of multiple planets and moons in a target exoplanetary system. It allows for a reduction in integration time, to account for target's temporal variability, to "remove the cloud cover". Clearly, as long as spacecraft are healthy, the mission will yield more valuable data, thus increasing the image resolution and SNR, and studying life on that exoplanet.

The new architecture developed in this study uses smallsats (<100 kg) with solar sails to fly a trajectory spiraling inward toward a solar perihelion of 0.1-0.25 AU and then out of the solar system on a nearly radial-out trajectory at 15-25 AU/year. Our design goal is 25 AU/year to permit reaching the SGLF operating region in <25 years. A long time, but less than the time Voyager took to reach the heliopause – less than 1/5 the distance of our goal in the far interstellar medium. Today we are technologically ready to size the unprecedented opportunity of using the SGL with a mission transit time of only ~2.5 times longer than the transit time of New Horizons to Pluto.

As a result, our baseline mission design assumes ~25 years of transit time to the operating distance in the focal region of the SGL moving at ~25 AU/year and then up to 15 years of science operations, thus covering heliocentric ranges from 650-1000 AU. With these enormous distances from the Earth our s/c must be injected on the trajectory with a very precisely defined and maintained position state vector enabled various navigational means and have significant autonomy capabilities relying on artificial intelligence and machine learning to be discussed in Section 3.

## 3 EXPLORING FURTHER & FASTER WITH A NEW MISSION ARCHITECTURE

The innovative features of our work rely on the recognition that many of the technologies needed to reach the SGL and to operate in the vicinity of its instantaneous FL are already at hand. Progress is required for the long duration spaceflight to reach the large heliocentric distances. Various technological aspects have been demonstrated by the Voyager 1 and 2, Pioneer 10 and 11, and New Horizons missions. With further progress in the development of highly capable small spacecraft, electric propulsion techniques, optical comm, foldable optical-quality mirrors, exoplanets research would gain imaging and spectral information by these carefully planned space missions.

### 3.1 Conceptual designs for the SGL mission

Many studies investigated the science objectives and the technological feasibility of missions beyond the solar system, including several NIAC studies. Our work benefited from these earlier studies by allowing us to focus on the SGL-specific features. Clearly, a mission design to the FL region of the SGL presents a set of interesting challenges. Table 1 presents the summary of the mission architecture issues and the required mission tradeoffs.

We considered the trades between a single telescope vs. a segmented imaging approach based on smallsats. The latter option opens the possibility of sending multiple spacecraft (Table 1). Therefore, two different classes of mission architectures were to be studied: 1) a single flagship-class spacecraft or smallcraft that relies on solar sail technology, 2) a multiple spacecraft mission architecture (with/without solar sail technology). These options allow us to explore the entire trade space leading to a more optimal architecture (Turyshev et al., 2018).

#### 3.1.1 A flagship probe approach

Given the long mission durations to the SGL, radioisotope power (i.e., RTG) is required, for instance advanced segmented modular RTG (SMRTG). The SMRTGs are the proposed next-generation of vacuum RTGs, capable of providing almost 5X more power at the end-of their lives over the Mars Curiosity MMRTG and 2X more power over the Cassini GPHS RTG. They take advantage of the skutterudite technology which is already being matured for the eMMRTG and use multifoil insulation and aerogel encapsulation to achieve high efficiency and low degradation rate.

During Phase I/II, we formulated an SGL mission concept capable of achieving an escape velocity of ~20 AU/yr. This objective is particularly driving, requiring a ΔV of >10 km/s at perihelion. Arora et al. (2015) suggest that achieving this requires moving beyond the traditional solid rocket motor (SRM) or bipropellant rocket engine. Among the two viable propulsion candidates – nuclear-thermal propulsion (Larson et al., 1995) and solar-thermal propulsion (STP) (Layman et al. 1998) – we selected an STP-based architecture as the most viable option for SGL mission.

Defining a baseline concept (via a 2017 JPL Team-X study), lead to a feasible point design: the baseline launch stack for this mission concept consists of a spacecraft (~550 kg wet mass), perihelion maneuver stage ($H_2$ tanks, a bipropellant system), and STP system, including the heatshield, the heat exchanger, 12 engine nozzles. The concept is optimized to achieve ΔV over 11 km/s at the solar perihelion using an STP system. The STP-based concept for solar-system escape relies on using cryo-cooled $H_2$ as propellant, which is heated due to spacecraft's proximity to the Sun, using a heat exchanger, acting as part of a larger heatshield to protect the spacecraft.

The mission concept requires a perihelion ΔV of ~11.2 km/s. The burn time must be is <1.5 hrs. The probe uses an RTG-powered EP system providing an additional ~2.4 km/s of ΔV. The escape velocity achieved is ~19.1 AU/yr (~90.5 km/s). Given the high-ΔV requirements, the STP mission concept is very sensitive to the mass of the $LH_2$ tank, ISP, mass and support structure mass.

Alkalai et al. (2017) have considered a mission to the SGL that will be able to reach ~550-700 AU and deploy a 2-m telescope for multipixel imagining of an exoplanet. The spacecraft reaches 600 AU in < 40 yrs. Factoring in the time required to build energy in the inner-solar system for a type 2 trajectory, this results in an escape velocity of > 20 AU/yr. A baseline SGL mission will use advanced low mass and power technologies, onboard autonomy, and STP propulsion stage.

Table 1. SGL mission architecture and design tradeoffs required.

| System | Technology | Benefits, Costs, Requirements, Tradeoffs |
|---|---|---|
| Propulsion | Chemical | Big solid rocket burn very close to Sun; Massive shield; Exit velocity limit ~18 AU/year |
| | Solar Sail | Lightest weight option, requires A/m beyond state of art 300x300 m sail with 100 kg sc -> 25 AU/y; with REP 30 Au/y |
| | Nuclear Electric | Expensive, heavy spacecraft, programmatic challenges |
| SC Size | 200-500 kg (conventional) | Allows state of the art design; Likely flagship mission development; Probably required for chemical spacecraft; Accommodates bigger optics. |
| | <100 kg (smallsat) | Will lower cost, consistent with multiple s/c. May be mission enabling Consistent with "string of Pearls" architecture; Required for sail; Requires new technology: e.g. low mass RTG; Imaging system, communications capabilities and system reliability TBD |
| ConOps | Single spacecraft | Straightforward, minimizes ops complexity |
| | Multiple spacecraft | 1) Creates more flexible mission design and data collection; 2) Consistent with "string of pearls" architecture and other relay or distributed data schemes; 3) Open architecture allows for incremented improvements in technology; 4) Provides important redundancy |
| Comm | Radio | May be enabled by use of sail as antenna |
| | Optical | Likely better performance & implementation for given mass and power. Allows for optical ranging measurements |

As a result, we see that there already exists a feasible mission architecture capable of reaching the focal region of the SGL. The single-spacecraft mission concept has some obvious challenges, but it offers a baseline mission concept that may be used to further evolve the design in Phase II.

### 3.1.2   A solar sail mission concept

We considered a solar sail mission architecture. New technologies enabling such missions are smallsats (<100 kg with power, comm, precision navigation) and solar sails. One interplanetary sail has already flown towards Venus (JAXA's IKAROS) (van der Ha et. al., 2015) and another to a near-Earth asteroid is being developed by NASA (NEA Scout) (McNutt et. al., 2014).

Heliocentric distances >500 AU can be achieved in practical flight times with solar sails flying toward the Sun with a perihelion of 0.1–0.2 AU. Although the required spacecraft area-to-mass ratios (A/m) are larger than the current state of the art, the requirements are consistent with those studied and considered in prior NASA and ESA studies. Other relevant technologies—an Oberth maneuver (Stone et al., 2015) and electric propulsion—but solar (or possibly electric) sails appear to be both of greatest performance potential and of nearest term readiness.

Friedman & Garber (2014) studied solar sail requirements to reach velocities of >20 AU/yr. Garber (2017) considered the A/m requirements to reach an exit velocity of ~40 AU/year. Sail A/m ratios of 900 m²/kg yield 25 AU/year and 40 AU/year requires A/m = 2,550 m²/kg. A (300×300)-m sail with a 100 kg s/c could fly out of the solar system at ~25 AU/yr, reaching the SGL in <25 years.

Friedman & Turyshev (2017) assumed that a 200×200 m sail might achieve a solar system exit velocity of 25 AU/year. That could be achieved with a notional mass of 30 kg for the spacecraft bus; 13 kg for a radioisotope power system providing 100W of electric power and possibly a small maneuvering capability; and 1.6 kg sail with a density of ~0.4 g/m² (equivalent to 0.25-micron dielectric based boron nitride nanotubes). It remains to be determined if the radioisotope system can be smaller, or if it can contribute to the exit velocity with a propulsive boost.

The major advantage of the solar sail concept is that it offers a high solar system escape velocity and, thus, a fast transit time to the SGL region. It also enables an interesting tradeoff for using a swarm of small spacecraft (with onboard telescopes of ~50 cm) and rely on a common external coronagraph (i.e., starshade, similar to the one designed for WFIRST) placed on a separate spacecraft to block solar light. A formation flying aspect of operating the swarm is important for this architecture; it will be investigated further.

### 3.1.3 A "String of Pearls" (SoP) approach

Based on the results of a single spacecraft mission design, we investigated the "string of pearls" architecture (initially introduced and studied in Phase I). Each pearl within the string consists of a group of 10-20 mass-producible small to mid-size s/c that fly as a group to perform all needed functions with redundancy and adaptability. Each "pearl" (group of s/c flying in formation) is followed by another pearl at ~1-year intervals to the "string".

By flying pearls at 1-year intervals, with a solar system escape speed of ~20 AU, year, the pearls are separated by 20 AU, and each arrives at the SGL of their target parent star in ~25 years. We envision ~10 annual pearl launches. The first pearl would reach the FL of the SGL in ~25 years, and flies along the FL for ~8 years – the final pearl enters the SGL zone ~10 years later and culminates the data acquisition portion of the mission 8 years after that.

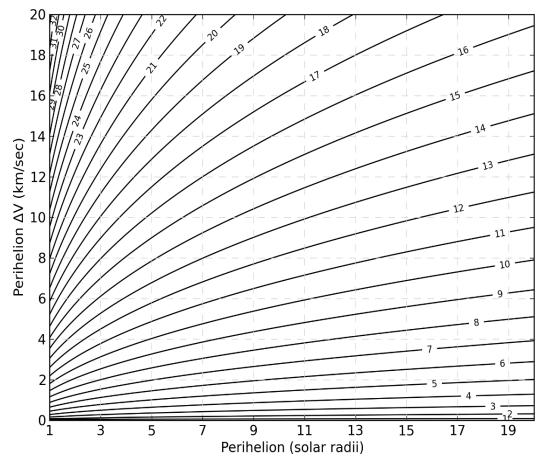

Figure 24. Required delta-V profile as a function of the solar distance at perihelion.

The cluster concept mitigates risk of a single point failure. Further advantages of this approach are that it relies on a mass-producible spacecraft; employs an open architecture design where any space-faring organization can contribute spacecraft/component; continuously enhances the design by learning from previous pearls - thus spreading the cost over decades while capitalizing on emerging technology.

We deliberately wanted to investigate such an open, flexible, evolving architecture as it allows us to explore the boundaries of the entire mission trade space envelope. Also, by loosely constraining the mission architecture with a common repeating component – the pearl – it is possible to study a new class of missions that today would be discarded solely based on the perceived high cost.

The technology requirements were of a particular interest. As many space systems are now being designed in anticipation of radical changes, focusing on resiliency, adaptability and disaggregation, we expect major progress in these areas in the near future, benefiting the SGL mission. We approached this study with the topics revolving around autonomy, positioning, navigation, communication, and onboard processing. We will focus on the mission architecture and application of small spacecraft. This includes operations in the vicinity of the FL to image the Einstein ring over a decade, possibly with cooperative sub-satellite small spacecraft at multiple locations.

The trade studies performed during our Phase II effort, favored the SoP approach and therefore the remaining sections of this Section will address only the SoP approach.

## 3.2   Solar sailing propulsion for the SoP mission architecture

### 3.2.1  *Inadequacy of the current chemical/nuclear propulsion*

The challenge of reaching >600 AU on a timescale that is not generational (i.e., less than 30 years), requires propulsion capable of velocities over 20 AU/yr. Current propulsion technologies are significantly challenged by these velocity requirements due to required order-of-magnitude advances in materials, structure, and power storage to enable the mission. Preliminary mission designs that utilize chemical and solar thermal require flybys of the Sun within 3 solar radii while nuclear electric propulsion options result in a maximum of a 40-year operating life resulting in a limited mission duration at the SGL. More exotic propulsion schemes, such as the electric sail, can achieve relevant velocities but require the deployment and control of 10s of km of tethers.

As a result of in-depth analysis of all the known propulsion options, we elected to focus on solar sails. This choice was made for several reasons outlined below:

(i)      Solar sailing requires nearly no propellant, which leads to very lightweight and lower cost spacecraft design.
(ii)     Solar sailing is the only propulsion that is currently on a technology roadmap (i.e., Breakthrough Starshot[8]) for interstellar flight.
(iii)    The solar sail – smallsat architecture of this study could be ready for flight project implementation within a decade, at a lower cost than other proposed concepts for reaching the SGLF within 30 (or even 50) years.
(iv)    Another advantage is a near-term (~3 years) would be a low-cost technology demonstration flight in the inner solar system that would produce the fastest spacecraft ever, capable of intercepting a newly discovered interstellar object (ISO) passing though the solar system.

The successful interplanetary flight of IKAROS (Interplanetary Kite-craft Accelerated by Radiation of the Sun)[9] solar sail from Earth to Venus by the Japanese space agency and recent successful orbital demonstration of LightSail-2[10] flight by The Planetary Society raise the confidence in (and the TRL of 7) solar sails. Indeed, two of NASA interplanetary missions, NEA-Scout[11] and Solar Cruiser[12], are planned for the near-future.  The alternative propulsion options are:

---

[8] https://breakthroughinitiatives.org/initiative/3
[9] https://space.skyrocket.de/doc_sdat/ikaros.htm
[10] https://www.planetary.org/explore/projects/lightsail-solar-sailing/
[11] https://www.nasa.gov/content/nea-scout
[12] https://ntrs.nasa.gov/archive/nasa/casi.ntrs.nasa.gov/20190032304.pdf

i) Chemical, using a low-perihelion large rocket burn (Oberth maneuver): Rejected because of the impractical low perihelion required for > 20 AU/year – approximately 0.2 solar radii (cf. Figure 24) and the consequently large spacecraft with massive shield and rocket motor;

ii) Solar electric rejected since it too requires too low a perihelion for operating solar cells;

iii) Nuclear electric in principle could achieve desired exit velocity, but the only design to date is a hybrid of a 30-kW electric system and 20-kW nuclear reactor. Even it would still take 40 years to reach SGLF and the reactor would be costly and politically prohibitive;

iv) Solar and nuclear thermal are being considered in a NASA new technology program, and in principle could satisfy the mission requirements. But the technology program is focused on a multidecadal, large budget spacecraft that with political uncertainty (of nuclear reactors in space) even larger than the budget uncertainty;

v) Electric sails – a promising new concept (on paper only) but unlikely to be applicable beyond 200 AU because of the sheer dimensions of the tethers required;

vi) Laser electric: A beamed laser powering electric propulsion is being studied in another NIAC study. This would give terrific performance, but it depends on a ~100-GW laser system on the back side of the Moon – beyond the scope and cost of anything we want to consider;

vii) Fusion rocket launch: Another NIAC study is considering a fusion rocket. Their design would reach the SGLF in 12 years. Unfortunately, it relies on the development of a practical fusion rocket, which is another idea beyond our scope of consideration.

Actually, our solar sail focus uses hybrid propulsion since either chemical or (more likely) electric propulsion is required for deep space navigation and maneuvers in the SGL focal region. The total propulsion system will still fit within the smallsat specification proposed in the mission architecture. It can also be ready for flight test and demonstration within a few years and for implantation on an SGLF mission in the current decade.

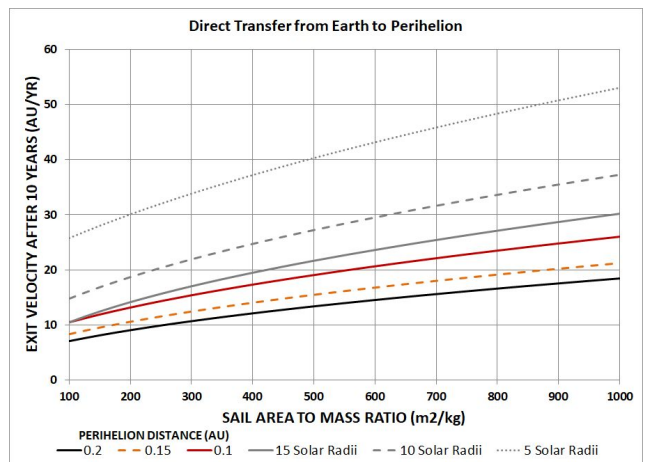

Figure 25. Sailcraft parameters to achieve high exit velocity.

The sailcraft parameters to achieve high exit velocity are shown in Figure 25. The relevant parameter is A/m, where A is the area in square meters of the sail, m is the total mass of the sailcraft in kilograms (spacecraft + sail), and the perihelion of the trajectory as it flies around the Sun before achieving solar system escape velocity. As can be seen from the graph, for a given perihelion the larger is the A/m ratio the higher is the exit velocity. Consequently, large area sails and low overall mass space spacecraft are needed. To ensure that the payload mass fraction is high, sail materials with very small aerial density would be required. The perihelion, in turn, defines how much sunlight power impinges on the sail – the lower the perihelion the higher is the sunlight flux and hence higher radiation pressure and thrust.

The solar sail perihelion maneuver is similar in nature to the so-called Oberth maneuver, when an impulse of a rocket engine fires at perihelion to gain highest delta-V in the most efficient way. Such a maneuver results in the raise of the aphelion (furthest distance on the trajectory), and if enough energy is gained during the maneuver, the aphelion goes to infinity – that is the elliptical orbit becomes a hyperbola, extending on a nearly straight line out of the solar system.

Figure 26 shows an example trajectory starting from Earth orbit, spiraling in toward the Sun and then achieving hyperbolic speed to the outer solar system and beyond. The figure shows a trajectory capable of catching and even rendezvousing with an interstellar object as it passes through the solar system – at a speed ~6 AU/yr. As seen in Figure 25 above we can achieve that even with a perihelion of 0.3 AU (i.e., Mercury orbit) and an A/m= 100 $m^2$/kg. These are parameters that can be achieved with today's sail technology.

As we want as low a perihelion as possible, and this defines the third parameter of the sailcraft design: ability to withstand high radiation flux emanating from the Sun. This will be determined by the solar sail material – plastic, metallic, ceramic, dielectric – whatever can be manufactured into large thin sheets. Figure 25 shows the parametric tradeoff of area, mass, perihelion distance to achieve a given high exit velocity. Perihelion distance is especially important for achieving the higher velocities. In Section 3.2.2 we discuss sail materials and their temperature properties – these will determine what perihelion can be achieved. Both laboratory and flight testing of sail materials are proposed for further study.

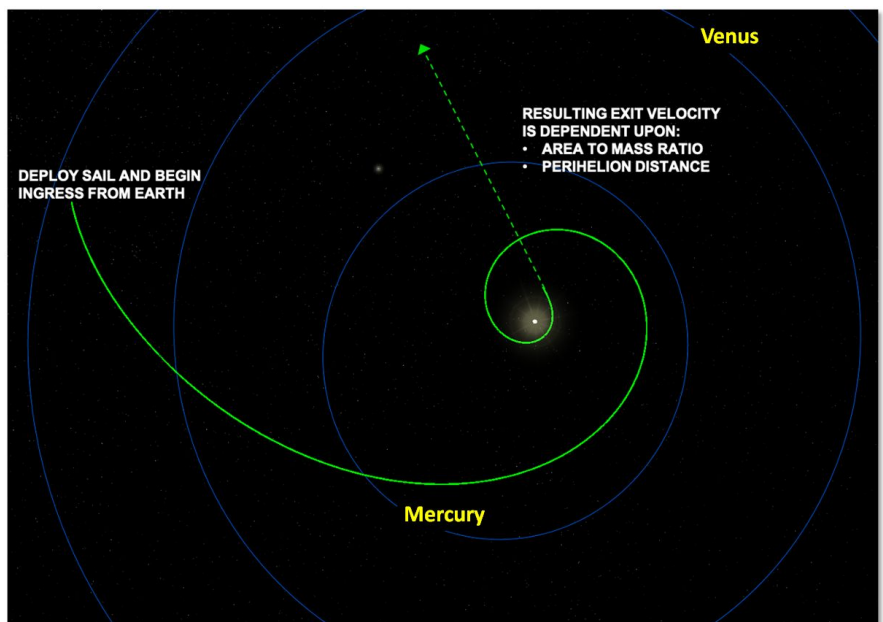

Figure 26. Sailcraft example trajectory towards the SGL.

The challenge for design of a solar sail is managing its size – large dimensions lead to unstable dynamics and difficult deployment. In this study we have consider a range of smallsat masses (<100 kg) and some of the tradeoffs of sail materials (defining perihelion distance) and sail area (defining the A/m and hence the exit velocity – see above figure). As an example, for the SGLF mission, consider perihelion distance of 0.1 AU (20$R_{sun}$) and A/m=900 $m^2$/kg; the exit velocity would be 25 AU/year, reaching 600 AU in ~26 years (allowing 2 years for inner solar system approach to the Sun). The resulting sail area is 45,000 m$^2$, equivalent to a ~212×212 m$^2$ sail.

To cope with that large dimension, a new design for solar sails was invented (and patented) by L'Garde and NXTRAC companies derived from the vane concept proposed for square solar sails. The new design, called SunVane, consists of several square panels aligned along a truss to provide the necessary sail area (Figure 27.) The truss can be extended to accommodate many such panels or vanes. The spacecraft (electronics, instruments, etc.) will be configured along the truss. The sailcraft design is discussed in Section 3.2.3.

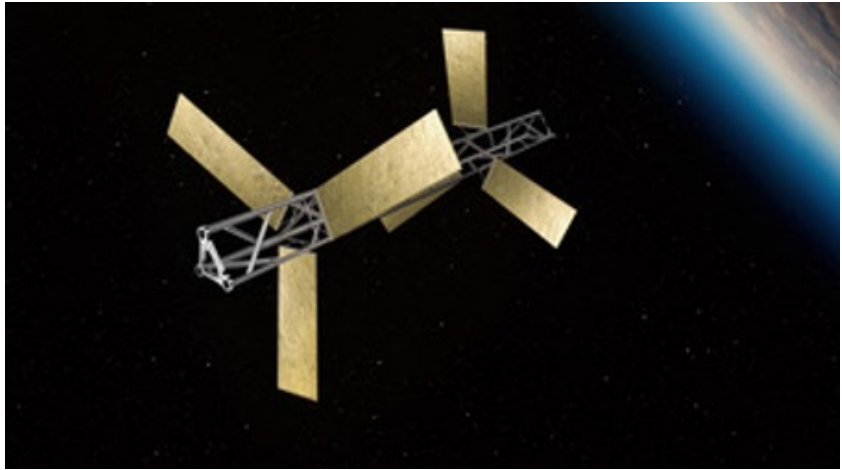

Figure 27. The SunVane sailcraft concept.

The trajectory design, depicted in Figure 25, enabled by the smallsat sail craft, to reach the SGLF can also be used in a more general mission design for solar system applications. Specifically, we note that one could design missions to stay in low circular orbit and then targeted to the interstellar object, or to any of the outer planets, their moons, or Kuiper Belt objects (KBO). This innovative new architecture for fast trajectories for outer solar system missions is a serendipitous result from the current study. In Section 5.2 we will propose a technology test of the smallsat-sail mission design, on such a trajectory – a low cost, near term test flight.

Interplanetary smallsats are still to be developed – the recent success of MarCO brings them perhaps to TRL 7. Solar sails have now flown – IKAROS and LightSail-2 already mentioned, and NASA is preparing to fly NEA-Scout. Scaling sails to be thinner and using materials to withstand higher temperatures near the Sun remains to be done. As mentioned above, we propose to do this in a technology test flight to the aforementioned 0.3 AU with an exit velocity ~6 AU/year. This would still be the fastest spacecraft ever flown. See Section 5.2 for discussion of the technology test which we have roughly estimated could be done within three years at a cost less than $40 million – and using a rideshare launch to approximately GEO.

### 3.2.2 Sail material

A close perihelion approach that will propel a sailcraft to >20 AU/year toward SGLF, as mentioned earlier, necessitates novel sail materials that will withstand high solar radiation flux. Present solar sails are thin aluminized polyamide films: ~100nm thick aluminum deposited atop of few microns' thick polyamide (e.g., Mylar$^{TM}$ or Kapton$^{TM}$). Such solar sails are not suitable for missions with close perihelion approaches. Aluminum, as well, as polyamide absorb a significant fraction of the solar radiation spectrum (>10%) resulting in a significant heating of the sail material. Ultrathin sails made of conventional materials are ideally suited for technology test demonstration missions at the early stages of the SGL program. With a melting temperature ~700º C such sails can get no closer than ~25 $R_{sun}$. Notably, conventional sails are well posed to perform missions at and slightly beyond the Mercury orbit, enabling such scenarios as interstellar asteroid intersection.

In this study, we have considered various potential sail materials, including metals and ceramics that can withstand high temperatures. We find that refractory metals (such as for example tungsten with 3,690K melting point) are better suited as sail materials in comparison to aluminum. However, metals possess high material density (>15g/cm$^3$) and relatively high solar absorbance (~30%), implying that such refractory metal sails would constitute a large fraction of the entire spacecraft (i.e., small payload fraction is expected).

Other material alternatives include ceramic dielectrics, such as silicon nitride and silica. They too possess high melting temperatures ≥ 2,000K. In addition, these materials have low mass density ~2-3 g/cm$^3$ and low solar absorptivity (<10%). Our models predict that sails made of ceramic dielectrics may reach 0.1 AU (~20 $R_{sun}$) perihelion without heating above their melting point, thus providing a robust pathway for high exit velocity missions. Stand-alone ceramic films however are transparent and contribute little to radiation pressure momentum transfer. We have identified several options for nanophotonic and metamaterial design that may lead to thin films with high reflectance and thus boost the radiation pressure based thrust. Our analysis shows that reflectivity in excess of 80% across the solar spectrum may be obtained in nanostructured sub-micron thick films. Further research is needed to elaborate this technology further.

Table 2. Area-to-mass ratios for known sailcraft.

| Spacecraft | A/m |
|---|---|
| IKAROS | 1.3 |
| Nanosail-D | 2.2 |
| Cosmos-1 | 5.7 |
| LightSail | 7.0 |
| Lunar Flashlight/NEA Scout | 8.0 |
| Sunjammer | 22.3 |

### 3.2.3　*Solar sail (current status, design, anticipated maturation)*

In contrast to the designs requiring large fuel tanks and rocket nozzles, the simple solar sail holds the key to achieving extremely high exit velocities from the solar system by simply exploiting radiation pressure. Current solar sails are simple 2D planar sheets of aluminized Kapton$^{TM}$ that are held rigid by a series of booms and the entire control of the vehicle is body steered to align the sail with the sun. Multiple sails of this configuration have successfully flown in orbit about the Earth (e.g. LightSail 1 & 2, Nanosail D), while others such as IKAROS have flown to and returned from an asteroid. The key solar sail performance metric is the area to mass ratio of the final integrated vehicle. The larger this number, the higher the net acceleration the vehicle can achieve while aligned with the Sun. Table 2 shows the area-to-mass ratios for various designed and flown sails.

Leveraging solar radiation pressure to provide the propulsion to 500 AU, requires a sail of a specific area-to-mass ratio to achieve the needed escape velocity to reach the SGL in 20 years. A plot of velocity versus sail area to mass ratio is detailed in Figure 28 for varying perihelion distances.

Highlighted line in green shows the minimum 20 AU/yr required to reach the SGL on a relevant timescale. As illustrated in Figure 28, the closer to the Sun the vehicle makes it perihelion the larger the acceleration as radiation pressure increases as a square of the distance from the Sun.

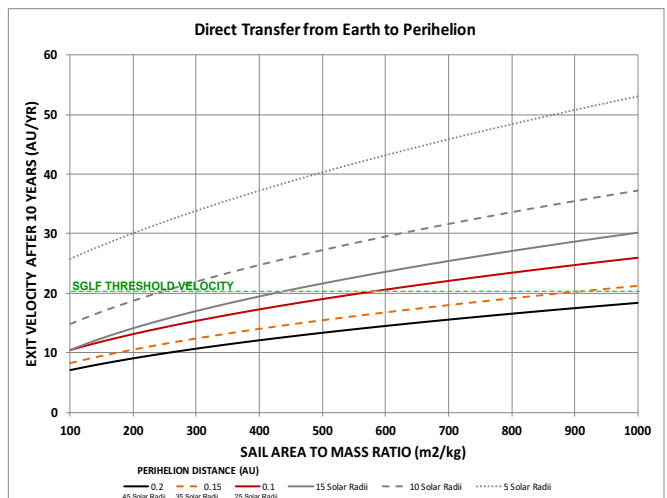

Figure 28. Solar system exit velocity for a sailcraft as a function of sail parameters (i.e., A/m).

Using the LightSail vehicle as an example, its sail would need to increase in size by a factor of 15 without changing the mass to even appear on the plot on Figure 28. This scalability challenge for large planar sails is even more daunting when coupled with the thermal challenges at close perihelion distances to achieve the 20 AU/yr velocity. Additional challenges with planar sails include packaging and deployment, control the center of gravity vs. center of pressure, durability and the inherent single degree of freedom the body steered planar sail must accomplish multiple mission objectives with competing attitude constraints such as power generation, communication, navigation and maintaining the trajectory.

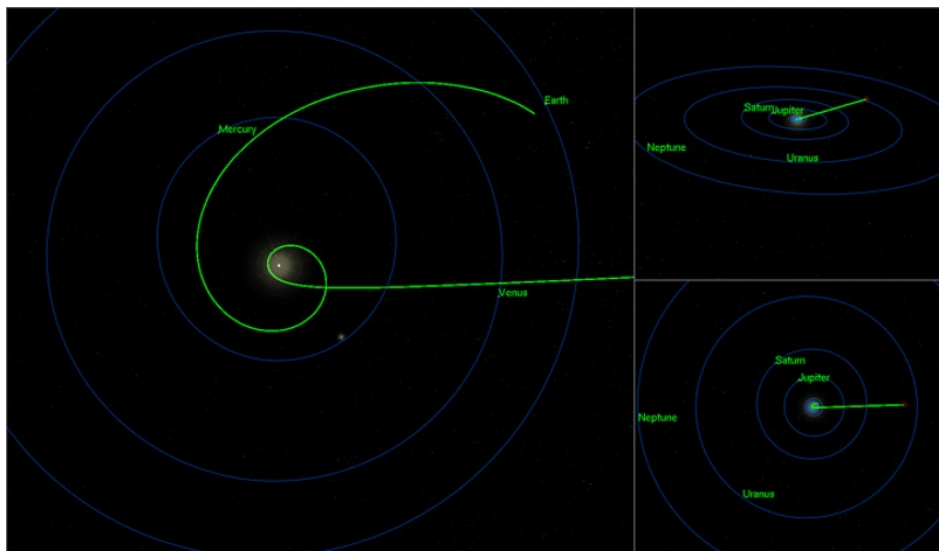

Figure 29. Sailcraft trajectory towards the SGL for TRAPPIST-1.

To address all of these design and operational issues and achieve the large area to mass ratios necessary to reach the SGL in less than a lifetime, the SunVane vehicle depicted in Figure 27 was developed using key technologies from the SunJammer project. The SunVane fractionates the sail area into multiple articulated vanes across a lightweight rigid structure such as a carbon fiber truss. By distributing the area in this manner, the other light weight subsystem components can be hosted more effectively on the vehicle instead of being confined to a small volume near the center of gravity. The vanes are multifunctional structures used for power generation, antenna and sails.

There is no complex packaging or deployment with the Vanes. Each vane is stowed and deployed in a similar fashion to solar panels. The articulation is accomplished with simple shape memory materials since the angular rate for any maneuver is slow (on the order of a minimum 30 minutes for 180 degrees of slew). The SunVane design is easily manufacturable today with an area to mass ratio of up to 400 m$^2$/kg, which would require a perihelion passage at 20 solar radii (higher than the Parker Solar Probe) to reach the required 20 AU/yr exit velocity.

As an example target for the SunVane, the SGLF for multiplanetary TRAPPIST-1 system was selected using a 200 A/m as a design point of departure to detail the mission design and CONOPS of the SunVane as it transfers from the Earth, to perihelion and ultimately on a radiant to the selected SGLF. Assuming a C3=0 launch, the SunVane with an area to mass a ratio of 200 m$^2$/kg is capable of transferring from the Earth's sphere of influence to perihelion in 120 days by orienting the sails to decelerate the vehicle as it spirals in towards the Sun. Similarly, the same SunVane can change its inclination with respect to the Sun at 1 AU 5 degrees every 50 days. Within the orbit of Mercury, the same 5-degree change can be accomplished in hours. The incredible maneuverability of the SunVane via solar radiation pressure removes any constraint on launch date since the SunVane can vary the requirement maneuver magnitudes to ensure alignment with the required exit trajectory as illustrated in Figure 29. For this example, the SunVane has targeted a stressing 5 $R_{sun}$ perihelion for a resulting 21 AU/yr exit velocity to examine the dynamics of the flyby.

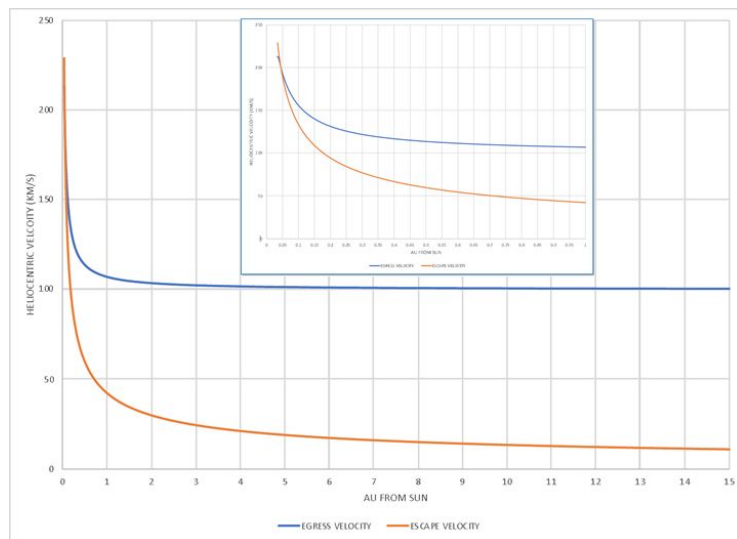

Figure 30. Egress Velocity Post-Perihelion.

The 5 $R_{sun}$ case is optimistic with respect to the thermal capabilities of the SunVane, but it was devised to show that there is no exquisite timing or control logic required to align with the correct exit radiant post perihelion passage. The vanes align themselves edge on to the Sun, until the perihelion point is reached which at that point the Vanes are directed face on to the Sun. During the close approach the vehicle has over 40 minutes to complete the alignment and the highest angular rates are less than 5 arcsec per sec for RA and DEC. As the SunVane accelerates away from the Sun, solar gravity is a significant perturbation on the vehicle. As depicted in Figure 30, the initial 200 km/s velocity rapidly drops to the final egress velocity of 100 km/s which far exceeds the hyperbolic exit velocity for the solar system as shown in orange.

### 3.3 New smallsat mission architecture: the "string of pearls" concept

The smallsat SoP architecture we propose is radical, specifically tailored to the mission requirements for the collection of many years of data. That is best accomplished by flying a number of s/c flying along the SGL focal lines of planetary systems with the possibility of flying concurrent missions to several candidate exo systems. This provides the needed CONOPS flexibility described in this report. It also spreads the risk of any single point catastrophic failure, particularly during the perihelion phase and the long mission duration thereafter. It also opens up the possibility of distributed funding.

This can be enabled by developing the smallsats with the ability to group themselves in space in order to "grow" to the needed capability – this is being extensively studied by the U.S. Space Force, DARPA and others as the future space operations – involving launching of components that are amalgamated in space and that then can be redistributed and re-purposed as needed functions change and requirements evolve. We propose to apply that wave to interplanetary spacecraft and to define the minimum possible weight of each separate component, and the methodology of rendezvous, docking, and connectivity[13].

The two biggest obstacles to minimizing the smallsat mass to enable this construct are (i) power – a lightweight RTG for the relatively small power required, and (ii) the coronagraph size – fitting the necessary optical system into the smallsat design. RTG approaches are defined in this report that suggest mass <50 kg is realistic; future technology work looking specifically at low power requirements will examine the requirements to reduce that mass by half. Figure 32 describes the process of the technology studies to meet the mass and size needed to enable meeting SGLF mission requirements.

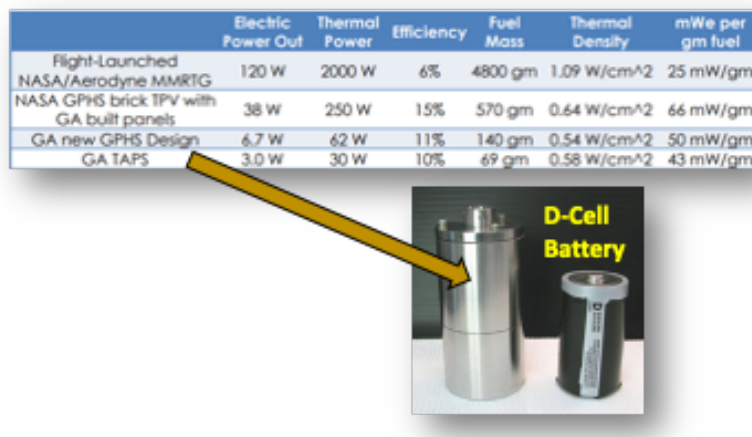

Figure 31. A small functioning RTG developed by General Atomics.

To reduce the weight of the power sources, we have been exploring two possible approaches. Technologies are beginning to emerge to reduce the size of an RTG into battery size units (e.g. D cell size), the concept there would be to distribute very small power sources where it would be needed, and by reducing the heat shielding, the released thermal energy could be used to warm other parts of the satellite (e.g. batteries which are proposed in option). By distributing very small RTGs then it becomes possible to integrate it with batteries to handle peak power loads. Figure 31 shows a functioning RTG version developed by General Atomics (GA)[14]. The GA TAPS design is shown next to a D-Cell battery which produces 3We. Another approach that would be beneficial is the explore radioisotope power systems that are assembled using then films of radioactive material as opposed to cakes (i.e., those built by Zeno Power Systems[15]). This concept enables us to repurpose a large area (possibly a solar sail) with a battery+ thin film radioisotope power system.

---

[13] https://www.satelliteconfers.org
[14] http://www.ga.com/

Earlier sections in this report concluded that the optics for the coronagraph/telescope imaging system will be 1-2 meters in diameter. Clearly, using conventional glass optics will not permit smallsat accommodation. We will need to apply approaches involving deployable structures, adaptive optics, new materials and advanced image processing software to define the imaging system for our smallsats. In particular, the in-space telescope assembly would allow us to achieve the required optical performance while operating at the SGL focal region.

For example, by using adaptive optics within a telescope it would be possible utilize a light collecting system (e.g. solar sail) where the optical wavefront is severely aberrated but the aberration is compensated by an adaptive optical system (which is a kind of deformable mirror). This is now routinely done to correct phase front error in laser material interaction studies where the focal properties are dependent on incident wavefront control. Variants of this are also found in some telescopes (e.g. Starlight Xpress Active Optics Guiding System, or Dynamic-optics.eu). One commercial adaptive lens developed today (27mmx116mmx63mm) can correct astigmatism to 2.5 waves (RMS @633nm), Coma to 0.7 waves (RMS @633nm). We expect significant enhancements in these types of devices because of their value in precision laser imaging/exposures. Another alternative to explore is what has been developed in the field of "foldable optics" by DARPA (Project Montage, ca 2007) and the future if this technology. A foldable lens has been developed by UCSD, that is 5 mm thick and can focus an image 2.5 m away (Tremblay et al., 2007).

Fortunately, much of this work is already going on – the roadmap for including it our design remains to be done. We may be able to include a prototype lightweight imaging system this in our early heliocentric technology test mission defined in this report. Our design goal is to fit this into our smallsat, e.g. in a 12U CubeSat or equivalent. It is noted that the SunVane design for the solar sail, permits distribution of the instruments along a truss.

An instrument that fits into component parts will be a typical challenge. Having power systems that are either distributed among s/c or are provided by a "central power spacecraft" to power each spacecraft in a pearl by plug-in or rf/laser transfer is another, as is the communications architecture that may use dedicated comm-sats for pearl to pearl relay or low bit rate data transfer to Earth. Current Earth orbit communications satellites are now being deployed in swarm configurations. Our approach is to adapt this for the interplanetary and interstellar medium exploration. It is a challenge, but the payoff, of having individual building blocks small enough to be propelled to 25 AU/yr by a reasonably sized solar sail, is worth the effort.

We predicate our mission on the ability of solar sail driven s/c to achieve the needed solar system exit velocity -- this leads us to miniaturization of each individual component in a pearl to make the solar sail size reasonable. During the solar sail downward spiral we have a primary navigational spacecraft that uses DSN data to adjust its trajectory towards perihelion and that provides the needed tracking beacon for all the others that use relative navigation so that the swarm of s/c emerge from the perihelion with the right trajectory and the right relative state vectors. As we move away from the sun, this process continues, with DSN tracking the primary navigation s/c and the others following its beacon so that the solar sails can trim the trajectories, to remove injection errors, and improve the relative state vectors of the other component s/c to allow subsequent joint operations. Once the solar sails are ejected, the swarm of component s/c are in free flight for some 20 years -- during this time they reconfigure themselves by rendezvous and docking into the needed functional s/c that comprise the pearl and perform all needed functions. This reconfiguration can be done as many times as needed to meet mission needs as we approach and fly along the SGFL. The needed instruments for example would each be composed of a number of individual elements that would be assembled into fully functional instruments on the fly.

---

[15] https://zenopowersystems.com/

This makes our NIAC a truly innovative approach to space research -- using techniques that are being tested in flight now by civil and governmental projects -- and that can be applied to many NASA science projects in the future. Below, we describe our approach in more details.

### 3.3.1 Spacecraft

The string-of-pearls architecture driven by solar sails, both reduces and spreads the cost. It also reduces mission risk by flying multiple, redundant and re-programmable spacecraft that is agnostic with regards to target location. To return data in a reasonable time (arbitrarily defined to not exceed two human generations), the mission must achieve a solar system exit speed of >20 AU/yr. This forces the satellite mass into the range of 10–30 kg so that the solar sail size and material composition become practical and perform at perihelion distances of 5–10 $R_{sun}$. Currently, it is possible to achieve area to mass ratios (A/M) of 100–200 m$^2$/kg with current sail material technology, ratios of 400–600 m$^2$/kg it can produce higher exit velocities, begin to get impractical. However, we anticipate that by incorporating newer advanced materials that include nano-photonic meta-materials and thin-film ceramics it will be possible to drive down the area to mass ratio while maintaining the high exit velocities. Analysis based on current materials show that solar orbit injection with a perihelion distance of 10 solar radii (approximate orbit of the NASA's Parker Solar Probe[16]) generates exit velocities of 15–18 AU/yr for A/M ratio of 100–200 m$^2$/kg, and 25 to 30 AU/yr for A/M of 400–600 m$^2$/kg.

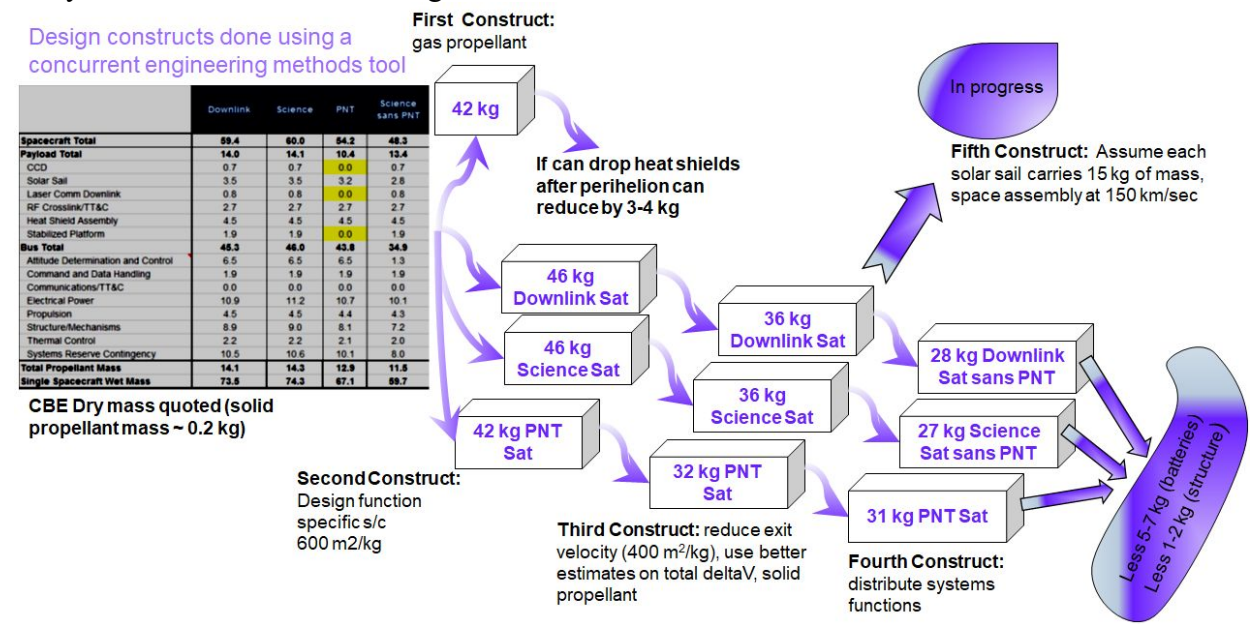

Figure 32. The process of the technology studies to meet the SGLF mission requirements.

We applied an Aerospace concurrent engineering methods (CEM) tool, used for designing satellites, to explore the spacecraft design space with the goal to drive down mass by distributing system functionality among the pearl satellites, exploring different forms of power (e.g. RTG only, smaller

---

[16] https://www.nasa.gov/content/goddard/parker-solar-probe

RTG and batteries), various power-scheduling CONOPS (e.g. science, downlink/crosslink, thrusting), different COMM systems and finally exploring the concept of in-space aggregation after achieving exit-velocity. We also evaluated the extent at which anticipated technology developments could further reduce mass. Figure 32 shows the CBE dry mass of four design-construct runs using the modified CEM tool. On the left is a table that lists the individual satellite subsystem mass values that were included in the trade study depending on whether the satellite is a primarily downlink COMM sat, a science payload satellite or a PNT satellite. Other merged combinations were also tried. The first construct produced an average mass in the range of 40+ kg (46 kg for downlink-sat, 46 kg for science-sat, 42 kg for PNT-sat). A configuration where the satellite ejected some heat shields after perihelion approach, a reduction to 34 kg can be achieved.

The second design-construct explored a spacecraft with A/M ratio of 600 $m^2$/kg. That information is not in the figure because the net physical area of the solar sail was deemed impractical with today's materials manufacturing technology. The third design-construct shown in the figure is based on a A/M ratio of 400 $m^2$/kg, better estimates on the total delta-V requirements (e.g. 200–300 m/s) and the use of solid propellant. The CBE mass values drop to 30+ kg range (Downlink-sat 36 kg, Science-Sat 36 kg and PNT-Sat 32 kg). The figure shows a fourth construct where additional re-distribution of the subsystem functions produces a CBE mass that "hovers" in the high 20 kg range. Our battery subject matter experts anticipate that battery mass would be 5–7 kg less within 10 years. New materials could further remove another 1–2 kg from the carbon fiber composite structure. Consequently, within 10 years the CBE mass of a single SGL satellite could be in the low 20 kg range.

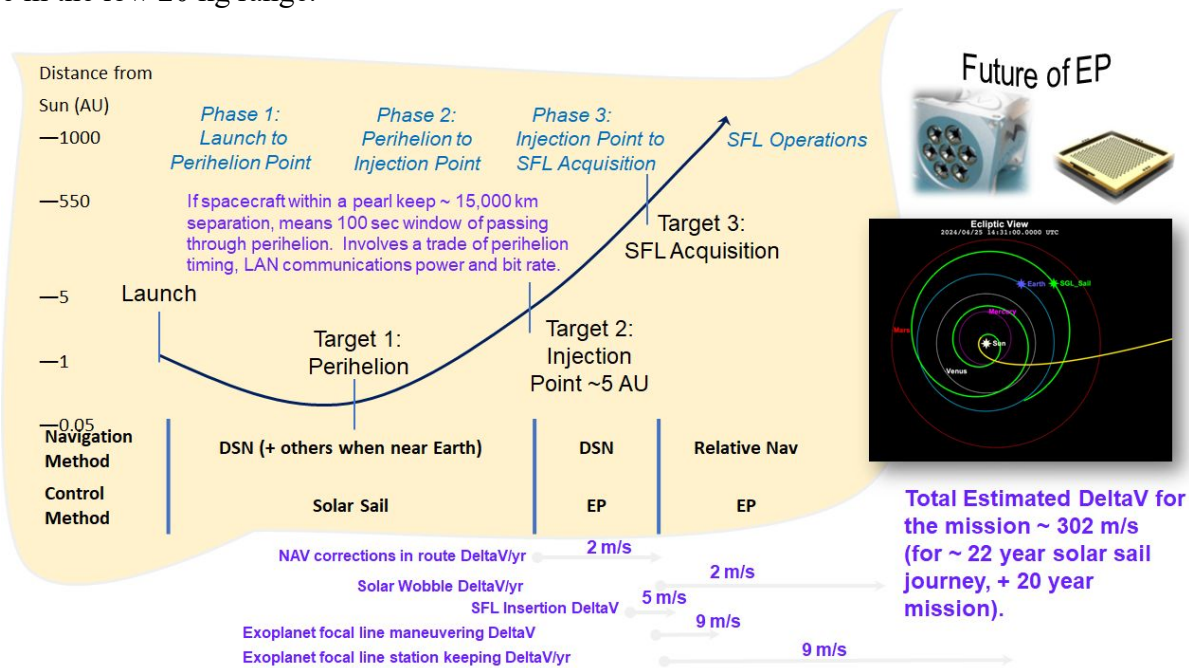

Figure 33. An overview of the navigation concepts and the calculated delta-V requirements for each segment of the mission to the SGL.

Figure 32 shows a fifth design-construct that is in early stages of progress and this would be the focus of our post-NIAC Phase II work. In this construct, the SGL s/c comprises an assembly of 15 kg mass units. Each 15 kg mass unit (comprising of minimal but necessary satellite components) carries a solar sail. After achieving exit velocity (~150 km/sec), some of the units in the cluster dispose of the solar sail and using low thrust propulsion (EP) dock with other units to form larger, more integrated, and higher performing systems. The aggregated system becomes one of the SGL mission's satellites in the cluster.

Figure 33 presents an overview of the navigation concepts and the calculated delta-V requirements for each segment of the mission. The total delta-V for the mission, discounting the obtainment of exit-velocity, is estimated to be ~300 m/s. We have defined a 22-year solar sail journey followed by a 20-year data acquisition mission. The SGL mission can be delineated into 3 phases: Phase 1: launch to perihelion point, Phase 2: perihelion to injection point and Phase 3: Injection point to SGL focal line and its acquisition. Each phase has its own navigation and propulsion requirements.

In Phase 1 a rideshare rocket takes aloft several canisters each containing multiple SGL spacecraft to GEO or cislunar orbits. We call this location the "gathering point". The canisters open, releasing SGL spacecraft which deploy solar sails. This ensemble comprises a single pearl. Using solar sails and NASA's DSN network for navigation they are guided toward solar perihelion point (Target 1) for a subsequent trajectory insertion toward the SGLF.

Figure 33 shows one calculated trajectory to Target 1 viewed from above the ecliptic. Phase 2 begins at the perihelion with the injection of the sailcraft toward an imaginary location, ~5 AU distant (Target 2), but on a near parallel path toward the SGLF of the chosen exoplanet. Individual spacecraft within a pearl will maintain <15,000 km separation, the time window for passing through perihelion is less than 1 hour. Within this time window, minute orbit corrections could be instituted by controlling the solar sail and this information could be shared among all the s/c within the pearl to keep the ensemble together. Consequently, there then results a trade between the amount of time for perihelion passing, the available LAN communications and bit rate transfer.

The pearl spacecraft use solar sail for propulsion and DSN for navigation while traversing from Target location 1 to 2. From Target location 2 to Target 3, which defines the beginning of the Phase 3 operations (~650 AU), the pearl spacecraft continue to use the DSN for navigation for most of the crossing (~20-year journey) but with increasing reliance on relative navigation consisting of tracking celestial objects. Miniature electric propulsion technology (e.g. Morpheus[17] NanoFeep Thrusters (1–20 N dynamic thrust, at 3,000–8,000 s specific impulse) will be used for navigation corrections en route, for injection into the SGLF and for operations and maintenance while traversing the SGL. Figure 33 describes the SGL mission phases and shows images from two possible miniature EP thruster vendors. The figure also provides the estimated delta-V requirements based on trajectory analysis to an exoplanet (e.g. TRAPPIST-1 system). The delta-V requirements for navigations corrections in route from Target 2 to 3 amount to 2 m/s per year. Compensating for the solar wobble is also 2 m/s per year which starts at Target location 3 (~650 AU) until the end of the mission. Delta-V for SGLF insertion is estimated to be 5 m/s which occurs at Target 3. This is the thrust required to find the focal line of the parent star which is part of the CONOPS to identify the exoplanet within that solar system. We estimate exoplanet focal line maneuvering to cost ~9 m/s delta-V and another 9 m/s for exoplanet station keeping. The total delta-V is estimated to be ~315 m/s which is enough for the EP thruster modules being considered.

We have considered the types of devices that will be required for navigation (i.e., positioning) and attitude control. The SGL spacecraft will use star-trackers for initial attitude control. Given that a pearl cluster comprises several satellites (e.g. 10), we estimate a number of these to have star trackers and the information to be shared among all free-flying units to get additional coordinate refinement. Nevertheless, during the SGL data acquisition phase, the attitude control requirements are exceptionally tight (i.e. nanoradians). However, we have developed a CONOPS that will rely on the use of photosensitive detectors, the SGL focused light from the parent star and the fact that there is multiple spacecraft in the cluster which can be used for "scouting".

---

[17] http://morpheus-space.com/

Table 3. SGL link budget analysis for the UHF, Ka and V-bands.

| Parameter | Units | Intra Pearl (UHF) | Intra Pearl (Ka-band) | Intra Pearl (V-band) | Intra Pearl (V-band) | Intra Pearl (V-band) |
|---|---|---|---|---|---|---|
| Data Rate | kbps | 0.1 | 10 | 1000 | 10000 | 100000 |
| Frequency | MHz | 401 | 28000 | 80000 | 80000 | 80000 |
| Distance | km | 50000 | 50000 | 20000 | 6000 | 2000 |
| **Transmit Station Parameters** | | | | | | |
| Antenna Diameter | m | | 0.20 | 0.20 | 0.20 | 0.20 |
| Antenna efficiency | % | | 65% | 65% | 65% | 65% |
| Wave length | m | 0.75 | 0.011 | 0.004 | 0.004 | 0.004 |
| Antenna Gain | dBi | 2 | 33.49 | 42.61 | 42.61 | 42.61 |
| Transmit Power | W | 1 | 1 | 1 | 1 | 1 |
| Line Loss | dB | -1 | -1 | -1 | -1 | -1 |
| Transmit EIRP | dBW | 1.00 | 32.49 | 41.61 | 41.61 | 41.61 |
| **Propagation Parameters** | | | | | | |
| Free Space Loss | dB | -178.5 | -215.4 | -216.5 | -206.1 | -196.5 |
| Atmospheric Loss | dB | 0 | 0 | 0 | 0 | 0 |
| Polarization Loss | dB | -0.4 | -0.4 | -0.4 | -0.4 | -0.4 |
| Tx Pointing Loss | dB | 0 | 0 | 0 | 0 | 0 |
| Isotrapic Signal Level at Rx Station | dB | -177.88 | -183.27 | -175.31 | -164.85 | -155.31 |
| **Receiving Station Parameters** | | | | | | |
| Antenna Diameter | m | | 0.20 | 0.20 | 0.20 | 0.20 |
| Antenna efficiency | % | | 65% | 65% | 65% | 65% |
| Antenna Gain | dBi | 2 | 33.5 | 42.6 | 42.6 | 42.6 |
| System Noise Temperature | K | 140 | 140 | 140 | 140 | 140 |
| Noise Figure | dB | 3 | 4 | 5 | 5 | 5 |
| Receiver Loss | dB | -1 | -1 | -1 | -1 | -1 |
| Receiver G/T | dB/K | -20.4 | 9.2 | 16.8 | 16.8 | 16.8 |
| Rx Pointing Loss | dB | 0 | -2 | -3 | -3 | -3 |
| Boltzman constant | dBW/Hz/K | -228.6 | -228.6 | -228.6 | -228.6 | -228.6 |
| Received C/No | dB-Hz | 30.28 | 52.57 | 67.09 | 77.55 | 87.09 |
| **Carrier Parameters** | | | | | | |
| Modulation order | | 1 | 1 | 2 | 2 | 2 |
| Code Rate | | 0.5 | 0.5 | 0.6 | 0.6 | 0.6 |
| Symbol Rate | ksps | 0.2 | 20 | 833.3 | 8333.3 | 83333.3 |
| Noise Bandwidth | dB-Hz | 23.0 | 43.0 | 59.2 | 69.2 | 79.2 |
| Received Es/No | dB | 7.3 | 9.6 | 7.9 | 8.3 | 7.9 |
| Received Eb/No | dB | 10.3 | 12.6 | 7.1 | 7.5 | 7.1 |
| Implementation Loss | dB | -2 | -2 | -2 | -2 | -2 |
| Received Eb/No | dB | 8.3 | 10.6 | 5.1 | 5.5 | 5.1 |
| Required Eb/No at BER=10^-7 | dB | 8 | 8 | 2 | 2 | 2 |
| Margin | dB | 0.3 | 2.6 | 3.1 | 3.5 | 3.1 |

Our calculations show that the photon flux from the parent star (at 30 pc) is on the order of $\sim 10^{11}$ photons/sec, which is brighter than the solar corona $10^8 - 10^9$ photons/s ($\lambda \sim 1$ μm, sensor aperture 1 m). Consequently, the parent star focused light could be used for navigation to refine the SGL spacecraft trajectory along the SGLF (Turyshev et al., 2019). Further refinement of the SGLF satellite trajectories will come from the CONOPS and the formation flying 10+ SGL satellites which comprise a pearl. This will be investigated further in our post-Phase II work.

For the Phase II study we considered various telescope sizes in the range 0.5-2.0 m. Clearly, the larger the telescope, the better the system performance as the images are less affected by the diffraction within the telescope. Nevertheless, a telescope aperture of $d = 1$ m was chosen as the baseline for the mission studies (Turyshev & Toth, 2020ab, 2020c).

The design for the SGL coronagraph was discussed in (Zhou, 2018). It is able to reject sunlight with a contrast ratio of $\sim 10^7$. At this level of rejection, the light from the solar disk is completely blocked to the level comparable to the brightness of the solar corona. The optical design of the payload incorporates the coronagraph as was discussed in (Turyshev et al., 2018).

### 3.3.2 Communication

The SGL mission comprises a string-of-pearls architecture that uses two types of communication. A laser based optical system to transfer information between pearls for uplink/downlink and to Earth and an isotropic RF system that operates within the pearl or among the formation flying cluster. Regardless of the COMM system used, given the extent of the mission the onboard clocks have to periodically be synchronized to compensate for drift. Current space-qualified miniature clock technology is represented by the MicroSemi Inc. chip scale atomic clock (CSAC)[18]. At 35 g and < 17 cm$^3$ volume, it is currently radiation tolerant to 20 krad (SEU tested to 64 MeV cm$^2$/mg, LET for cosmic ray 10-100 MeV cm$^2$/mg) and draws <120 mW. The Allan deviation values suggest that a 10 MHz CSAC will exceed the < 1ppm drift value (a desired mission requirement) in 4 months. We anticipate a 100X improvement in the Allan deviation values for such devices over the next decade. If feasible, synchronization will not be required for 33 years. The current technology limitation is controlling the inherent thermal drift.

Table 4. RF link budget model in V-band for a pearl-to-pearl distance of 20 AU.

| Parameter | Units | Pearl to Pearl (V-band) |
|---|---|---|
| Data Rate | kbps | 0.1 |
| Frequency | MHz | 80000 |
| Distance | km | 2992000000 |
| **Transmit Station Parameters** | | |
| Antenna Diameter | m | 4.00 |
| Antenna efficiency | % | 65% |
| Wave length | m | 0.004 |
| Antenna Gain | dBi | 68.63 |
| Transmit Power | W | 10 |
| Line Loss | dB | -1 |
| Transmit EIRP | dBW | 77.63 |
| **Propagation Parameters** | | |
| Free Space Loss | dB | -320.0 |
| Atmospheric Loss | dB | 0 |
| Polarization Loss | dB | -0.4 |
| Tx Pointing Loss | dB | 0 |
| Isotrapic Signal Level at Rx Station | dB | -242.79 |
| **Receiving Station Parameters** | | |
| Antenna Diameter | m | 4.00 |
| Antenna efficiency | % | 65% |
| Antenna Gain | dBi | 68.6 |
| System Noise Temperature | K | 140 |
| Noise Figure | dB | 5 |
| Receiver Loss | dB | -1 |
| Receiver G/T | dB/K | 42.8 |
| Rx Pointing Loss | dB | -3 |
| Boltzman constant | dBW/Hz/K | -228.6 |
| Received C/No | dB-Hz | 25.63 |
| **Carrier Parameters** | | |
| Modulation order | | 2 |
| Code Rate | | 0.2500 |
| Symbol Rate | ksps | 0.2 |
| Noise Bandwidth | dB-Hz | 23.0 |
| Received Es/No | dB | 2.6 |
| Received Eb/No | dB | 5.6 |
| Implementation Loss | dB | -2 |
| Received Eb/No | dB | 3.6 |
| Required Eb/No at BER=10^-7 | dB | 1.5 |
| Margin | dB | 2.1 |

---

[18] https://www.microsemi.com/product-directory/embedded-clocks-frequency-references/5207-space-csac

The CSAC technology is currently at a TRL 7-8. However, we anticipate another technology to overtake the CSAC. Currently at a TRL 2-3 the clock is based on an optical frequency comb and would meet the < 1ppm drift requirement for 1300 years (for a 10 MHz clock) and 32 years when running at 500 MHz (Hisai et al., 2019). A 10X improvement in the Allan Deviation would allow a 53-year synchronization period for a 3 GHz clock.

The issue for the optical frequency comb clock is not the fabrication of the "comb" which is small enough to be held at the tip of a finger but miniaturizing the supporting electronics. Optical frequency combs can be purchased today (e.g. Menlo Systems) but the Allan deviations need to improve. There is altogether alternative approach that NASA has been considering which is the use of X-ray pulsars (i.e. NICER). X-ray pulsars can not only assist to synchronize on board clocks, but with triangulation and against a starfield could also permit autonomous navigation. China has launched the XPNAV-1 satellite which has an X-ray pulsar sensor and Nav instrument on board. The XPNAV-1 instrument is on the order of ~ 50 kg. If it could be miniaturized by a factor 10, then it would serve much of the needs for the SGL regarding PNT.

Table 5. SGL laser link with PPM (pulse position modulation) coding.

| | LASER COMM | | 20 AU | | | | 40 AU | | | | 100 AU | | | | 200 AU | | | | 500 AU | | | |
|---|---|---|---|---|---|---|---|---|---|---|---|---|---|---|---|---|---|---|---|---|---|---|
| | Cross-Link Parameters | Variable | Value | Units | dB | Units | Value | Units | dB | Units | Value | Units | dB | Units | Value | Units | dB | Units | Value | Units | dB | Units |
| **Satellite Transmitter** | Transmission Rate | $R_T$ | 4000.0 | bps | 36.0 | dB | 1000.0 | bps | 30.0 | dB | 160.0 | bps | 22.0 | dB | 40.0 | bps | 16.0 | dB | 6.3 | bps | 8.0 | dB |
| | Code Rate | r | 0.5 | | | | 0.5 | | | | 0.5 | | | | 0.5 | | | | 0.5 | | | |
| | **Information Rate** | $R_i$ | 8000.0 | bps | | | 2000.0 | bps | | | 320.0 | bps | | | 80.0 | bps | | | 12.5 | bps | | |
| | Modulation Format | M | 4 | | | | 4 | | | | 4 | | | | 4 | | | | 4 | | | |
| | **Transmit Wavelength** | λ | 1.55 | μm | | | 1.55 | μm | | | 1.55 | μm | | | 1.55 | μm | | | 1.55 | μm | | |
| | **Tx Power** | $P_{Tx}$ | 5.0 | W | 37.0 | dBm | 5.0 | W | 37.0 | dBm | 5.0 | W | 37.0 | dBm | 5.0 | W | 37.0 | dBm | 5.0 | W | 37.0 | dBm |
| | Tx WDM Loss | $L_{TxWDM}$ | 100.0% | % | 0.0 | dB | 100.0% | % | 0.0 | dB | 100.0% | % | 0.0 | dB | 100.0% | % | 0.0 | dB | 100.0% | % | 0.0 | dB |
| | Tx Fiber Coupling Loss | $L_{Txfiber}$ | | | 0.5 | dB | | | 0.5 | dB | | | 0.5 | dB | | | 0.5 | dB | | | 0.5 | dB |
| | Output Backoff | OBO | | | 0.0 | dB | | | 0.0 | dB | | | 0.0 | dB | | | 0.0 | dB | | | 0.0 | dB |
| | **Telescope Diameter** | $D_{Tx}$ | 40.0 | cm | | | 40.0 | cm | | | 40.0 | cm | | | 40.0 | cm | | | 40.0 | cm | | |
| | Rx Telescope Efficiency | $\eta_{Tx}$ | 40.0% | % | -4.0 | dB | 40.0% | % | -4.0 | dB | 40.0% | % | -4.0 | dB | 40.0% | % | -4.0 | dB | 40.0% | % | -4.0 | dB |
| | **Tx Angular Beamwidth** | $\theta_{Tx}$ | 0.98 | arcsec | | | 0.98 | arcsec | | | 0.98 | arcsec | | | 0.98 | arcsec | | | 0.98 | arcsec | | |
| | Tx Telescope Gain | $G_{Tx}$ | | | 114.2 | dBi | | | 114.2 | dBi | | | 114.2 | dBi | | | 114.2 | dBi | | | 114.2 | dBi |
| | Tx Pointing Loss | $L_{Txpoint}$ | 0.2 | arcsec | 0.7 | dB | 0.2 | arcsec | 0.7 | dB | 0.2 | arcsec | 0.7 | dB | 0.2 | arcsec | 0.7 | dB | 0.2 | arcsec | 0.7 | dB |
| | Transmit EIRP | EIRP | | | 146.0 | dBmi | | | 146.0 | dBmi | | | 146.0 | dBmi | | | 146.0 | dBmi | | | 146.0 | dBmi |
| **Channel** | Elevation Angle | $\theta_{elev}$ | ± 1.0 | degree | | | ± 1.0 | degree | | | ± 1.0 | degree | | | ± 1.0 | degree | | | ± 1.0 | degree | | |
| | Slant Range (max) | $D_{SR}$ | 2.99E+09 | km | 20 | AU | 6.0E+09 | km | 40 | AU | 1.5E+10 | km | 100 | AU | 3.0E+10 | km | 200 | AU | 7.5E+10 | km | 500 | AU |
| | Path Loss | $L_{FS}$ | | | 387.7 | | | | 393.7 | | | | 401.7 | | | | 407.7 | | | | 415.7 | |
| | Total Atmospheric Loss | $L_{atm}$ | | | 0.6 | dB | | | 0.6 | dB | | | 0.6 | dB | | | 0.6 | dB | | | 0.6 | dB |
| | Net Path Loss | $L_{path}$ | | | 388.3 | | | | 394.3 | | | | 402.3 | | | | 408.3 | | | | 416.3 | |
| **Satellite Receiver** | **Rx Footprint Diameter** | $D_{foot}$ | 1.42E+04 | km | | | 2.8E+04 | km | | | 7.1E+04 | km | | | 1.4E+05 | km | | | 3.5E+05 | km | | |
| | **Rx Telescope Diameter** | $D_{Rx}$ | 250.0 | cm | | | 250.0 | cm | | | 250.0 | cm | | | 250.0 | cm | | | 250.0 | cm | | |
| | Rx Telescope Efficiency | $\eta_{Rx}$ | 50.0% | % | -3.0 | dB | 50.0% | % | -3.0 | dB | 50.0% | % | -3.0 | dB | 50.0% | % | -3.0 | dB | 50.0% | % | -3.0 | dB |
| | Rx Telescope Gain | $G_{Rx}$ | | | 131.1 | dBi | | | 131.1 | dBi | | | 131.1 | dBi | | | 131.1 | dBi | | | 131.1 | dBi |
| | Polarization Mismatch Loss | $L_{pol}$ | | | 0.0 | dB | | | 0.0 | dB | | | 0.0 | dB | | | 0.0 | dB | | | 0.0 | dB |
| | **Receiver Pointing Loss** | $L_{Rxpoint}$ | 0.2 | arcsec | 0.7 | dB | 0.2 | arcsec | 0.7 | dB | 0.2 | arcsec | 0.7 | dB | 0.2 | arcsec | 0.7 | dB | 0.2 | arcsec | 0.7 | dB |
| | Rx Fiber Coupling Loss | $L_{Rxfiber}$ | | | 0.5 | dB | | | 0.5 | dB | | | 0.5 | dB | | | 0.5 | dB | | | 0.5 | dB |
| | Rx WDM Loss | $L_{RxWDM}$ | 100.0% | % | 0.0 | dB | 100.0% | % | 0.0 | dB | 100.0% | % | 0.0 | dB | 100.0% | % | 0.0 | dB | 100.0% | % | 0.0 | dB |
| | Receiver Net Gain | $G_{Rx}$ | | | 126.8 | dB | | | 126.8 | dB | | | 126.8 | dB | | | 126.8 | dB | | | 126.8 | dB |
| | Received Power | $P_{Rx}$ | | | -115.5 | dBm | | | -121.5 | dBm | | | -129.5 | dBm | | | -135.5 | dBm | | | -143.4 | dBm |
| | Equiv System Noise Temperature | $T_{sys}$ | 9292 | K | 39.7 | dBK | 9292 | K | 39.7 | dBK | 9292 | K | 39.7 | dBK | 9292 | K | 39.7 | dBK | 9292 | K | 39.7 | dBK |
| | Equiv Receiver Figure-of-Merit | G/T | | | 91.4 | dB/K | | | 91.4 | dB/K | | | 91.4 | dB/K | | | 91.4 | dB/K | | | 91.4 | dB/K |
| **Demodulator** | Planck's Constant | h | 6.63E-31 | mW-$s^2$ | -301.8 | dBm-$s^2$ | 6.63E-31 | mW-$s^2$ | -301.8 | dBm-$s^2$ | 6.63E-31 | mW-$s^2$ | -301.8 | dBm-$s^2$ | 6.63E-31 | mW-$s^2$ | -301.8 | dBm-$s^2$ | 6.63E-31 | mW-$s^2$ | -301.8 | dBm-$s^2$ |
| | Available C/No | $C/N_o$ | | | 43.4 | dBm-s | | | 37.4 | dBm-s | | | 29.5 | dBm-s | | | 23.4 | dBm-s | | | 15.5 | dBm-s |
| | Available PPB | $\mu_b$ | 5.5 | Pho/b | 7.4 | dB | 5.5 | Pho/b | 7.4 | dB | 5.5 | Pho/b | 7.4 | dB | 5.5 | Pho/b | 7.4 | dB | 5.7 | Pho/b | 7.5 | dB |
| | Ideal Required PPB | $\mu_{ideal}$ | 0.5 | Pho/b | -2.7 | dB | 0.5 | Pho/b | -2.7 | dB | 0.5 | Pho/b | -2.7 | dB | 0.5 | Pho/b | -2.7 | dB | 0.5 | Pho/b | -2.7 | dB |
| | Rx Implementation Loss | $L_{imp}$ | | | 5.0 | dB | | | 5.0 | dB | | | 5.0 | dB | | | 5.0 | dB | | | 5.0 | dB |
| | Net Required PPB | $\mu_{net}$ | 1.7 | Pho/b | 2.3 | dB | 1.7 | Pho/b | 2.3 | dB | 1.7 | Pho/b | 2.3 | dB | 1.7 | Pho/b | 2.3 | dB | 1.7 | Pho/b | 2.3 | dB |
| | **Link Margin** | L-M | | | 5.1 | dB | | | 5.1 | dB | | | 5.1 | dB | | | 5.1 | dB | | | 5.2 | dB |

We have used satellite-design COMM models to generate link budgets for the SGL architecture. The link budget calculations show that a V-band (80 GHz) COMM system should be able to supply enough data rate (1Mbps at 20,000 km range) with a 3db+ margin. Table 3 shows the link budget analysis for the UHF, Ka and V-bands. The table shows the link budgets for various ranges ($5\times10^4$; $2\times10^4$; $6\times10^3$ and $2\times10^3$ km) for UHF (401 MHz), Ka (28 GHz), V-band (80 GHz). All the transmit powers at set to 1 W, antenna efficiencies of 65% are chosen and the Tx/Rx antenna diameters are 20 cm. These results demonstrate that RF for intra-pearl comm will be feasible. We anticipate that the cluster of satellites within a pearl are not likely to be separated by more than 10,000 km. Table 4 applies the RF link budget model in V-band for a pearl-to-pearl distance of 20 AU. With 2.1 dB margin, the data rate is 100 bps. While this data rate is enough for sending "results", it is not enough for exchanging information among the oncoming pearls to allow optimization of the CO-NOPS protocols (i.e., to apply machine learning methodologies among the pearls).

A second link budget model was developed that is based on an optical COMM system. It employs a 5 W laser operating at a wavelength of 1.55-microns. Aerospace have recently flown such a laser COMM system in a CubeSat class satellite and have tested the downlink and more recently crosslink capabilities (Welle et al., 2017; Janson et al., 2016).

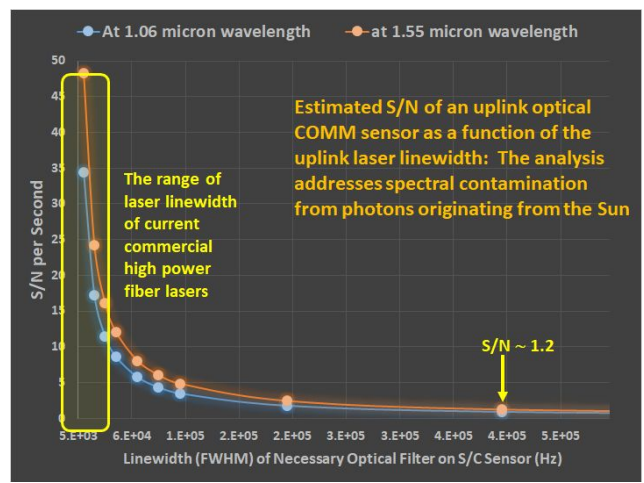

Figure 34. SNR analysis for wavelengths at 1.06 um & 1.55 um.

The link budget analysis is shown in Table 5 for PPM (pulse position modulation) coding. The analysis for DPSK (differential phase shift keying) coding was found to have a lower bit rate for information transfer. The analysis is a function of distance between repeating stations (at 20, 40, 100, 200, and 500 AU). For this analysis a 40 cm diameter "receive" telescope is used to get the sense of possible data rates. The results in the figure show that information transfer 8 kb/s is possible at a range of 20AU (link margin of 5.1 dB). The information transfer rate drops to 12.5 b/s at a range of 500 AU, which would be the information transfer from the last pearl (i.e. the caboose) to Earth. However, the bit rate for the caboose pearl could easily be increased by increasing the laser power and by increasing the telescope diameter. While this analysis was done for a 40 cm diameter telescope with precision optics, the SGL spacecraft design uses a 6 m diameter primary mirror but with less precise focusing capabilities. The ratio of diameters (i.e. gain) is 15 and if we estimate, as result of poor optical focusing, a 50% loss in gain then the bit transfer rate from 500 AU could be close to 100 bps with a 5 W laser and double that with a 10 W laser. This would be the downlink rate. The uplink rate from Earth could be substantially higher for communicating with the caboose-pearl.

The downlink optical communication is constrained by the available laser power on board and range. The uplink communication has a constraint that is unique for the SGL mission. The laser photons emanating from Earth at 1.5 μm arrive against a background of solar radiation photons at 1.5 µm. Consequently, this reduces the SNR at the spacecraft sensor. Figure 34 shows a SNR analysis for the laser wavelengths at 1.06 µm and 1.55 µm. The analysis is based on a 10W laser COMM system communicating between two pearls that are separated by 200 AU against the background photons arriving from the Sun located at 500 AU. The analysis describes the situation where a pearl has arrived at 500 AU and needs to communicate with a pearl that is 200 AU distant and closer to the Sun. The intent of the analysis is to show that if the laser linewidth is broad (~400 kHz FWHM) then the SNR is roughly 1.2 for both 1.55 and 1.06 laser wavelengths. However, if the linewidth has been reduced to 60 kHz (FWHM) or filtered at the Rx sensor, then the S/N ratio becomes 8 and 6 for the 1.55- and 1.06-micron wavelengths respectively. Currently, the range of laser linewidths of commercial high-power fiber lasers range from 5 – 50 kHz so S/N ratios in the range of 10-30 should be possible with the 1.55-µm laser being slightly better.

### 3.3.3  *Intelligent autonomy and data processing*

A mission to the SGL presents a complex set of contradictory requirements, centered on reliability, survivability over a mission duration in excess of 40 years, vs. the required autonomy in light of the enormous distances and an associated light round trip light travel time of over one calendar week, and the limited communication bandwidth.

**AI as the Enabler – Functional Reconfiguration**

- Autonomous Navigation
  - SGL mission satellites will be unable to communicate with ground operations in a timely enough manner for acceptable command and control
  - Individual satellites will need to learn from the observations of the swarm to make the necessary observations
  - Particle swarm optimization (PSO) can be used to minimize an objective function iteratively using spatially dispersed particles of varying velocity
  - Satellites will need propagated ephemeris to determine their location and then PSO could act as the ground station to optimize where the spacecraft should go, what measurement to take, or any other critical function.
  - In this architecture each satellite acts as beacon directing the other satellites in navigation and control
  - Goal Oriented commands should be used. This means the satellites will need to be programmed on what the mission requirements are, but left up to themselves to determine best course of action to maximize mission success.

- Autonomous Fault Management
  - Satellite needs to monitor abnormalities, determine if they are anomalies, AND determine what should be done to recover
  - In addition to enabling the mission, this would reduce power consumption via reduced communication with ground

Figure 35. AI functional requirements for a mission to the SGLF.

Requirement specifications that must drive the planned informatics architecture of the SoP type mission must include: 1) Deciding exactly what data processing will take place on board (e.g., data processing used for navigation, data processing to reduce and prefilter data for bandwidth-limited transmission); 2) Selecting a robust, existing, *proven* set of algorithms to autonomously set up and maintain a mesh network with variable latency and unreliable nodes and communication links; 3) Developing algorithms for autonomous cluster navigation; 4) Development of adaptive algorithms to data collection, aimed at obtaining low-resolution images of the primary target as early as possible, use of additional data acquisition to refine the image, and respond autonomously to the loss of a spacecraft or partial malfunctions; and 5) Design of a distributed operating system that is capable of delivering mission goals so long as a single spacecraft remains functional (i.e. autonomous fault management).

Examples of intelligent autonomy drivers include long periods of blackout (where the exoplanet is blocked by the parent star), complex maneuvers during perihelion passage where AI manages the GNC, and the managing of power usage in route. Given the length of the mission, autonomy also produces a cost savings with reduced ground station management (data show that ~90% cost savings is captured by employing autonomy to support ground operations of constellations, per Jeroen Rotteveel, CEO of ISISpace[19]) and the fact that computing/analytics is done at the edge amongst the pearls and across the string.

The temptation to use the latest innovative software technologies must be tempered by the requirement for the system to remain reliably operational over the full duration of the mission, and for its behavior to remain predictable and modellable. In-flight software upgrades must be considered as part of mission design, as advances in autonomous software technologies are to be expected over the course of a long-duration mission. Nonvolatile storage used for system firmware must be capable of reliable operation without data loss over the anticipated duration of the SGL mission.

For successful comm and data transfer between nodes, a low latency network must exist between s/c nodes that supports mesh characteristics of ad hoc forming and self-healing to provide robust data processing and analytics.

For data engineering, algorithms should optimize data storage, management, redundancy and eviction policies. Balancing tradeoffs between centralized vs. distributed processing, local vs. remote processing, and load balancing are all happening in the constellation, not the ground, due to distance between the constellation to Earth. Engineering design considerations are shown in Figure 35. and these designs could be implemented using the upcoming neuromorphic computing technology that allows machines to think and learn more like people.

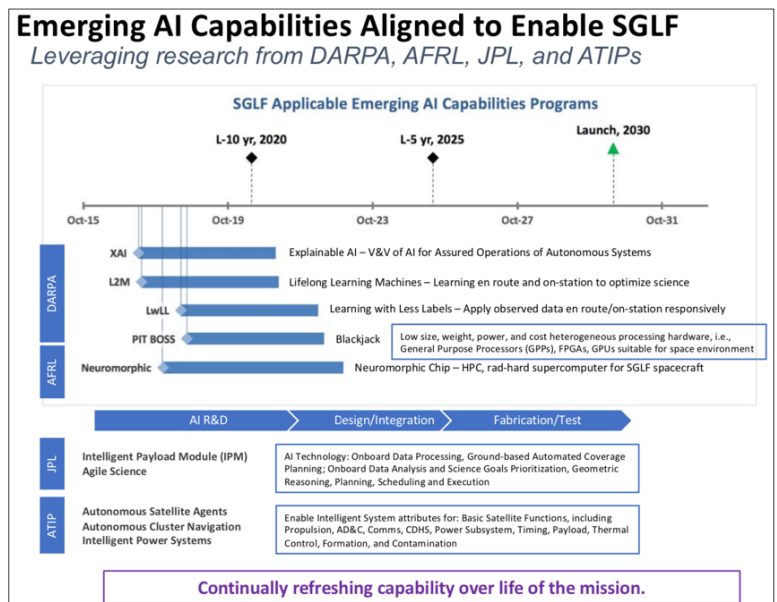

Figure 36. Emerging AI capabilities for a mission to the SGLF.

In terms of system integration, the processing module would be mounted on each satellite electronically situated between the payloads and the spacecraft bus providing electrical and network connectivity for each payload. Functionally, the processing module routes data between payloads and networked spacecraft nodes. Upcoming new computation platforms include CPUs (for traditional computing) but using less power, GPUs (for neural networks and complex imaging analysis) and FPGAs for flexibility. An example is the high-performance spaceflight computing (HPSC) architecture being developed by AFRL, JPL, GSFC and Boeing. It is designed for a small sat C&DH with a board size of 70x70 mm$^2$, a mass of 1.9kg and usage power between 0.5-7W depending on operational mode. Test and characterization of the HPSC is in 2021.

---

[19] https://www.isispace.nl/

To support acquisition over a long mission period of multiple strings and ensure continual technology refresh, the strings will employ a rapid spiral approach, where each successive string will be more advanced than the last. Technologies to support AI architecture software, hardware, and integration are being developed by the national laboratories, DARPA, DoD, and The Aerospace Corporation for a technology roadmap supporting the design. Figure 36 summarizes AI capabilities that are currently being developed.

### 3.3.4 Multiple spacecraft

There are constraints that uniquely define the SGL mission that do not vex most space missions of record: A mission duration lasting 40+ years in which the science mission begins at mark year 20, a target that is not a physical entity to fly towards, but an empty location in space, the need to arrive within two generational time frames and communicating from distances possibly as far as 1,000 AU. The effort that is necessary to accomplish the SGL mission along with a long wait for results almost forces a zero-risk tolerance, which of course is not possible. Consequently, risk mitigation steps must be built in the architecture that will guarantee a high degree of success.

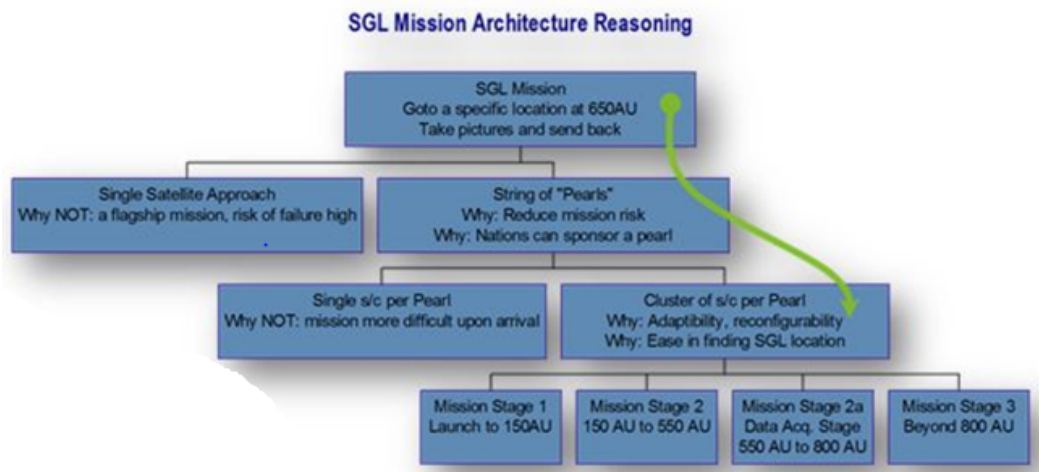

Figure 37. Design logic to identify the SGLF mission architecture.

Figure 37 shows a block tree type diagram that incorporates the reasoning used to arrive at the architecture as presented. The "top of the tree" starts with a basic concept of the mission, "go to the SGL location, take pictures and send back". At the next tier down, we address whether the mission should be via a single satellite approach or a string of pearls. We argue against the single satellite approach because of high risk of failure and accept the string-of-pearls concept because it can spread the cost among multiple nations. The third-tier addresses whether a pearl is comprised of a single spacecraft or multiple spacecraft. We argue against the single spacecraft because it places an undue burden upon arrival, that of finding the SGL focal line and tracking exoplanet, while a cluster of spacecrafts in a formation-flying mode allows for adaptability and reconfigurability in addition to lowering the risk of locating the SGL focal line.

The CONOPS for finding the SGL focal line with multiple spacecraft includes formation flying (a spatial distribution of spacecraft) in which there is one located in the center and other distributed about it. The outer satellites conduct raster scan maneuvers in the radial direction in hopes of crossing the strong interference region of the SGL. The fourth and final tier of Figure 37 parses the totality of the mission into 3 stages or phases. Within each mission stage we list the major goals. However, not listed in mission stage 2 (the long cruise phase at high transit velocity) is the possibility of accomplishing inflight assembly or aggregation in which multiple s/c join (i.e. docked) to form a more mission capable spacecraft or the repurposing of structure (e.g. solar sails as COMM receiving "antenna") to suite the mission goals of mission stage 3 and beyond.

## 3.4 CONOPS

### 3.4.1 Getting into orbit via rideshare or low-cost alternatives

The SGL mission must be affordable for the exploration of not only one, but a number of candidate exoplanetary systems. One of the enablers of an affordable space mission is the minimization of launch costs. The flight components are small and can be launched at any time (no highly restricted launch windows). We thus can avoid any dependence upon dedicated launch vehicles, and instead, use the many rideshare opportunities that will be available during the NASA Moon-Mars initiative and associated government and commercial launches to cislunar space and beyond.

Rideshare is already gaining widespread application for mission to LEO and GEO. The use of the ESPA ring and ESPA-Grande for secondary or proliferated payloads is now routine. The next decade will see this concept expanded to cislunar missions, in which the very large (several tons) cislunar payloads can easily accommodate numerous SGL sailcraft, each of which will weigh less than 50 kg and can support launches into any cislunar trajectory, from which they will independently manage their further trajectory.

### 3.4.2 Data management and TT&C

The essential feature during the flight into perihelion is DSN tracking and resultant guidance updates to (1) aim each sailcraft to the exact location and time of perihelion and, (2) to command the onboard trajectory management system to exit perihelion passage with the right state vector to reach the SGLF of the selected stellar system. So far this is routine operations.

Next phase is the early fly-out in which, again, DSN tracking is used to command the on board trajectory management system to null out injection errors so that by the time each sailcraft reaches the distance from the sun where the sails are no longer effective, the trajectory error ellipsoid is reduced to the uncertainty of the DSN tracking. The sailcraft then becomes a spacecraft with the ejection of the solar sails, except for those that may be used for repurposing (such as for future use as communications antennas).

The spacecraft then enter into the hibernation period (similar to New Horizons) with occasional small trajectory corrections to be made as far out as the DSN can track. Given the light-trip time to the SGL (round trip) ~1 week or more complex onboard navigation becomes an essential part.

### 3.4.3 Targeting the amplified light of the parent star

The trajectory of the SGL mission starts by using guidance and navigation in an Earth-based coordinate system in which the DSN manages the early flight phases (spiral in, acceleration at perihelion, and flight to ~100 AU or more). After a period of transition, the flight enters into an SGL-based coordinate system, in which the system navigates itself to capture the exoplanet's SGL. The planetary SGL is however a very narrow and a dim target that would be difficult to acquire by the onboard instrumentation. Conversely, the SGL of the parent star is very bright and broad – so we use it as the initial "guidepost" to be acquired by the navigator of the pearl. Once the navigator acquires the star SGL, it moves onto it and then guides the other spacecraft within the "pearl" into alignment as well. From there it is merely a matter of orbital mechanics to transition to the SGL of any of the planets in the exoplanetary system. Similar techniques will be used during the 20+ year sojourn along the SGL, in which the SGL axis for the parent star will be the baseline, from which excursions will be made to examine each solar system object of interest (planets, moons, etc.) as viewing conditions and science objectives dictate.

The general mission CONOPS entails, the pearl cluster of satellites passing through perihelion and egress out toward the SGL's optical axis of the target, they acquire the parent star light, establish a navigation baseline, and begin science operations. The two-decade journey between the insertion point at about Jupiter's orbital distance from our Sun where the solar sail effectivity wares off to the target SGL at ~650 AU, will require frequent precise tracking and monitoring to maintain low position and velocity uncertainty limits. To achieve the tens to low hundreds kilometer level accuracy required for this mission, the navigation solution will be a combination of measurements that include: NASA one-way ranging and Doppler, two-way laser communication with repeaters, intra-cluster (between satellites) and inter-pearl (between clusters) two-way ranging, and consideration of X-ray pulsar navigation.

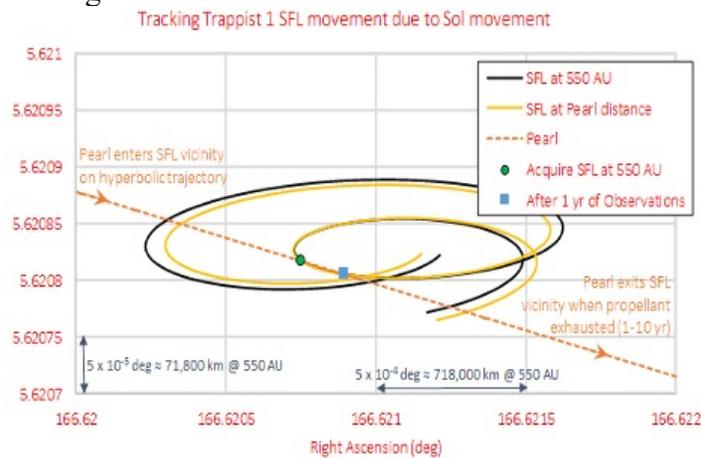

Figure 38. Solar FL movement relative to the hyperbolic.

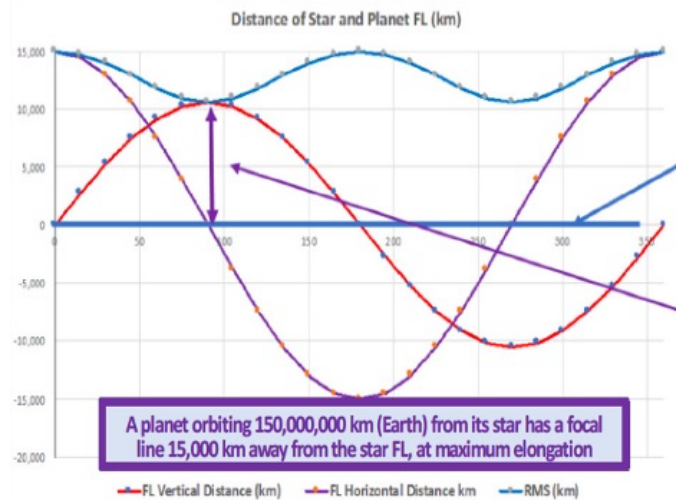

Figure 39. Star-planet focal line geometries, 45° inclined exoplanetary orbit.

The DSN interferometric tracking uncertainty is about one nanoradian in the plane of the sky, yielding ~82 km accuracy at ~650 AU (Lichten, 2018)[20]. This implies that our navigation solution for each pearl cluster will be enough not only for the inner solar system, but all along the trajectory to the target SGLF starting at ~650 AU. Autonomous navigation is a necessity for the pearl cluster of satellites as it gets further away from Earth and becomes less practical to have a human in the loop (communications transit from the SGL is ~4 light days).

---

[20] Dr. S. Lichten, JPL, DSN Manager for Special Project, Personal Communication, November 13, 2018.

A real-time autonomous navigation system onboard a satellite requires an accurate timing source, such as a deep space atomic clock. Development of deep space atomic clocks is ongoing but requires further research for the technology to mature. Navigation status messages will be sent back to Earth with health and safety reports as part of a regular schedule of maintenance and monitoring. If it is determined that the pearl cluster or individual satellites are off course the electric propulsion system will be available for small maneuver corrections along the way. A possibility exists for repurposing the solar sail structure as a radio antenna among other uses instead of jettisoning it.

As the pearl cluster of satellites approaches ~550 AU, the transition between the cruise phase and SGLF operations begins by acquiring the parent star's focused light. SGLF acquisition would be similar in scope to a New Horizon's approach to Pluto/Charon system, except the satellites will not be able to observe the amplified light from the parent star upon approach and will have to detect it when they are within it. The uncertainty of the exact location of the SGLF is aided by a distributed satellite architecture to cover the distance around the uncertainty of the predicted location. Once the satellite cluster is within the theoretical location of the parent star's light, each satellite will use its respective telescope to measure the amplitude of the signal to determine the cluster's precise location relative to the center of the SGLF region to establish a base of operations. If the signal is not detected at first, the cluster as a whole would maneuver to search the surrounding space at a relatively low cost of delta-V (~1-5 m/s). Once SGLF acquisition is confirmed, a handover to intra-cluster ranging relative to the parent star focal line will be used to determine and control relative spacecraft positioning.

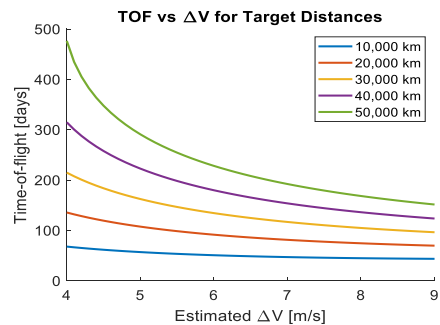

Figure 40. Time of flight vs. delta-V.

The more challenging planet acquisition strategy to look for a thin and dim focal line is undertaken after the parent star's focal line is acquired. The parent star focal line serves as a baseline for science operations for navigation and maneuvering to and from the planetary focal lines. Electric propulsion will be required to maintain the path along the parent SGLF as the Sun wobbles. Figure 21 depicts the movement of the center of the Sun's average position over many decades. The picture is split into four quadrants with a series of dots that spiral around the center representing each month from year 1944 to 2020 (see Figure 21). The dots begin in the lower right-hand quadrant in 1944 and spiral counterclockwise closer toward the center 180° by 1950. Then spiral outward almost toward the right center edge of the figure another 180° by 1956. This spiral pattern is repeated, getting closer to the center, then further away until year 2020. This is due to the gravitational influence of the large planets, mainly Jupiter, in our solar system on the Sun. This wobbling motion will cause the projection of the SGLF to move along this path out at ~550 AU and beyond.

Consequently, the pearl approach trajectory must mimic the wobble during the observation period. The hyperbolic trajectory from the Sun can be tailored to reduce the required delta-V over the many years of operation. This is accomplished by selecting the correct perihelion point at the beginning of the mission that targets the location of the SGLF at the time of arrival. Figure 38 illustrates an example of this relationship between the incoming hyperbolic trajectory and the motion of the SGLF. The plot axes are Right Ascension and Declination that encompass the Sun's wobble motion at ~550 AU. Note that this solar focal line is projected out hundreds of AUs and the motion of this light can be considered moving right to left or vis-versa depending on the actual SGLF target as oppose to the Sun's barycentric wobbling.

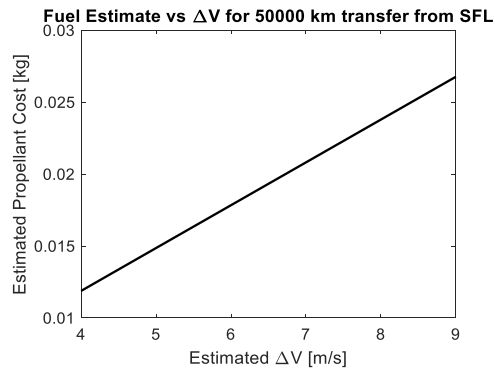
Figure 41. Propellant cost vs. delta-V.

The trajectory in the figure intersects the SGLF at the time of acquisition and remains along the SGLF for 1-year of observations. At this point, the SGLF will have moved due to the wobble and will require a small delta-V (~2 m/s/year) for the pearl of satellites to remain within the SGLF. The amount of time spent within the SGLF is somewhat fictitious as the Sun's motion is constant.

The design of the trajectory could be tailored such that it follows the motion of the focal line at the same rate of the solar wobble and therefore the pearls stay within the SGLF for a much longer period time with less delta-V used. The SGLF motion is bounded and reverses direction (Figure 21) where the Sun's position moves from the right-most edge to the left-most edge of the plot and back again. The incoming trajectory into the SGLF would need to follow the motion of the optical axis until there is a direction reversal requiring a larger delta-V (~15 m/s) to continue following the SGLF. The time of reversal occurs depends on when the pearl cluster acquires the SFL in its solar wobble cycle.

Strategies for moving from the parent star SGLF to planetary focal lines (PFL) are depicted in the Figure 39. The figure shows the two-dimensional movement of the PFL relative to the SGLF. The y-axis represents ±15,000 km distance from the SGLF at zero, and the x-axis is angular measurement of the planets orbital motion from 0° to 360°. The plot shows the vertical and horizontal distance of the PFL for a fictitious Earth orbit inclined 45°. There are many different orbital configurations of exo planets along with their solar system orientation which generate different sinusoidal patterns. In this example, the vertical distance of the PFL starts at 0 degrees and is 15,000 km above the SGLF and moves to -15,000 km in 180° and back to +15,000 km at 360°. The horizontal motion starts at 0 km and osculates between ±10,000 km over 360 degrees. The root-mean square of these orbital components osculates between +15,000 to +10,000 km above the SGLF over the 360° angular motion. This means that to acquire the PFL the satellites would maneuver to the minimum distance of 10,000 km relative to the SGLF to follow the planet's focal line pattern. The delta-V cost to maneuver to these PFLs are dependent on geometry of the extra-solar planet and operational scheme to collect the science data over the length of the mission.

The estimation of delta-V for planetary focal line maneuvering used a rectilinear model as a first order approximation. An Electrospray thruster was used as a close approximation of what will be used on the satellites with an $I_{sp}$ of 1,200 seconds and thrust of 0.1 mN. The model assumes constant gravity from Sun and a constant thrust from the engine. Each of the estimates start with zero initial velocity. The burn direction is against the Sun line for a time, then immediately turn around and burn in the opposite direction such that vehicle is at zero velocity again at desired distance. Figure 40 shows the results of this model as the time of flight versus the estimated delta-V to various PFL distances. There are five curves that represent the PFL distances ranging from 10,000 km to 50,000 km from the SGLF. To transverse the largest distance of 50,000 km it takes between 200–500 days at 4–9 m/s of delta-v respectively. The shortest distance of 10,000 km takes 50–75 days at 4–9 m/s of delta-V respectively. Note that these delta-V are only the amount to reach the PFL location but require approximately twice this amount to follow the planetary motion trajectory. Figure 41 depicts the fuel consumption versus the estimated delta-V for a 50,000 km transfer from the SGLF. The fuel consumption spans between 0.012–0.027 kg at a delta-v cost of 4–9 m/s respectively. This plot provides a way to approximately size the amount of fuel needed to create a delta-V budget for the science operations.

### 3.4.4 Taking and processing data

The AI system CONOPS employs Intelligent Autonomy software to support constellation operations and complex maneuvers. To accomplish this, the AI system learns as it flies, with payload tasking and prioritization, the AI architecture supports adaptive processing of mission data and employs various levels of Intelligent Autonomy that is dependent upon mission needs (Figure 42).

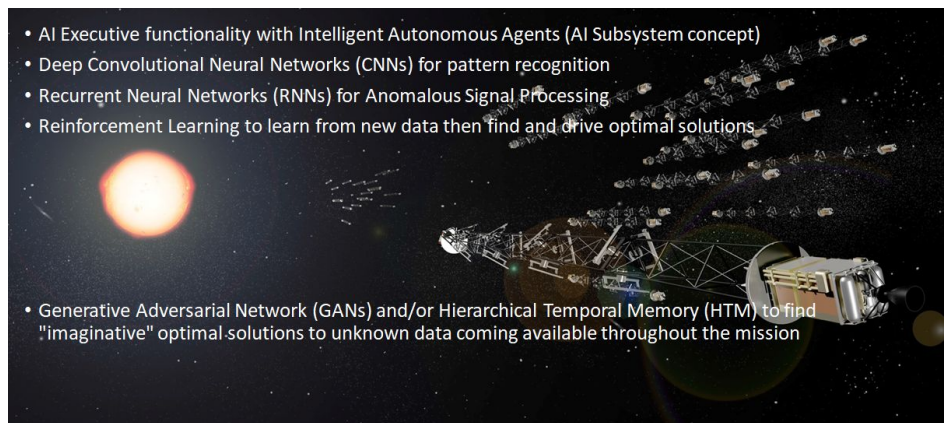

Figure 42. AI algorithm design for a mission to the SGLF.

Autonomy functions are designed to be hierarchical. For example, the system must reprogram and reallocate functionality to optimize consumables, data collection, and analytics to remove "spurious" photons. On station and in real time, the AI must drive relative navigation for optimal positioning and refine the science mission collection by adjusting for distortions of in the SGLF (e.g., corona effects). Each step for mission planning will be calculated by AI to optimize mission performance (see Figure 43).

High level autonomy at the system level could be similar to a traditional MAPE loop for mission tasking – Monitor-Analyzer-Planner-Executive. Each node would have a MAPE loop running. JPL, NASA Ames, and Aerospace have done a lot of work in this area. There are trades between allocating functionality at the local node vs. system level. Levels below this might use more advanced techniques depending on type of mission, e.g., optical sensor data processing might incorporate a Convolutional Neural Network (CNN) for object classification, mesh networks might have their own algorithms for routing, General Fuzzy Inference Engine (GFIE) or Reinforcement Learning for constellation orbit maintenance, Recurrent Neural Networks (RNNs) or Hierarchical Temporal Memory (HTM) for signals intelligence, and Generative Adversarial Networks (GANs: a class of machine learning systems where two neural networks contest with each other in a game-theory sense on a given training set. The outcome is new "perceptions" and characteristics of the data) to find unique solutions to unknown data obtained during the mission (Figure 42).

Figure 43. AI as a mission enabler for SGLF.

There will likely be some sort of neural network or statistical process to detect and counter attacks in the cyber domain. Lifelong Learning should feed experience back into the system, so it can retrain itself to improve over time. For instance, given a certain condition, the AI calculates courses of action (COAs), selects best COA, executes it, then the system should take the result and feed that back to learn for mission optimization.

Autonomy levels use a range of algorithms from deterministic to stochastic, from mission autonomy software to enable collaboration among constellation nodes and for long-term operations without human interaction to creative solutions to unforeseen problems. The AI architecture will be designed to manage and mitigate degradations and latencies across the constellation with automated failure recovery and auto-scaling.

Trusted AI and autonomy will form the foundation of the autonomous decision making. Methods to achieve this will range from traditional agile verification and validation for deterministic conditions to more esoteric methodologies developed for nondeterministic conditions, where the AI will learn on station, then act on what it learned. Trust in this dynamic condition could be established by using methods similar to those used for testing human performance, like evidence-based licensure. Aerospace and JPL are studying this range of methods to determine how best to proceed with Trusted AI.

## 3.5 Benefits of the SoP architecture

### 3.5.1 Science

The SoP architecture allows for faster image completion, "clouds removal", observation of multiple targets in the exosolar system, etc. Our analysis (Turyshev & Toth, 2020c) suggests that for $d$ = 1m it could take up to $\sim 3 \times 10^3$ sec of integration time per pixel to reach the SNR$_R$=7 for an image of N=100×100=10,000 pixels. For $z$ = 650 AU, this translates ~ 1 year of total integration time needed to recover the entire 100×100-pixel image of an exoplanet at 30 pc. Using for this purpose a larger telescope, say $d$ = 2 m, the per-pixel integration time drops to 390 sec, reducing the integration time required to recover an image with the same number of pixels to < 1.5 months of integration time. Use of a 5~m telescope implies a per-pixel integration time of ~ 150 s on a 250×250-pixel image, for a total integration time of ~110 days.

The SoP architecture allows sampling higher temporal frequencies relevant to target evolution. Thus, collecting more, redundant data will allow us to account for the diurnal rotation of the exoplanet and its variable cloud cover. To compensate for the diurnal rotation, we may also benefit from a multi-telescope architecture that can reduce the total integration time (Turyshev et al., 2018; Turyshev & Toth, 2020c), while matching the temporal behavior of the target. However, if the direct spectroscopy of an exoplanet atmosphere is the main mission objective, this can be achieved with a single spacecraft. We emphasize that direct imaging and spectroscopy of an exoplanet at such resolutions are impossible using any of the conventional astronomical instruments, either telescopes or interferometers; the SGL is the only means to obtain such results.

The primary innovation of this mission concept comes from the ability of using a well-positioned sensor at the SGL to image and raster scan the image of a promising exoplanet. Other innovations include: i) the recognition and intentional use of the SGL to magnify with unprecedented angular resolution, ii) the CONOPS of acquiring the information pixel-by-pixel with calibrated and time synchronous s/c motion, iii) deconvolving the image using the known PSF. Given the state of current astronomical telescope developments, including JWST and TESS, our proposal is timely and relevant. It takes advantage of the likely discovery of numerous exoplanet candidates by TESS and further assessed by JWST. While these missions may provide "hints", one image from the SGL could "seal-the-deal". The SGL mission offers a unique means for determining exoplanetary atmospheric chemical composition and defining the propensity for habitability of life.

### 3.5.2 Technical

Each spacecraft is based on the same bus structure but may contain different subsystems. The common bus allows international organizations to participate in the mission by producing satellite bus structures at the very least. Each pearl consists of as many satellites as needed to provide high functional reliability (estimates are 10). There is a factor of 2 redundancy built into the architecture with 10 spacecraft. Component or subsystem reconfiguration and reuse is applied to repurpose on board assets to drive down weight. Also, the solar sail material is re-used to collect SGL photons and serve as an optical uplink antenna dish. The string-of-pearls concept provides for an inter pearl distance of 20 AU, but COMM link budgets close at 40 and 60 AU and even with 200 AU separation (albeit with a much lower bit rate transfer). The proposed architecture relies on the fact that early pearl arrivals to the SGL will be able to pass down, to approaching pearls, information. This approach permits a "learning" mode to be implemented among the approaching pearls. Because of the string-of-pearls concept and the ability to pass PNT information among the pearls, it is estimated that it would easier to keep the pearls along the SGL "vector" line.

### 3.5.3 Economies of scale, multiple partners

The SGL offers a unique means for imaging exoplanets and determining their habitability. Theoretical considerations (Turyshev & Toth, 2017, 2020abc) are promising, both for getting there and for capturing high-resolution images/spectra of potentially habitable exoplanet. The mission concept has the potential of being the most (perhaps only) practical and cost-effective way of obtaining km-scale resolution of a habitable exoplanet, discovering and studying life on other worlds.

The analysis shows that the highest trajectory adjustment occurs during the outbound phase where corrections on the order of 95 m/s are needed over a 5-year timeframe. One efficient thrusting approach is to produce the ΔV over a period of 1 year or less for each 5-year trajectory update. Providing 95 m/s over 1 year requires a thrust of only 3 µN per kg of spacecraft mass. Power and thrust are proportional to s/c mass. During the data acquisition phase, the yearly ΔV requirement drops to 20 m/s per year. Ultimately, as mentioned before, on-going technological advancements could lead to better performance, reliability, and lower size, weight and power which often translates to lower cost. This opens a path for many partners to participate in the mission.

### *3.5.4 Impact on the entire space industry*

The design and development of a space architecture for deep space missions capable of autonomous, adaptive, and self-learning operations, establishes the criteria of how future spacecraft will be developed and expected to operate. This study will develop: (1) A plausible mission concept that could deliver a healthy/capable spacecraft to distances beyond 650 AU from the Sun, place it on an actively controlled trajectory to form a telescope for megapixel *direct* imaging/spectroscopy of an exoplanet; (2) A CONOPS for the precision positioning, navigation and timing of s/c outside the solar system that will detect, track, and study the brightness of the exoplanet's Einstein ring around the Sun; (3) A "must-have" set of remote-sensing instruments and onboard capabilities for unambiguous detection and study of life on another planet; (4) A list of technologies that could enable autonomous, small s/c, energy harvesting materials and propulsion approaches for fast exit from the solar system. Assessment of technologies needed to survive the long journey (i.e., clocks, sensors, electric propulsion) and to operate upon arrival (i.e., reliability and resiliency).

Learning about the existence of alien life present on an exoplanet will be a transformational event in human history, resulting from the proposed mission's delivery of direct multipixel images and high-resolution spectroscopy enabled by the SGL. This possibility could become the focal point for the multiple efforts in the entire space industry.

## 4 TECHNOLOGY GAPS ANALYSIS, RISKS AND MITIGATION STRATEGIES

### 4.1 Current status in various technology areas

Figure 44 presents a table of the CONOPS of the SGL mission. The CONOPS is predicated on the desire to reduce mission risk and cost. The table presents a series of mission ideas or approaches that were studied and its effect on architecture, propulsion, COMM, PNT, structural materials, computation and sensors/payload. The concepts speculated range from a single s/c mission to a series of s/c solar sail powered to achieve solar system exit velocity that apply self-assembly in flight to form larger integrated s/c. We start with the estimated TRLs of the various subsystems.

- Propulsion:
    - We believe solar sail technology is at a TRL level 5 (given the space missions, IKAROS and LightSail 2) however the building of larger sails is currently at a TRL 3, but the folding concepts proposed by JPL's Starshade[21], provide hope that larger sails can be developed. Solar sail materials that could be survivable for a close perihelion passing is a TRL 1, but the use of new materials such as metamaterials, thin film ceramics could be a "game-changer".

---

[21] https://science.jpl.nasa.gov/projects/Starshade/

- o We believe that ion thrusters, for minute orbit/trajectory corrections, have a TRL 7 or higher given that there are multiple companies developing and have flown small ion thrusters (e.g. Morpheus, Millennium space[22], etc.).
- COMM:
  - o For intra-pearl cross link communications, the suggested V-band COMM technologies are well advanced to TRL 8 or higher given that a part of this frequency band is now used in automotive systems for collision avoidance. NASA have developed an experimental campaign for collecting V-band propagation data using a Geo-Sat (Acosta et al., 2012). W-band RF would also work, part of the choice would depend on which frequency band would most easily be integrated into a small package and have the higher electrical efficiency).
  - o For inter-pearl down/uplink COMM, we have recommended the use of optical laser COMM. NASA is now evaluating the use of laser comm for deep space communications to add to the capabilities already present via the DSN (Biswas et al., 2016). More recently Aerospace have flown laser comm system in the CubeSat form factor (Rose et al., 2016). We estimate that miniature-packaged laser comm is currently at a TRL 5.

| Mission ideas studied ⇒ Affect on ⇓ | Single s/c | Multiple medium size s/c | String of Pearls | Multiple s/c in a pearl | Appropriate functions to multiple s/c | Rendezvous orbit prior to mission start | Use of Ride-Sharing | Use of Solar Sail | Repurpose structure | Apply Self Assembly in flight |
|---|---|---|---|---|---|---|---|---|---|---|
| Architecture | Large, costly, large risk | Still costly, but reduces single point failure | Allows high bps downlink, self learning, share cost, hardware upgradeability | Reduces risk during Mission, | Reduces net mass per s/c | Allows on orbit testing prior to leave, removing "dead" s/c | Reduced cost | Exit velocity if s/c low mas | Reduce s/c mass | Allows the build of larger s/c after achieving exit velocity. |
| Propulsion | Heavy, no launcher available, slow | Same limitations as single s/c | Smaller ion thrusters for cross track drifts (>1500 s⁻¹, solid fuel) | Forces multiple propulsion schemes on board | Reduces s/c mass and total DeltaV, allows for "fuel tankers" | Install teaming principle, "pattern flying" during high thrust phase | Dispenser size defines s/c size | DeltaV > 20 AU/yr. | Convert structure to solid state fuel. | Allows for "in space fueling |
| COMM (Down and Cross link) | Requires 50-100 W optical laser for downlink | Same as single s/c with less mass V band crosslink 1Mbps, 2E5 km | Permits optical downlink with repeaters 5-10W laser | Downlink only on limited s/c, RF crosslink V band 1 Mbps at 2E5 km, today | Must have high bps cross link | NA | NA | NA | Solar sail to optical downlink& payload telescope, | Allows larger downlink s/c by connecting a laser COMM s/c with battery-s/c |
| PNT | DSN to 500 AU, star tracker, clock updates | DSN to 500 AU, star tracker, clock updates | trajectory revise via best fit 500AU baseline, star tracker | Can assign NAV s/c (e.g. star tracker sat), redundancy | Forces group synchronous processes | NA | NA | Forces sail control & momentum dumping | NA | Possibility of building a larger telescope or downlink |
| Structure Materials | Heavy, high mass materials | High mass materials | NA | Scaffold design graphite bars merged with propellant | NA | Can use containerized shipping | Can | | Include propellant fuel in structure | Move structure or function where necessary |
| Computation | Traditional flight comp. | Distributed computation, | Availability of machine learning for pearl in route | Distributed computation, and distributing search for SGL | Possible use of high power GPU and FPGA | NA | NA | Fast analysis at perihelion | NA | Can provide addition power to increase computation |
| Sensors payload | All on one s/c | Data speed increase with multiple sensor | Each Pearl has same sensors | Downlink COMM s/c, Science s/c | Reduces redundancy, increases means | NA (key sensors in container ship) | NA | High speed control at perihelion | NA | Provide higher power (batteries, UPTIME), larger capture area |

Figure 44. Concept of operations (CONOPS) for a mission to the SGL.

- PNT:
  - o *Clocks and clock synchronization*: The current technology of the chip scale atomic clock (CSAC) offers a TRL 8 but requires near continual synchronization at nearly 4-month intervals (for < 1ppm drift). We anticipate a factor of 10 to 100 improvement in less than a decade pushing the synchronization requirements to decades. We estimate that frequency comb-based clocks (currently TRL 4) to supplant the CSAC within 10 years and given this, clocks synchronized on earth will not need further corrections for the duration of the mission.
  - o *Star trackers*: Current star trackers are at TRL 9, miniaturized versions of these units are now commercially available (e.g. BST ST200, ST400) but current star tracker accuracies are on the order of microradians (e.g. Terma corp. the HE-5AS), a factor of 10 to 100 improvement would help. One advantage in the SGL mission is that in the cruise phase (Phase 2 and 3) there is time for long duration exposures which should increase the accuracy (e.g. the Terma Corp.[23] HE-5AS quotes < 4 microradian accuracy with 3 second exposure).

---

[22] https://millennium-space.com/

- *Pulsar sensors*: Currently at TRL 5 with space relevance, they need to be miniaturized with a significant reduction in mass (i.e. 50 kg → 5 kg). With multiple pulsar sensors there would be no need for start trackers nor worry about clock synchronization.
- We believe that NASA's DSN will continue to be upgraded such that its utility for positioning could be extended beyond 150 AU (e.g. Voyager 1-2 missions).

- Structural Materials: Carbon fiber nanocomposite materials are currently at TRL 9. The current design of the SGL spacecraft attempts to imbed within the structure, sensors (e.g. structural, temperature, vibration) and power sources (e.g. batteries, microRTGs). We believe the TRL of this composite material is either at a TRL 7-9 for imbedding sensors (e.g. temperature, strain, etc.) and 1-2 for embedding power sources.

- **Technologies that should be closely tracked:** *to drive down weight, risk and cost*
  - Solar sail materials; − Solar sail propulsion control;
  - Higher speed computers and rad hard computers; − Contamination buildup.
- **Relevant & Anticipated Developments in the next 10 years:**
  - **Batteries:** Anticipate battery density (J/kg) to increase by factor of 2 to 4 → removing ~ 5-7 kg of mass per SGL s/c.
  - **On-board clock:** A factor of 100 improvement in chip scale atomic clocks, 33 yrs for 1 Hz drift (0.1 ppm).
  - **NAV:** Star trackers with high resolution data from Gaia mission should reach 1 uas angular resolution.
  - **Ride Share services:** e.g. Arianespace dedicated rideshare mission to GEO in 2022 (GO-1).
  - **RF antenna:** mmWave D-band antenna arrays for more efficient RF crosslink and lower SWaP (e.g. NuvoTronics).
  - **COMM/Downlink:** NASA's Terabyte Infrared Delivery (TBIRD) Program: Large-Volume Data Transfer from LEO11
  - **COMM/Crosslink:** NASA Inter-spacecraft omnidirectional optical communicator (ISOC), Honeywell's Optical Pointing and Tracking Relay Assembly (OPTRAC) 10Gps optical intersatellite COMM.
  - **Rad Hard electronics**: upcoming commercial "mega constellations" will be a training ground for small sat electronics and autonomous systems (e.g. Cobham Advanced Electronics Solutions), RadFxSat mission radiation effects on advanced nanoelectronics.
  - **Solar Sail Technology:** New companies developing technology (e.g. Roccor).
  - **Space-based COMM relays:** commercial "mega constellation" development will also spawn new companies to move data (e.g. Analytical Space, Addvalue).
  - **Propulsive ESPA rings for "last mile" orbit insertion:** (e.g. Momentus via Vigoride platform, Moog via Orbit Maneuver Vehicles, JPL via UEBER, Firefly via Orbit Transfer Vehicle).

Figure 45. Relevant technology developments.

- Computation: Recent work in integrating CPUs with GPUs suggests that a high-performance computing platform with multiple capabilities is in the design phase which will allow onboard AI and machine learning algorithms to be run.[24] One driver in system design is reliability over a mission duration in excess of 40 years; as onboard data processing requirements are modest and the required level of autonomy does not depend on high performance computing, a simpler, proven, more robust architecture may be the solution of choice. Software technologies that are essential to mission success, such as adaptive networking and distributed data processing, are readily available. Selection of hardware with proven reliability over the course of 40+ years is a critical part of mission preparation.
- Sensors/payload: We believe the sensor technology for the science mission is already at a high TRL. It is the coronagraph technology that needs to move up the TRL (currently at TRL 4).

---

[23] https://www.terma.com/
[24] Please see Intel's announcement: https://newsroom.intel.com/news/intel-named-darpa-project-focused-machine-learning-artificial-intelligence/

Moreover, if the repurposed solar sail material is to be used as an optical surface for directing SGL photons on to the sensors, a TRL improvement (currently TRL 2) will be necessary to maintain the surface shape.

Figure 45 summarizes the relevant technology developments for a mission to the SGL.

## 4.2 Gaps identified for SGLF

Many of the needed technologies will be developed by the many ongoing space programs that are moving away from single large, costly, and high-risk platforms, towards proliferated constellations, on orbit reconfiguration, autonomous operations, and most importantly reduced s/c mass.

The technology gaps are in two categories – the first is nav and guidance during the flight phase the extends beyond DSN tracking capability to the acquisition of the stellar SGLF by the navigator spacecraft. DSN enhancements already planned may make the issue moot, if this interregnum can be eliminated. Otherwise the application of pulsar nav techniques might be applied, already flying at TRL 8-9, but in need of major weight reduction to be applicable to the SHL mission.

The second category is the solar sail technology, now at TRL 8-9 for missions that have been flown but at TRL 2 for the multivane configuration described for this mission.

In somewhat more specific detail, the major technology gaps that have been identified include,
a) Solar sail materials that can withstand a close perihelion pass (e.g. 5-15 solar radii) that can be manufactured and packaged to enable surface area to mass ratios of 200-400 $m^2$/kg,
b) Processes and means for accelerated testing of reliability for components, subsystems, systems (i.e. 50-year operation),
c) Technology that will maintain and or remove molecular adsorbate contamination from critical surfaces over the mission duration,
d) High performance computation platforms that can operate at low power but allow for artificial intelligence and machine learning algorithms to be continuously run for optimizing and adapting to unexpected scenarios,
e) Radiation shielding of critical components. It is estimated that 20% of the anomalies occurring in satellites are due to the space environment (Bourdarie & Xapsos, 2008). It will be necessary to protect from both galactic cosmic rays (GCR) and solar particles from coronal mass ejections (CMEs). The former originates outside the solar system (87% protons, 12% alphas, 1% heavy ions) can have energies up to $10^{11}$ GeV (nominal values are 1 GeV/nucleon) with fluxes 1-10 $cm^{-2}s^{-1}$. The CMEs are not continuous, but harmful and are comprised of 96.4% protons, 3.5% alphas and 0.1% heavy ions and have ~GeV/nucleon energies with fluxes $10^5$ $cm^{-2}s^{-1}$. The critical components of the SGL spacecraft will be placed in a housing with cross section of 1$m^2$. Simple calculations suggest that over the course of the mission, the housing will be bombarded by ~$1.6 \times 10^{14}$ very high energy particles. The housing will have to shield these GeV particles, while they are just few, they do carry significant energy/momentum and over the course of the mission, the probability of a nonrecoverable single event effect goes up.

## 4.3 Technology roadmap

Given the technologies that underlie the SGL mission, we have evaluated the NASA Space Technology Road Maps (NASA Space Technology Roadmaps-STR: for the fourteen technology areas/Technology area strategic roadmaps (TASR) Technology area breakdown structures (TABS)). The STR is mapped out to 2030-2035-time frame. Some of the key SGL technologies are discussed using the NASA STR document as guide.

- *In-Space Propulsion*: we are happy to see EP technology having started in 2010 with colloid micropropulsion and electrospray propulsion with flight demonstrations ST7, Falcon Sat in the 2015-time frames. The "Miniature Ion Thruster" having been started in the 2015 timeframe has a technology pull missions Grace II and Exoplanet Finder, while the electrospray technology (MEMS device) has a technology pull by DARPA and also the Exoplanet Finder mission. NASA also started work on solar sail materials <10 g/m$^2$ sail areal density in the 2013-time frame and estimates materials with areal density at < 1g/m$^2$ in the 2023-time frame with a technology push for a space demo > 40,000 m$^2$ solar sail in 2027. We estimate the SGL solar sail area to be <10,000 m$^2$ (and have the acceleration properties of a 400 m$^2$/kg area to mass ratio). The SGL mission requires not only large solar sail area but also the materials that that can withstand close solar perihelion pass by (further discussed in the Sections 3.3 and 4.)
- *Power generation*: NASA has been exploring radioisotope technology both advanced-Stirling and 10 W-radioisotope for the past 10 years. Other relevant energy harvesting technologies will be explored in the 2022+ time frame, for example Nanothermoelectrics (High Z-T) which is intended to lead into a wireless micro-power bus.
- *Robotics & autonomous systems*: NASA has a number of ongoing projects relevant to the SGL, such as auto rendezvous and docking with adaptive systems technology slated for the Europa and Mars missions (late 2020's), both are technology pulls with strong overlap to the SGL mission. NASA estimates swarming nanosatellites to be a technology available in the 2025-time frame.
- *Communications & PNT*: NASA estimates advances in the DSN (both downlink and uplink) by factors of 100X and 1000X by 2024 in support of Mars-22, Mars-24, … NASA also, anticipates X-ray NAV to be available by 2028-time frame.
- *Electronics*: NASA already has a road map for hardening electronics to 3Mrad by the 2020s. We anticipate this to be a continual effort for further hardening for the longer missions on the NASA logs.
- *Materials/Structures*: NASA estimates the development of cyber physical systems starting in the 2020s. These would be systems that can sense and adapt as necessary leading to autonomous action.

As a result, we conclude that the technology needs for the SGL mission are already under development by NASA because of its desire to extend exploration within the solar system using autonomous systems.

## 4.4 Long-duration project

The length of the SGL mission poses an issue of sensor contamination via self-produced outgassing. While not a showstopper, it will have to be addressed. Technology on contamination control has advanced recently such that key segments can be protected from contamination build up by a) proper choice of materials, b) proper venting and c) heating. Using data from ESA's Rosetta mission satellite which was designed with contamination control in mind and had contaminant particle monitors on board, we estimate a ~9 μm of buildup over a 40-year SGL journey. Figure 46 is a plot that relates the amount of contaminant monolayers formed as a function of contaminant particle density (#/m$^3$). The estimation is based on gas kinetic rates given a spacecraft temperature of 80K, an adsorbed layer sticking coefficient of 1, the average mass of the contaminant is 12 amu (i.e. carbon), but it does *not* include UV/VUV photochemistry which could increase buildup by radical and charged species formation. From the fitted line in the figure, for a contaminant particle density of $5\times10^{12}$, it is estimated that $\sim1\times10^5$ monolayers to be formed over a 40-year journey.

This amount of buildup (typically of polymers) would be detrimental to the optical/imaging sensors. Rosetta implemented sensor/heating technology and proper spacecraft venting designs and an open spacecraft structure. All of which are implemented in the SGL spacecraft design. Additional approaches might be necessary to maintain key sensors free of all forms of contamination. One possible solution (currently at TRL 2) is the use of ultrasonic excitation to enhance molecular diffusion on critical surfaces. A 19-fold enhancement in molecular adsorbate diffusion has been measured (4500-fold estimated) for gold clusters on silicon surfaces (Shugaev et al. 2015).

## 4.5 Risk and risk mitigation strategies

The SGLF mission utilizes a telescope with high-TRL sensors and electronics on a low-TRL small sailcraft. The highest-risk, low-TRL aspects of the mission are associated with the sail materials, GNC of a sailcraft, long-term autonomy, onboard data processing and communication. In the mission error budget, the largest error source is corona photon noise, but the observing scenario is carefully designed to maximize the independence of measurements on any given target during the course of the mission. The detector is large enough that it allows variation of the placement of the Einstein ring on the focal plane. For these and other reasons, a significant portion of the error reduction is through averaging a large number of errors. Therefore, a single aspect of the mission working with degraded performance is not enough to have a disruptive effect on the science performance. Finally, at the present time, we have testbed and modeling results at JPL that assess the performance in each of the risk areas, and the expected performance exceeds requirements in all.

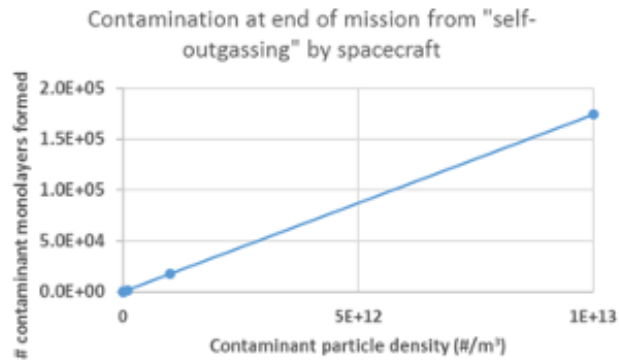

Figure 46. Contamination due to long duration flight.

The areas of greatest technology risk for the mission are 1) sail materials and survival near the sun, 2) precision GNC of the swarm, 3) telescope/coronagraph optics, 4) the telescope structural resilience, and 5) onboard processing and autonomy, in order of importance. Some of these will be retired during this Phase III effort, some will be matured in the course of upcoming test missions. The evolutionary collaborative nature of the SoP architecture enables mitigation – i.e., this approach distributes the risks broadly to multiple stakeholders and the pearl provides resiliency against failures due to reliability. The multiyear launches provide opportunities to address design or material shortfalls assuming we can design in adequate fault detection, identification and recovery capability into the avionics to enable us to learn from any anomalies experienced during flight.

One of the architectural risks that must be mitigated is resolving the inherent conflict between minimizing the weight of each sailcraft launched through perihelion (to maintain the size of the solar sail within achievable parameters) vs. the need to have a large-enough set of instruments-coronagraphs to effectively collect the exoplanetary photons. In short – how can such small s/c support such heavy instruments?

We meet this challenge by first looking to biology: a cell, while a functioning unit, does not a human make. Moreover, we apply the technologies being developed for on-orbit assembly and reconfiguration. Recent publications, such as the report by Aerospace Corporation Center for Space Policy and Strategy (Piskorz & Jones, 2018): *"On-orbit assembly is an important step toward the proliferation of highly adaptable and capable space infrastructure. Space capabilities for both traditional and new stakeholders will be revolutionized by the ability to assemble modular building blocks into a functional and complex infrastructure. To succeed, these building blocks must be compatible and interoperable, with some level of autonomy; however, there is currently no governance for establishing standards in key areas that enable on-orbit assembly (e.g., mechanical, electrical, power, thermal, and data interfaces)."*

NASA has studied this approach in application to large space telescopes (Stahl et al., 2005; Mukherjee et al., 2019). We are looking at it for the case of multiple small telescope segments that are brought in contact during the SGL cruise to assemble and reconfigure a functioning telescope during the SGLF operational phase.

We will study this from the technical and scientific points of view. The technical challenge will be to parse the functioning SGLF s/c into a prefabricated set of subsystems, that have some self-operative capability to survive to the cruise phase for re-aggregation. The swarm of spacecraft in each "pearl" will be designed to perform rendezvous and docking to share power, fuel, communications and other resources. This will include the prefabricated components to build the instrument-coronagraphs.

This in-flight s/c systems aggregation approach creates an advantage for the science mission. For example, the aggregation of a s/c that has high power for downlink or data processing. Similarly, an aggregation that forms a large aperture coronagraph. Many science missions are sent to physical objects (moons, planets, asteroids, etc.) in which the characteristics of the targets are known and so the system design can be "frozen" early in the development phase. The SGLF mission is very different, it is a voyage to a "stream of photons" in which the geometry, intensity, and temporal variability of the signal (the Einstein ring) and the sources of interference such as the solar corona, are currently only known and are well-established by our extensive recent modelling efforts (Turyshev & Toth, 2020abc). Given the distances and the decades-long time needed to arrive there, we do not have the luxury of preliminary exploratory flights to the SGLF to design and build future missions – to get it right, we have to be able to adapt and not just in the software.

The scenario is as follows. Upon arrival of the first pearl, we will analyze by initial observations to determine the right dimensions and capabilities of the set of science instruments and coronagraphs. Alterations or adaptations to the instrument configurations could be handled by the arriving pearl and the information passed down to the subsequent pearls in transit. It may be that a single design will be best, or a range of designs to meet collection needs, or an evolving design to meet conditions driven by the impact parameter increase as the s/c progress along the SGLF. The same flexibility applies to the data collection, storage and transmission protocols employed among the s/c in each pearl and the from pearls to Earth in order that the most important science data is sent over the vast distances with minimal expenditure of power.

For each SGLF and over a period of ~ 6 months, individual subsystems are launched as sailcraft, using rideshare opportunities. Then solar sail propulsion will be used to spiral down to perihelion at a common place and time to allow subsequent group flight to the image cylinder in the strong interference region of the SGL. For example, if the s/c all pass perihelium within 100 seconds of each other, they would emerge, spread out, by some 15,000 km, well within the capability of the sailcraft to reduce the separation during the next phase of flight.

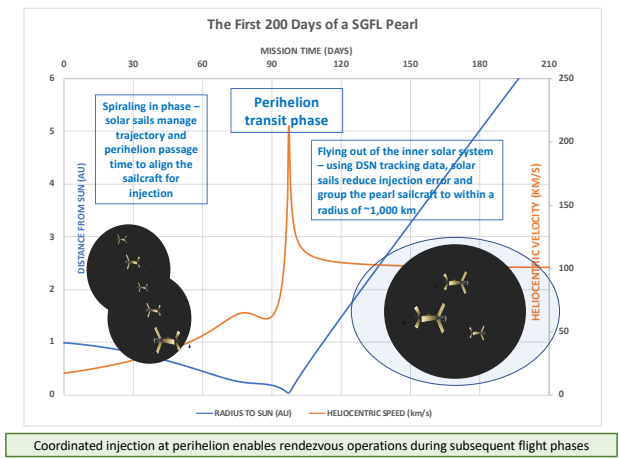

Figure 47. Velocity and distance from Sun of sailcraft to and after perihelion transit.

The perihelion delta-V is used to propel the sailcraft at ~150 km/sec towards the chosen parent star's SGL focal line During the first hundred days of the fly-out from the sun (days 100 – 200 in the example shown below) the solar sails are used, with DSN tracking, to reduce the injection error ellipse and to group all of the sailcraft in the pearl into an integrated "swarm" with common state vectors accurately aimed at the SGLF and position them within an outward flying envelope of ~1,000 km in radius.

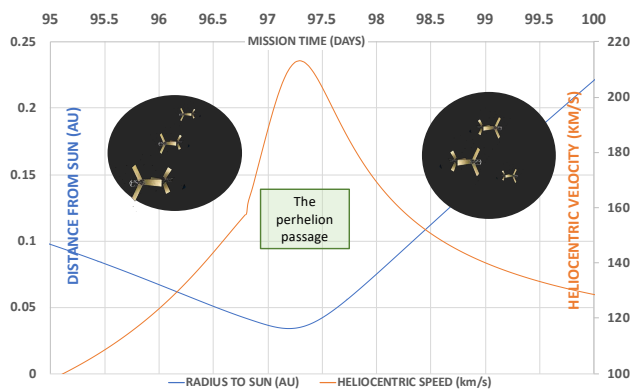

Figure 48. Expanded view of velocity and distance from Sun of a sailcraft during critical perihelion transit passage.

Once the solar sail propulsive effectiveness is lost due to distance from the sun (at Jupiter's orbit the solar sails have only 4% of their effectiveness at 1 AU) the sails are ejected to reduce weight (except for material that may be later repurposed for uses such as contamination protection, thermal control, etc.). The outward flight of the pearl is controlled by the navigation spacecraft that manages trajectory trimming and the positional relationships of the swarm components. The process is illustrated in the next two figures (Figure 47 and Figure 48).

The pearl then flies outward for the next ~20 years with the navigational s/c of the swarm commanding occasional small delta V corrections to intercept the SGLF of the parent star. The parent star SGLF is the relatively bright and broad beacon that acts as the "glide slope" to bring the pearl to the exoplanet SGLF. During this process the individual spacecraft can be brought together with micro-thrusts using <1 m/sec of total delta V over the entire mission, as time is not of the essence. Even though the absolute speed of the pearl is very high, we realize that the space during the cruise phase is benign. Consequently, minute relative speed alterations among the spacecraft on the order of 1 micro-meter/sec are possible with negligible fuel consumption. The technologies needed for group flight and precision rendezvous and docking are being developed and tested to TRL 8-9 by government and commercial projects. The support of the Space Force in flight proofing these technologies will be particularly helpful in implementing the needed SGLF CONOPS.

The use of AI/machine learning to assist in this process along with laser ranging are essential, as well as how the s/c components are parsed among the cluster of formation-flying satellites that form the pearl. This approach to the reconfigurable space-based design and assembly process and the learning from pearl to pearl, will give each successive pearl entering into the SGLF more scientific capability. This allows comprehensive surveillance of each object in an exosolar system and compensates for the time variability of the targets (their evolution around their star, their rotation, and possibly, their weather patterns).

## 5 SUMMARY AND RECOMMENDATIONS

### 5.1 SoP architecture to revolutionize planetary science

The science objective is not just to get an image of one planet but to have the opportunity to image and explore in a great depth many exoplanets orbiting stars in our stellar neighborhood and to do it affordably. The uniqueness of the SGL is that the flight distance to the target (650+ AU) is independent of the distance of the star, only the celestial latitude and longitude need to be dialed in and everything else is the same. This leads to the economies of scale that permit multiple concurrent flights to different candidate stars.

The SoP architecture meets the science requirements of enabling flights to one or more exosolar system SGL FLs – and within a mission to a given exosolar system, allows the examination of all objects of interest in search of signs of intelligent life – for example signs of distribution of life throughout a solar system, not just on a "home planet".

Figure 49 shows the overview of the SoP architecture. We start (lower left) with rideshare deployments of sailcraft into cislunar space. We define a sailcraft as a smallsat and its associated solar sail. They become spacecraft once the tasks of the solar sails are fulfilled, and the solar sails are ejected (for some spacecraft).

We "loadmaster" deployments depending upon the availability of rideshare missions, during which multiple sailcraft may be deployed for each rideshare mission (center top of Figure 49) – once in cis lunar space, the sailcraft spread their solar sails and use their internal capability to begin the downward spiral towards the perihelion. The rideshare principle is particularly attractive in that it operates on a noninterference basis with the host, can be deployed anywhere and at any time in cislunar space, and can then "fend for themselves". The sailcraft are then aggregated into "pearls" during the spiral down phase (center of Figure 49).

During the spiral down phase, the inclination of the heliocentric orbit is adjusted to match the celestial elevation of the target parent star SGL FL. The time of the perihelion passage is adjusted to provide the right ascension aim point of the target. In this way the pearls can be aimed at any target determined by the science community with no changes in the design or CONOPS. This allows for a development program and prototype flights to proceed well prior to the selection of a target – a major distinction from conventional explorations in which the target and mission must be clearly defined very early in the program planning phase.

At perihelion, the solar sail accelerates the sailcraft to ~150 km/sec and aims it at the selected parent star SGL rendezvous point and time. During the fly out to Jupiter, DSN data is used to further refine the trajectory, using the solar sails as the propulsion. The lower right side of the figure show the cruise phase (~20 years) during which the pearls "hibernate" except for occasional (annual) health and status transmissions to Earth (at bit rates ~100 b/sec to conserve power).

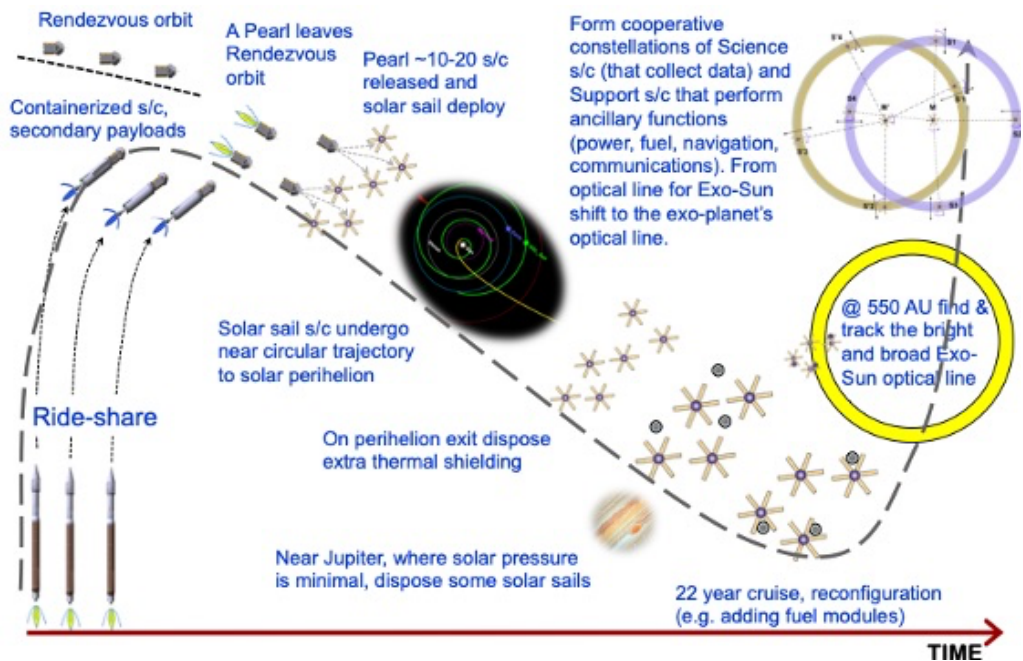

Figure 49. An overview of the SoP architecture for the SGL.

The upper right of Figure 39 shows the arrival of the pearls at the SGL focal region – first using the SGL of the parent star as a beacon, and then moving to the SGL of each of the candidate objects (planets, moons, etc.) of the solar system. Once there, the pearls follow the star SGL with small lateral course corrections due to motion in BCRS (<5 m/sec/year). Figure 39 shows the process of moving from the star to the planet SGL – made simple by the natural motion of the planet around the star, and by the geometric effects of the scale factor.

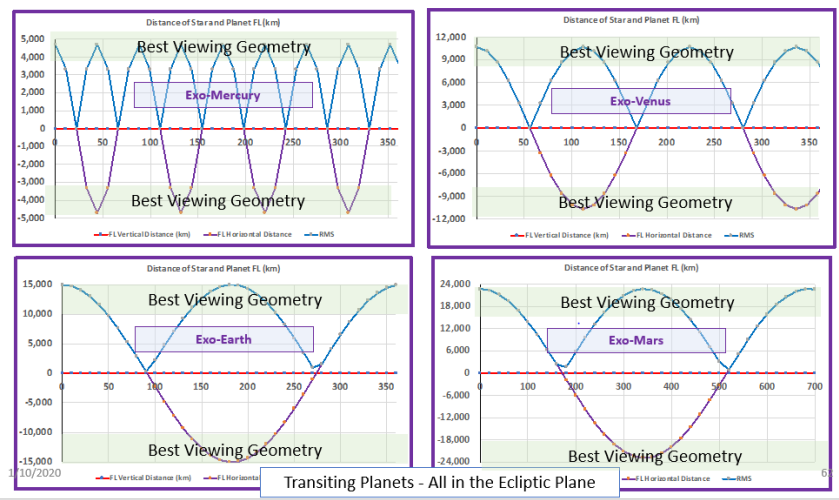

Figure 50. Repositioning of the sailcraft to explore other targets.

Figure 50 shows how a spacecraft flying along the SGL focal line can be repositioned to follow other objects. As there are many s/c in a pearl the CONOPS will be planned to optimize viewing by the ensemble of instruments, depending upon the position and illumination of various objects of interest. Thus, when a planet is in conjunction with or opposition to the star, it is not viewable, and viewing time could be employed looking elsewhere.

The significance of our work are its potential payoffs: exoplanet imaging, fastest s/c ever, ISO intercept, and a paradigm shift of carrying out the advanced outer solar system objectives with a swarm of interacting s/c which are suitable for investigating different targets and distributing technical requirements.

The concept uses smallsats with solar sail propulsion flying first to low perihelion and then rapidly on trajectories exiting the solar system (e.g. 25 AU/year). Requirements include low mass, small RTG power, micro-electric propulsion thrusters, optical communications all to fit into smallsat spacecraft and work reliably over decades of flight time. We propose to build a breadboard system with spacecraft components, solar sails with advanced materials as the first stage of development for a technology flight demonstration in 3-5 years. The breadboard will include functional simulation of communications, power, navigation and integration with the sail design.

Specific technologies or recent scientific developments:

1. Interplanetary smallsats, the premise of which was demonstrated in the recent success of JPL's MarCO; requirements associated with longer flight times, greater communications distances and higher power requirements.
2. Development of solar sails to achieve very fast trajectories to fly through and exit the solar system; sailcraft have been successfully demonstrated on the Japanese IKAROS mission, the ongoing Planetary Society LightSail-2 mission.
3. Observations, including imaging of KBOs, moons, ISOs with very fast flyby speeds under low mass, low power payload constraints.

Several mission concepts have been proposed to the far reaches of the solar system. Most of these missions have high propulsion demands, long flight times and require sophisticated autonomy, with new communications challenges, making them too expensive to be practical. The swarm of Light-Sailing smallsats overcomes this problem.

### 5.2 Low-cost, early technology demonstrations

Our Phase II effort opened a new paradigm to develop missions through the solar system that might enable exploration of distant regions of the solar system to occur decades sooner and faster than previously considered. The paradigm involves potentially multiple smallsats and ultrafast trajectories that create a series of increasingly ambitious missions moving us closer to the stars. In the next section candidate science missions with intermediate targets in the outer solar system are suggested. But the credibility and feasibility of our new paradigm rests on the near-term ability to prove smallsats suitable for solar system transit and sails to propel them to high velocity. Both are new capabilities.

Our preliminary design of the SunVane solar sailcraft has led us to conclude that we could build and launch a test spacecraft for $20-40 million. This is affordable both for a NASA technology test and potentially for a public-private partnership (see section 5.4) taking advantage of potential public interest in the interstellar precursor. A smallsat could be launched on a rideshare to high Earth orbit (e.g. GEO). There are readily available. From there, the solar sail could gain energy to enter heliocentric space and begin to spiral in toward the Sun. This would already be a meaningful test of the control and design of the sailcraft and the capabilities for interplanetary flight. Spiraling in for about a year to reach a perihelion <0.3 AU would test the sail control and sail materials for a flyby near and around the Sun. After that flyby the spacecraft would achieve a high

escape velocity from the solar system, e.g. 6-7 AU/year. This would already make it the fastest spacecraft to leave the solar system. Even if this mission flew for only one year, it would test the smallsat-sail design. Flying several such spacecraft (at little incremental increase in cost) would add the capability to test and learn on the swarm architecture. The test spacecraft would not have RTGs of course, but it could test the low power design we propose for the SGLF mission, as well as communications, navigation and control. Such a technology demonstration mission could pave the way for further capabilities and for science missions described in the next section.

For the technology test mission will examine new technologies and mission architectures for an SGLF mission. We will consider science opportunities and technology requirements using fast solar system trajectories like those required by a SGLF mission, including:

1. A test of the technologies with an inner solar system mission of moderate cost, necessary to enable smallsats to fly relatively quickly to the SGLF;
2. A mission that could quickly catch, intercept, and even rendezvous with an ISO as it passes through the solar system (Currently, an ISO is at its perihelion at ~2AU. With our mission idea, we could be ready to explore the next such visitor);
3. Fast flybys and observations of outer solar system small bodies, including moons of the outer planets, dwarf planets and KBOs;
4. A potential rendezvous with an ISO flying through the solar system. Of course, the ultimate, highest priority goal is to create a mission to the SGLF. If such a mission is proven practical, it will enable high-resolution observations of Earth-like, potentially habitable exoplanets. Our study will address the science strategies and evaluate possible missions.
5. Heliophysics observations of the solar corona (in the inner solar system) and of the interstellar medium (ISM) as it encounters our solar system medium; and
6. Hyperbolic trajectories to reach heliocentric distances >10 AU to enable measurements of (a) the extragalactic background light (EBL), (b) the cosmic IR background (CIB), (c) tests of the gravitational $1/r^2$ law up to 100 AU and beyond. Other test ideas exist.

Specific solutions to the technological challenges of s/c stability, control, comm and power are required to enable long-duration, long-distance outer solar system missions. Maturing these solutions in technology test missions will be addressed in our study

## 5.3 Science missions enabled by fast transit through the solar system

Science objectives in regions close to the Sun and then in the outer solar system will be considered. Since a goal of the technology test mission is to advance development of the SGLF mission, we will consider related science goals even though the test mission will not go to such distances. The technology test mission must advance the technologies for spacecraft operating over very long distances from Earth and for long times of flight. Science opportunities connected to the high velocity and fast flybys enabled by such a mission will be identified, including the rather unique goal of possible close-up observations of an ISO.

Several proposed concepts could significantly advance technology readiness level. For example, extremely high delta-V (8 AU/year) low cost, small spacecraft capability to rapidly investigate objects across the entire solar system in less than a decade is made possible thanks to L'Garde's innovative SunVane technology. A reconnaissance, rendezvous and sample return mission to the recently discovered transient interstellar objects (ISO) A/2017 U1 (Oumuamua) or C/2019 Q4 (Borisov) would best illustrate the strengths and capabilities of the solar sail technology and could serve as a training ground for the SGL campaign.

After the technology demo mission described in the previous section, further missions in the solar system might be considered – before or concurrent with the SGLF goal. The first goal might be a flyby of Sedna or another trans-Neptunian dwarf planet, to measure its mass and other properties characteristic of Kuiper Belt objects (KBO). The second goal would be a follow-up to the Voyager discoveries, by traveling through, and just beyond, the heliopause to study its complex environment and conduct heliophysics and astrophysical experiments.

Key technologies being developed to drive down weight risk and cost include solar sail materials, solar sail propulsion control, higher speed computers and rad-hard computers. Relevant developments in the next 10 years anticipate battery density (J/kg) to increase by factor of 2 to 4, removing about 5-7 kg of mass per SGL s/c. Onboard clocks can foresee a factor of 100 improvement in chip scale atomic clocks and 33 years for 1 Hz drift (0.1 ppm). Star trackers with high resolution data from Gaia mission should reach 1 microarcsecond angular resolution.

Millimeter-wave D-band RF antenna arrays could provide for efficient RF crosslink and lower SWaP (e.g. NuvoTronics). NASA's Inter-spacecraft omnidirectional optical communicator (ISOC), or Honeywell's Optical Pointing and Tracking Relay Assembly (OPTRAC) 10Gps optical could enable intersatellite communication. And for large-volume data transfer downlink communication NASA's Terabyte Infrared Delivery (TBIRD) Program could be utilized.

Upcoming commercial "mega constellations" will be a training ground for smallsat radiation-hardened electronics and autonomous systems. Commercial "mega constellation" development will also spawn new companies to move large amount of data. And new companies are developing advanced solar sail technology. Rideshare services can provide cost effective launch opportunities, e.g. Arianespace dedicated rideshare mission to GEO in 2022 (GO-1).

The prospect of getting an image of an exoplanet and to spectroscopically detect and characterize life being there is compelling. New coronagraph design, even if it is never used at SGL distances, would benefit the exoplanet search community. The mission strongly suggests the development an architecture that relies heavily on autonomous action, adaptability and the ability to "learn". These capabilities would benefit Mars missions or missions to the outer planets where comms are delayed. If this mission provided the indisputable spectroscopic proof of life on an exoplanet, then it would provide one of the most profound scientific discoveries ever in the history of humanity.

The results of this investigation could be catalytic. Our approach is radically different compared to existing ones arguing for instruments with ever larger aperture size, e.g. HST to JWST to LUVOIR to other 15-30 m diameter large space telescopes (ATLAST++). Because of its broad nature, any technology developed in support of the SGL mission will benefit the astronomical community. Our study could lead to novel designs of coronagraphs, high-precision navigation, autonomous, intelligent and adaptive systems, instruments designed to operate for decades. All these technologies are essential to other NASA missions. However, perhaps the more compelling benefit is how to implement missions to 500-1,000 AU over the next 40 years with the benefit of enabling investigations throughout our solar system, the Kuiper Belt region, and to the Oort cloud.

### 5.4 Towards the SGL mission implementation

We studied the modern generation of light-sailing smallsats as the means for an affordable standardized module for a disseminative approach to solar system exploration. We will consider an architecture consisting of a swarm of smallsats called the SoP that was developed at the Aerospace Corporation to provide the swarm's functionality as well as a robust and affordable architecture for deep space exploration.

The concurrent development of long-lived smallsats and solar sails suggests a fusion of these two technologies to create a new space-exploration building block: the light-sailing smallsats. This is a 3U to 6U smallsats equipped with a solar sail, capable of exploration anywhere in the solar system and beyond. Standardization and mass production drives down costs of the basic spacecraft and the operational costs of flight. Each could carry a hosted payload appropriate to the mission.

We are at a watershed of small satellite technology. Miniaturization of electronics, photonics and MEMS have enabled CubeSat class flight computers to be developed that are rad hard. Precision manufacturing has permitted the development of "micro" reaction wheels with design life of > 5 years and the fabrication of more efficient RF antennas (MarCo[25], Starlink[26]). Battery energy storage have increased significantly within the past 10 years and the development of new materials such as carbon fiber composites and its cousins now enable lighter satellite structures. Optical COMM has been demonstrated in space with smallsats along with miniaturized propulsion. Optical spectrometers, on a large chip, that rival their desktop size counterparts are a few years out.

Smallsats are no longer only a technology testbed but conduct real missions evidenced by the recent pioneering success of MarCo that provided "the communication link from Mars" and the 175+ nanosatellite Earth imaging constellation of Planet Inc[27]. We anticipate further miniaturization of satellite subsystems but of more importance is the increased reliability of smallsats. Solar sail technology has also advanced with demonstrations by JAXA, NASA (Sunjammer). The challenges to feasibility scales with the mass of the s/c. A solar sail surface area to mass ratio ($m^2$/kg) of 70:1 is viable today if the spacecraft mass is on the order of 10 kg but significantly less so when the mass is 100 kg.

The combination of long-lived smallsats and solar sails opens possibilities for rapid, affordable exploration of our solar system and beyond by a trajectory that spirals towards the Sun and accelerates outward at perihelion. All locations within the solar system become accessible including locations out of the ecliptic. Light-sailing smallsats could act as Discovery mission "scouts" to pave the way for more targeted but costlier medium and larger flagship missions. The 2003 NAS report (NAS, 2003) argues for a more balanced exploratory approach for solar system exploration and we believe the light-sailing smallsats to be a viable alternative and in need of further study.

The CONOPS of such a mission can be partially standardized. An ESPA-class shared booster deploys each s/c into cislunar space. The solar sail propulsion then spirals each s/c down to the perihelion where the solar sail accelerates the s/c onto the vector needed to rapidly reach the scientific target. Speeds of 10-25 AU/year appear reasonable. Target costs of $5-10M for an all up s/c plus payload and shared launch costs makes this approach financially attractive. Total time from concept to launch of 2 years makes the quest for scientific and technology experiments much more rapid than previous concepts. Moreover, calculations done at NASA/Marshall show mission elapsed times to visit near Earth asteroids, for example, vary from 1.5-3 years.

### 5.4.1 *Community development and public outreach*

Imaging of an exoplanet with the SGL brings a significant interest from young members of the science community and the public. This interest is both intellectual and technological. The pioneering spirit of the SGLF attracts young researchers and influences their professional interests.

We are working with a multidisciplinary group of engineers and scientists from various organizations to delve into the new approach of using smallsats for fast interplanetary missions. Specialists in s/c architecture, comm, control/stability, solar sail materials, mission and system analysis will interact with comet/asteroid scientists, heliophysics, exoplanet scientists, and pubic engagement specialists interested in the interstellar message possibility.

---

[25] https://www.jpl.nasa.gov/cubesat/missions/marco.php
[26] https://www.starlink.com/
[27] https://www.planet.com/company/

Technology steps to create the s/c capable of achieving the very high solar system exit velocities will be examined both for the applicability for interplanetary flight and for how they integrate with the longer-term vision of imaging a habitable world by enabling the SGLF mission.

An exciting element of our NIAC I/II efforts has been the enthusiasm with which the program has been received by the STEM community. This will be a significant feature of the NIAC III effort as the SGL mission is multigenerational by its very nature. Notable to date are the Texas A&M two semester student course project studying the engineering of a SGLF mission and the Caltech KISS workshops and lecture and student involvement.

### 5.4.2 Public-private partnership

Existing public interest is the reason for the development of our mission. On the technology frontier that enables the mission, the specific possibilities of interplanetary smallsats, low-cost exploration missions, fastest s/c ever, and connection to the ultimate life detection goal of exoplanet imaging are highly motivating to younger and newer members of the space science and engineering communities. Cornell Tech and The Planetary Society particularly, and the Breakthrough Foundation are also interested in the entrepreneurial aspects of a possible interstellar message and engaging "new space" interests.

Aerospace has been investigating novel ways towards securing the funding needed for the SGL mission. Rather than working towards a multibillion-dollar NASA investment, we are examining ways to organize a public-private partnership that would encourage many space faring entities and the public to participate. This would create a new way of funding long duration mission space programs that are of great interest to cultures worldwide. This affordability-driven approach is particularly appropriate for our distributed architecture in that it does not require billion dollar "chunks" for large complex s/c, big dedicated launch vehicles, costly technology development, or large ground-based mission management teams. So, the project will advance in pace with the breadth and scope of available funds from science and tech demos.

Most of the needed technology is already being funded by NASA, ESA, DARPA, USAF and international firms — for example the enormous investments in AI/machine learning and s/c miniaturization that are needed for SGL. Aerospace would manage the architectural standards and interfaces and assure protection of the participants' IP related to new technology. NASA would manage the program's science aspects.

In NIAC Phase II, we drove down mission costs by application of dual use technologies and planned prototype flights. There is great interest by Aerospace's military and other government customers in such dual use space technologies. USAF priority in increasing space investments was recently confirmed by the Nov 5-6, 2019 Space Pitch Day in which Dr. Roper (USAF assistant secretary for acquisition, technology and logistics) together with Aerospace CEO Isakowitz, and SMC commander Lt. Gnl. Thompson, offered several million dollars of awards to startups working on such technologies.[28] Dr. Roper declared, "We will do this every year, forever."

The creation of the Space Force will further accelerate space technology development, rapid prototype flights, and opportunities for rideshare missions. The Space Force focus on space software and AI, miniaturization, autonomous flights, satellite swarms, new sensors, minimal ground control, and low-cost launch is well aligned with the needs of the SGL. The SGL mission places a premium on reliability and long-term use. It would enable a new ecosystem of space systems design and manufacturing where "active systems" has the same longevity as passive systems.

---

[28] https://spacenews.com/air-force-awards-9-million-on-first-space-pitch-day-san-francisco/

## 5.5 Summary of Phase II results

Detection of signs of habitability via high-resolution imaging and spectroscopy of an exoplanet is the most exciting objective of a mission concept to the focal area of the SGL. The work during Phase I was directed at the development of the instrumentation and mission requirements, and also to study a representative set of mission architectures. Our Phase I study results demonstrate the feasibility of the SGL imaging mission, providing us with a solid foundation for this effort.

During Phase II, we explored the topics for a robust SGL mission, including refinement of the mission architectures by taking them through simulations and design trades. We have demonstrated that a mission to the SGL technically is challenging, but it is possible with most of the technologies being either already at hand or at a comfortably high level of maturity.

Our major results are summarized below:

1. We improved the knowledge of the optical properties of the SGL. Our analysis now includes a rigorous treatment of expended sources located at large but finite distance from the Sun. We investigated the process of imagining with the SGL for the faint sources in our galactic neighborhood. We investigated impact of the solar corona on the sensitivity of the imaging measurements with the SGL. We investigated direct deconvolution and its impact on the quality of the image reconstruction. We confirmed that a multipixel imaging with the SGL is possible.
2. We were able to identify key mission design drivers and developed a set of instrument and mission requirements. We studied mission requirements to deliver a spacecraft beyond 700 AU, to form an imaging system that could exploit the optical properties of the SGL. We identified the design trade parameters and considered several mission concepts, including single spacecraft, cluster of solar sail spacecraft, and cluster of mid-size spacecraft.
3. We investigated the technologies and mission architectures from the Phase I study in a finer detail to remove the unfeasible options. We considered: (i) optical comm with low mass, volume, and power. (ii) Smallsats with low power electric propulsion, (iii) mass producible small satellites in a cluster within a string of pearls (SoP) configuration compared to a single "flagship" craft. (iv) Gravity assisted trajectories, flight very close to the Sun, to achieve high solar system escape velocity and types of lightweight materials necessary for Sun protection, (v) Use of lightweight RTGs (studied in previous NIAC studies) enabling missions far away from the Sun and multifunctional solar sails, (vi) CONOPS for a single satellite, a cluster moving together, and a series of clusters in a SoP arrangement. (vii) CONOPS for tracking, identifying and positioning the sensor/s upon arrival to the SGL location.
4. We identified (i) technologies that pose the highest current risk with regards to reliability, (ii) mission processes with a need for autonomy and adaptability; (iii) navigation requirements for a craft in the image plane. (iv) For pointing, we studied the use of laser beacons to serve multiple functions (comm, guidance, navigation) for the SGL spacecraft.
5. We studied segmented mission architectures which break the journey into segments where i) control authority is provided by DSN to 200 AU, ii) periodic updates are provided by laser comm and astrometry of Jupiter, Saturn, and stars ~200-600 AU and beyond. We conducted trade studies to track not only power and mass but develop a utility or a value parameter, much like size, weight and power. Using such parameters, we refined the system requirements and identify the architecture trades. We studied preliminary design concepts and assess key mission, system, and operations technology drivers.

6. For the SoP architecture that showed the highest likelihood of success we developed mission models to investigate the tracking and imaging processes needed for acquiring high-resolution images, pixel-by-pixel. We explored potential failure modes and possible recovery approaches for the architecture. This approach allowed probing the spatial structure of the caustic as a function of time and has the potential to allow sampling, modeling, and removal of systematic errors due to possible radial, azimuthal, and temporal departures due to dynamical effects is the caustic's structure; it also using amplified star light for energy harvesting.

7. We investigated CONOPS for a spacecraft at the SGL for detecting, tracking, and studying the Einstein ring. We developed a CONOPS for acquiring the data & requirements for their onboard processing; same for onboard capabilities to learn by experience. We studied the image reconstruction and related attitude control requirements/options of the SGL spacecraft. We considered a 3-axis stabilized s/c with a few nanoarcsec pointing knowledge and stability. This capability is necessary to sample ~$10^5$ image pixels. We investigated impact from the dynamics is involved – proper motion of its parent star, orbital motion of the planet, and its rotation.

8. We studied the most promising mission architectures. We developed a model tracking key PNT parameters of the architecture to distinguish the different missions. We improved (i) flight system/science requirements; (ii) key mission, system, and operations concepts and technology drivers, (iii) description of mission & small-craft concepts with nav/system design to reach/operate at the SGL; (iv) study instruments & systems for the SGL including power, comm, nav, propulsion, pointing, and coronagraph. An output of our study is a set of recommendation on mission architecture with risk/return trades for single spacecraft vs a swarm.

9. We have identified a new interdisciplinary mission that may be proposed to expedite the development of the SGL mission. It will rely on the fastest ever spacecraft moving at velocity 20-30 AU per year! The mission relies on a set of innovative technologies that include solar sail designs with refractive film, low-power RTG and electric micro-thrusters, navigation and swarm architecture as well as radio navigation is feasible, communication will be optical.

We reached and exceeded all objectives set for our Phase II study: We developed a new wave-optical approach to study the imaging of exoplanets while treating them as extended, resolved, faint sources at large but finite distances. We designed coronagraph/spectrograph instruments to work with the SGL. We properly accounted for the solar corona brightness (Turyshev, & Toth, 2020c). We developed deconvolution algorithms; demonstrated feasibility of a high-quality image reconstruction. We identified the most effective observing scenarios and integration times.

As a result, we now are able to evaluate SNR for photon fluxes from realistic sources in the presence of the solar corona. We have proven that a multipixel imaging and spectroscopy of exoplanets up to 30 pc are feasible. By doing so, we were able to move the idea of practical applications of the SGL from a domain of theoretical physics to the mainstream of astronomy and astrophysics.

Our SGLF mission concept relies on a multiple spacecraft architecture deployed in a SoP configuration. To develop the CONOPS for this mission, we have applied analytical tools to its critical functions to demonstrate proof-of-concept feasibility. The derivation of the architecture was done by trade studies that utilized COMM link-budget tools (uplink/downlink via optical COMM (inter-pearl) and crosslink/intra-pearl COMM via RF and LAN) developed for space systems.

We used physics-based analytical tools to define the trajectories towards, about, and outward from the sun towards the SGL. Solar sail propulsion brings each smallsat to perihelion via an in spiral trajectory from Earth, accelerates the s/c towards the SGLF target, and is used to remove residual injection errors (done via the NASA's DSN) during exit from the inner solar system.

Each pearl is targeted to and aligned with the SGL that will exist when the pearl arrives at the focal point (>548 AU). Once the solar sails are no longer useful, they are ejected to reduce s/c weight. Subsequently the $\Delta V$ requirements (~315 m/s including barycentric motion) are provided by onboard propulsion that must be highly efficient, long-lived. We have identified commercial entities that have applicable technology (some at TRL9) that would be adapted for the SGL mission.

We addressed the PNT requirements by extending the RF DSN to include distributed onboard star-trackers and X-ray pulsars. The design of the s/c utilized a concurrent engineering methodology tool. The analysis went through 4 different constructs resulting with s/c units of ~30 kg in CBE mass (+15% contingency), a solar sail of 400 $m^2$/kg ratio and distributed satellite functionality where the downlink, science and PNT functions are distributed among the s/c within the pearl.

Two forms of power system designs have been evaluated (RTG only and RTG+ rad hard battery). The tradeoffs show that the RTG+rad hard battery offer the best option, given the continued development of rad-hard batteries. Analysis have also been done on the self-induced contamination and the effects on critical sensors given the ~50-year mission. We have established a technology roadmap for the evolution of the SW/HW needed for onboard computation. The analysis includes the AI/machine learning type modalities needed to accomplish the SGL mission. Finally, we have identified a list of the technology areas where improvement would further reduce the mission risk.

We concluded that most of the technologies for SGLF mission either already exist (rideshare/cluster launch, sailcraft, RF/optical comm, all at TRL9), or are in intermediate levels of readiness. Sail materials (TRL 4), thermal management in solar proximity (TRL 7), swarm operations (TRL 5), terabit onboard processing (FPGA/GPU, TRL 9/7), CONOPS (TRL 7). What is missing is the system approach in assembling all these technologies for autonomous operations in space (TRL 3). We have a solid plan to close this gap in the near future, maturing SGLF concept to TRL 4-5.

## 5.6 Towards a realistic SGL mission concept

As a result of our Phase II effort, the knowledge of the physical properties of the SGL much evolved. We have developed a much better analytical models for imaging of extended sources with the SGL. Our models now confirmed with numerical simulations. The solar corona is now fully accounted for in the SNR analysis. We have studied many scenarios of image reconstruction with the SGL. As a result, we have an improved set of mission-relevant parameters that allow us to study the i) detection sensitivity, ii) instrument size and performance, iii) the per-pixel integration time; iv) duration of imaging mission phase, v) impact of a number of spacecraft, as well as the vi) needed navigational precision. We were able to formulate mission requirements to deliver a spacecraft beyond 700 AU, to form an imaging system that could rely on the unique optical properties of the SGL to image distant faint targets. Based on this work we have investigated several possible mission architectures. Specially, we considered: i) single large spacecraft, ii) cluster of mid-size spacecraft, and iii) cluster of solar sail spacecraft. Our chosen architecture is that offered by the "string of pearls" approach which is based on solar sail propulsion. We have studied the CONOPS for spacecraft at the SGL to detect, track, and study the Einstein ring.

To build a $10^3 \times 10^3$-pixel image, we need to sample it in a pixel-by-pixel fashion, while moving with resolution of ~1 m. This can be achieved relying on a combination of inertial navigation and laser beacons s/c placed in 1 AU solar orbit whose orbital plane is co-planar to the image plane.

A mission to the SGL is challenging, but not impossible. Given the current state-of-the-art, several technologies may enable a meaningful step beyond our solar system to distances of 550–1,000 AU in 25–35 yrs (Stone et al. 2015; Alkalai et al 2017). We have considered a long-term technology development program with the following topics: i) Mission and trajectory design needed to achieve high escape velocity and shorter flight time, but also with small orbit injection errors; ii) Propulsion systems, such as nuclear-thermal, nuclear fusion, nuclear fission, solar thermal propulsion, laser-beamed energy, laser ablation, solar sails, electric sails and more; iii) Power systems, including nuclear power; v) Structures, such as lightweight multifunctional structures, deployable structures, etc. vi) Thermal design and stability, low-power, low-temperature systems, etc. vi) Telecommunications systems utilizing both RF and optical communications; viii) GNC, including spacecraft stability, pointing to Earth. ix) Avionics systems to support long term survivability and autonomous operations; x) Instruments and payload, including highly miniaturized solutions.

Following the miniaturizing of electronics, there have also been recent strides in miniaturizing spacecraft to include complex sensors for situational awareness and embedded analytical computation capabilities which imbues them with autonomous-like properties. Current space materials have also evolved from relying on heavy metal structures to the lighter and stiffer carbon composite material. Computer algorithms now exist that can adapt or "learn" from prior 'experience". These technologies enable adaptability in the SGL mission to ensure success.

Based on the work in Phase II we concluded that that are no major showstoppers for a mission to the SGLF. Imaging with the SGL is challenging, but feasible. We have developed a Technology Roadmap and a set of flight demonstrations in order to implement a mission to the SGL by 2032.

## 5.7 Benefits of the Investigation

We witness a progress in exoplanet research: by detecting a plethora of potential Earth-like exoplanets, Kepler has placed the possibility that another Earth-like world exists into the public consciousness. Follow-ups on Kepler candidates with other current exoplanet characterization technologies may yield unresolved images at low spectral resolution (typically $R < 100$). The next steps include TESS (2017), which will extend Kepler's work by performing an all-sky survey to identify additional exoplanet candidates, including Earth-like planets; JWST (2018), which will be used for targeted follow-up on candidate planets; and missions in formulation, such as the Exo-C (2015), Exo-S (2015) and LUVOIR (2015) concepts.

There is no concept for direct multipixel imaging of an exoplanet. All the exoplanet imaging concepts currently studied by NASA, will attempt to capture the light of an unresolved Earth-like exoplanet as a single pixel. The limiting factor is contamination from the parent star at a distance of ~0.1" from the planet. Even WFIRST with its telescope of 2.4-m is looking is only for Jupiter-like exoplanets at 10 pc. An SLGF mission opens up a unique possibility of direct high-resolution imaging and spectroscope of a habitable Earth-like exoplanet, not technically feasible otherwise.

The prospect of getting an image of an exoplanet and to spectroscopically detect and characterize life being there is compelling. New coronagraph design, even if it is never used at SGL distances, would benefit the exoplanet search community. The mission strongly suggests the development an architecture that relies heavily on autonomous action, adaptability and the ability to "learn". These capabilities would benefit Mars missions or other missions to the outer planets where communications are delayed. If this mission provided the indisputable spectroscopic proof of life on an exoplanet, then it would provide one of the most provocative pieces of scientific discovery ever!

The results of this investigation could be catalytic. Our approach is radically different compared to existing ones arguing for instruments with ever larger aperture size, e.g. HST to JWST to LUVOIR to other 15-30 m diameter large space telescopes (e.g., ATLAST). Because of its broad nature, any technology developed in support of the SGL mission will benefit the astronomical community. Our study could lead to novel designs of coronagraphs, high-precision navigation, autonomous, intelligent and adaptive systems, instruments designed to operate for decades. All these technologies are essential to other NASA missions. However, perhaps the more compelling benefit is how to implement missions to 500-1,000 AU over the next 40 years with the benefit of enabling investigations throughout our solar system, the Kuiper Belt region, and to the Oort cloud.

We have developed a new mission concept that delivers an array of optical telescopes to the SGL's focal region and then flies along the focal line to produce high resolution, multispectral images of a potentially habitable exoplanet. Our multiple smallsat architecture is designed to perform concurrent observation of multiple planets and moons in a target exosolar system. It allows to reduce integration time, to account for target's temporal variability, to "remove the cloud cover".

The SGLF CONOPS uses multiple small satellites in an innovative SoP architecture where a pearl consisting of an ensemble of smallsats is periodically launched. As a series of such pearls are launched (to form the "string") they provide the needed comm relays, observational redundancy and data management needed to perform the mission. For example, if pearls are launched annually, then they will fly outwards towards and then along the SGL at 20 AU intervals. By employing smallsats using AI technologies to operate interdependently, we build in mission flexibility, reduce risk, and drive down mission cost. This makes possible concurrent investigations of multiple exosolar systems by launching strings towards multiple exoplanet candidates.

This affordable architecture design reduces cost in many ways (Alkalai et al., 2017, Friedman & Turyshev, 2018, Turyshev et al., 2018): 1) It cuts the cost of each participant by enabling multiple players broad choices of funding/building/deploying/operating/analyzing system elements at their choice. 2) It delivers economy of scale in an open architecture designed for mass production to minimize recurring costs. 3) It drives down the total mass (and thereby both NRE and recurring costs) by using smallsats. 4) It uses realistic-sized solar arrays (~16 vanes of $10^3$ m$^2$) to achieve high velocity at perihelion (~150 km/sec). 5) It applies maturing AI technologies to allow virtually autonomous mission execution eliminating the need for operator-intensive mission management, (6) It reduces launch costs by relying on "rideshare" opportunities to launch the smallsats, avoiding the costs of large dedicated launchers (Friedman & Turyshev, 2019).

Other Innovations: To reliably accomplish a 50-year mission (~650–1,000 AU) with the needed reliability, the smallsats use AI/machine learning to flexibly distribute (and redistribute as needed) functionality among the s/c. By launching over a span of a decade, the designers will improve successive launches as technology improves and operational lessons are learned. The ability to "pass on" information, data and operational experience from the early arrivals to those following will constantly improve the CONOPS and the value of the science from many uses of the SGL.

Our concept enables entirely new missions, providing a great leap in capabilities for NASA and the greater aerospace community. It lays the foundation for fast transit (>20 AU/yr) and exploration of our solar system and beyond (outer planets, moons, KBOs, and interstellar objects/comets).

## 5.8   Anticipated paths after Phase II

Currently, there is interest in missions outside the solar system with many ongoing efforts (Alkalai et al. 2017). There is also strong public support for interstellar precursor missions (Friedman & Turyshev, 2017). The recognition that high-resolution imaging of an exoplanet may not be possible without the SGL is a sobering thought and requires placing more efforts on the study of the SGL mission. If this work proves the mission feasible, through the articulation of realistic engineering and scientific principles, then additional effort will arise from the community. We plan to motivate the science community by attending and hosting meetings, presenting position papers to the National Academy of Sciences and other scientific bodies.

Our Phase I work (Turyshev et al., 2018) was a joint effort of JPL, The Aerospace Corporation, three universities. NASA identified several missions to the very deep regions outside the solar system that emphasize the science objectives of exploring KBOs and imaging of exoplanets with the SGL. Relevant technology efforts are ongoing in NASA/JPL, MSFC, KSFC.

The Phase II effort provided us with a much deeper understating of the entire SGL mission design envelope and related technology requirements. With no showstoppers found, a mission to the SGL emerges as a complex, challenging, but a very exciting endeavor for the next several decades. This mission has the potential to serve as a major motivation to re-energize and focus deep space exploration efforts, relying on a significant public support.

Imaging of extrasolar terrestrial planets combined with spectroscopy is probably the single greatest remote sensing event one can contemplate in terms of galvanizing public interest and government expenditure for interstellar missions. Our work on the SGLF imaging provides a capability with a dramatically low cost and short schedule to see exolife not possible otherwise. It generates enthusiasm for a mission and potential for support within NASA and the greater aerospace community.

We consider a set of science missions intended to demonstrate critical technologies that could be undertaken by NASA and/or industry programs. The concept of clustered nanosatellites that form the pearl could be used to establish a set of warning sensors about the sun for harmful corona flares. This could be expanded to establish a near-permanent constellation of solar sailcraft around the sun to conduct heliophysics missions and to serve as a planetary defense warning system for all Earth space assets. A second test mission could provide rapid transit "probes" to a solar system target, extending out to the KBOs/dwarf planets. These would be low-cost flybys with a primary sensor on board. With solar sails and sun perihelion exit velocities of 20 AU/yr, Jupiter is reached in 3 months, Saturn in 6 months, Pluto in <2 years and the outer edge of the Kuiper belt in 3 years.

The efforts align with our roadmap for the SGLF capability build-up that we developed during Phase II. This roadmap has several critical steps. 2020-21: NIAC Phase III develops SGLF design/cost for and validates critical technology; 2021: SGLF mission concept & technology presented to industry; 2022: NASA-Aerospace form the nucleus of the public-private partnership to fund the SGLF program; 2023-4: Sailcraft flights (<$20M) to achieve TRL 9; 2026-8: Sun-accel. flights (10 AU/yr): confirm CONOPS; 2027: SGLF Project starts for a preselected target; 2032-42: Launch SoP (20 AU/yr) to the target; 2060: Discover life beyond the Earth/solar system. This roadmap is realistic; in < 40 years it could deliver results of tremendous value to our civilization.

In light of the profound significance of finding life on an exoplanet, we will work with private industry partners: Breakthrough, Blue Origin, Xplore, NXTRAC, NovaWurks, already involved in our efforts. Our primary partner for the SGLF mission will be NASA, where each SMD division (i.e., heliophysics, astrophysics, planetary) has its own interest in the mission. We are working with NASA leadership and the ongoing NAS Decadals to include SGL in their science strategy.

Phases I and II of our NIAC Study of exoplanet imaging with the SGL have made three remarkable innovations: (1) proven the feasibility of high-resolution, multipixel imaging of an habitable exoplanet; (2) devised a swarm architecture for smallsats to explore the interstellar medium; (3) designed a low-cost sailcraft to achieve the highest exit velocity from the solar system yet achieved. These innovations are "out-of-the-box", counter to the usual ideas for exoplanet imaging and for spacecraft to explore the nether regions of our solar system. To put them "into the box" of science mission development, a thorough system and mission design study must be presented to the NASA SMD and a compelling demonstration of technology readiness must be carried out by the NASA.

With our mission, a high-resolution exoplanet imaging could be achieved in the 21st century, and interstellar medium exploration will stop being a multibillion-dollar flagship dream, but instead it will become a flexible, low-cost, affordable SoP concept going faster and further into deep space.

*Acknowledgments:*

We would like to express our gratitude to our many colleagues who have either collaborated with us on this report or given us their wisdom. We specifically thank Roy Nakagawa, Nathan Strange, Phil A. Willems, John L. West, Stacy Weinstein-Weiss, who provided us with very valuable comments, encouragement, support and stimulating discussions while this report was in preparation.

We acknowledge fruitful collaborations with several our NIAC Fellows, especially with Stephanie Thomas, John Brophy, Philip Lubin, Robert Adams and Nickolas Solomey.

Several on-going collaborations with our industry partners, non-profit and academia contributed to our work, namely with NXTRAC, Explore, L'Garde, Firefly, Morpheus Space, Zeno Power Systems, Breakthrough StarShot, Space Venture Coalition, Texas A&M University, UCLA, Cornell Tech, MIT, University of Colorado in Colorado Springs, as well as NASA Ames Research Center, NASA Marshall Space Flight Center, and others.

Finally, we are especially thankful to NASA Innovative Advanced Concepts (NIAC) for their support for our work in the pursuit of an exciting objective – using the SGL for direct high-resolution imaging and spectroscopy of a potentially habitable exoplanet.

The work described here was carried out at the Jet Propulsion Laboratory, California Institute of Technology, under a contract with the National Aeronautics and Space Administration.

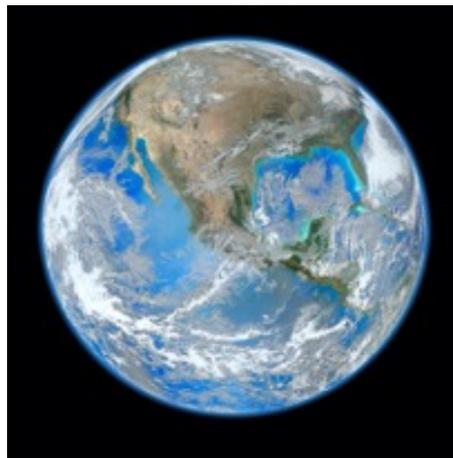

An image of our Earth in physical colors.
A mission to the focal region of the SGL may yield a similar image an exo-Earth.
(Original image credit: NASA/NOAA/GSFC/Suomi NPP/VIIRS/Norman Kuring.)

No major showstoppers have been identified during our NIAC Phase II effort.
Conclusion: Imaging with the SGL is challenging, but feasible.

# 6 References and Citations

# 7 Publications generated during the Phase II effort

We thank NIAC for the unique opportunity to study a tremendously exciting topic of direct multipixel imaging and spectroscopy of exoplanets with the SGL. Below, we provide the list of publications that were developed during our Phase I/II efforts:

## 7.1 Technical papers on the physics of imaging with the SGL

[1] Slava G. Turyshev, "Wave-theoretical description of the solar gravitational lens," Phys. Rev. D 95, 084041 (2017), arXiv:1703.05783 [gr-qc]

[2] Slava G. Turyshev, Viktor T. Toth, "Diffraction of electromagnetic waves in the gravitational field of the Sun," Phys. Rev. D96, 024008 (2017), arXiv:1704.06824 [gr-qc]

[3] Slava G. Turyshev, Viktor T. Toth, "Wave-optical treatment of the shadow cast by a large sphere," Phys. Rev. A 97(3), 033810 (2018), arXiv:1801.06253 [physics.optics]

[4] Slava G. Turyshev, Viktor T. Toth, "Wave-optical treatment of the shadow cast by a large gravitating sphere," Phys. Rev. D98, 104015 (2018), arXiv:1805.10581 [gr-qc]

[5] Slava G. Turyshev, Viktor T. Toth, "Diffraction of light by the gravitational field of the Sun and the solar corona," Phys. Rev. D 99, 024044 (2019), arXiv:1810.06627 [gr-qc]

[6] Slava G. Turyshev, Viktor T. Toth, "Optical properties of the solar gravitational lens in the presence of the solar corona," Eur. Phys. J. Plus 134: 63 (2019), arXiv:1811.06515 [gr-qc]

[7] Slava G. Turyshev, Viktor T. Toth, "Scattering of light by plasma in the solar system," Journal of Optics 21(4), 045601 (2019), arXiv:1805.00398 [physics.optics]

[8] Slava G. Turyshev, Michael Shao, Viktor T. Toth, "Putting gravity to work: Imaging of exoplanets with the Solar Gravitational Lens," IJMPD 28, 1950125 (2019).

[9] Slava G. Turyshev, Viktor T. Toth, "Imaging extended sources with the solar gravitational lens," Phys. Rev. D100, 084018 (2019), arXiv:1908.01948 [gr-qc]

[10] Slava G. Turyshev, Viktor T. Toth, "Photometric imaging with the solar gravitational lens," Phys. Rev. D 101, 044025 (2020), arXiv:1909.03116 [gr-qc]

[11] Slava G. Turyshev, Viktor T. Toth, "Image formation process with the solar gravitational lens," Phys. Rev. D 101, 044048 (2020), arXiv:1911.03260 [gr-qc]

[12] Slava G. Turyshev, Viktor T. Toth, "Image formation for extended sources with the solar gravitational lens," submitted, Phys. Rev. D, (February 2020), arXiv:1911.03260 [gr-qc].

[13] Viktor T. Toth, Slava G. Turyshev, "Image deconvolution with the solar gravitational lens", work in progress, to be submitted, Phys. Rev. D, (2020).

## 7.2 Papers and reports on the SGLF mission design

[14] Louis Friedman, Slava G. Turyshev, "Finding Earth 2.0 from the Focus of the Solar Gravitational Lens," 100 Years StarShip (100YSS), essay, a winner of the inaugural Canopus award, (2015), http://100yss.org/news/press.

[15] Travis Brashears, Philip Lubin, Slava Turyshev, Michael Shao, Qicheng Zhang, "Solar lens mission concept for interstellar exploration," Proc. SPIE 9616, Nanophotonics and Macrophotonics for Space Environments IX, 96160A (2015); doi:10.1117/12.2189019

[16] L. Alkalai, N. Arora, M. Shao, S. Turyshev, L. Friedman, P. C. Brandt, R. McNutt, G. Hallinan, R. Mewaldt, J. Bock, M. Brown, J. McGuire, A. Biswas, P. Liewer, N. Murphy, M. Desai, D. McComas, M. Opher, E. Stone, G. Zank, "Mission to the solar gravity focus: Natural High-ground for Imaging Earth-like exoplanets," NASA's Planetary Science Vision (PSV) 2050 Workshop, NASA HQ, Washington, DC on Feb 27- Mar 1, 2017.

[17] Leon Alkalai, Nitin Arora, Slava Turyshev, Michael Shao, Stacy Weinstein-Weiss, Merav Opher, and Seth Redfield, "A Vision for Planetary and Exoplanets Science: Exploration of the Interstellar Medium - The Space Between Stars", IAC-17-D4.4.1x41640, 68th International Astronautical Congress, Adelaide, Australia (2017).

[18] Slava G. Turyshev, Michael Shao, Leon Alkalai, Nitin Aurora, Darren Garber, Henry Helvajian, Tom Heinsheimer, Siegfried Janson, Jared R. Males, Dmitri Mawet, Roy Nakagawa, Seth Redfield, Janice

### 7.3 Science community engagement

White papers for the National Academy of Sciences (NAS) Astrophysics Decadal Astro2020:

# 8 Appendix: List of Acronyms

List of Acronyms used in the Proposal:

| | |
|---:|---|
| A/m | Area-to-mass ratio |
| ATLAST | Advanced Technology Large Aperture Space Telescope, http://www.stsci.edu/atlast |
| AU | Astronomical unit |
| BCRS | Barycentric Coordinate Reference System |
| CONOPS | Concept of operations |
| Exo-C | A probe-scale space mission to directly image and spectroscopically characterize exoplanetary systems using an internal coronagraph: https://exoplanets.nasa.gov/exep/about/exoc/ |
| Exo-S | A probe-class exoplanet direct imaging mission concept relying on an external starshade: https://exoplanets.nasa.gov/exep/about/exos/ |
| FL | Focal line of the SGL |
| FWHM | full-width half maximum |
| JPL | Jet Propulsion Laboratory |
| JWST | James Web Space Telescope, https://www.jwst.nasa.gov/ |
| LUVOIR | Large UV Optical Infrared Surveyor, https://asd.gsfc.nasa.gov/luvoir/ |
| ly | light year |
| mag | stellar magnitude |
| mas/uas/nas | milli-arcsecond / micro-arcsecond / nano-arcsecond |
| NASA | National Aeronautics and Space Administration |
| PSF | point-spread function |
| ps | parsec |
| RF | Radio frequency |
| RTG | Radioisotope Thermal Generator |
| s/c | spacecraft |
| SME | Subject matter expert |
| SGL | Solar Gravitational Lens |
| SGLF | Solar Gravitational Lens Focal region; also refers to a mission to the SGL's focal region |
| SNR | signal-to-noise ratio |
| STP | solar-thermal propulsion |
| TESS | Transiting Exoplanet Survey Satellite, https://tess.gsfc.nasa.gov/ |
| TPF | Terrestrial Planet Finder mission, https://en.wikipedia.org/wiki/Terrestrial_Planet_Finder |
| TRL | Technology Readiness Level |